\documentclass[12pt]{article} 
\usepackage{epsfig,cite}
\usepackage{amsmath}
\usepackage{amssymb} 
\usepackage{color}
\usepackage{graphicx}
\usepackage{verbatim,mathrsfs,subfigure,feynmf}
\usepackage{url}
\usepackage{ytableau} 
\include{epsf}
\newcount\timecount
\newcount\hours \newcount\minutes  \newcount\temp \newcount\pmhours
\hours = \time
\divide\hours by 60
\temp = \hours
\multiply\temp by 60
\minutes = \time
\advance\minutes by -\temp
\def\hour{\the\hours}
\def\minute{\ifnum\minutes<10 0\the\minutes
            \else\the\minutes\fi}
\def\clock{
\ifnum\hours=0 12:\minute\ AM
\else\ifnum\hours<12 \hour:\minute\ AM
      \else\ifnum\hours=12 12:\minute\ PM
            \else\ifnum\hours>12
                 \pmhours=\hours
                 \advance\pmhours by -12
                 \the\pmhours:\minute\ PM
                 \fi
            \fi
      \fi
\fi
}

\def\monthname{\relax\ifcase\month 0/\or January\or February\or
   March\or April\or May\or June\or July\or August\or September\or
   October\or November\or December\else\number\month/\fi}

\def\bold#1{\setbox0=\hbox{$#1$}%
     \kern-.025em\copy0\kern-\wd0
     \kern.05em\copy0\kern-\wd0
     \kern-.025em\raise.0433em\box0 }

\def\beq{\begin{equation}}
\def\eeq{\end{equation}}

\def\ga{\mathrel{\raise.3ex\hbox{$>$\kern-.75em\lower1ex\hbox{$\sim$}}}}
\def\la{\mathrel{\raise.3ex\hbox{$<$\kern-.75em\lower1ex\hbox{$\sim$}}}}
\def\gev{{\rm \, Ge\kern-0.125em V}}
\def\tev{{\rm \, Te\kern-0.125em V}}
\def\gyr{{\rm \, G\kern-0.125em yr}}

\def\slash#1{\rlap{\hbox{$\mskip 1 mu /$}}#1}%

\def\gappeq{\mathrel{\rlap {\raise.5ex\hbox{$>$}}
{\lower.5ex\hbox{$\sim$}}}}
\def\lappeq{\mathrel{\rlap{\raise.5ex\hbox{$<$}}
{\lower.5ex\hbox{$\sim$}}}}
\def\Toprel#1\over#2{\mathrel{\mathop{#2}\limits^{#1}}}

\def\m12{m_{1\!/2}}

\def\bea{\begin{eqnarray}}
\def\eea{\end{eqnarray}}

\makeatletter 
\def\slash{\@ifnextchar[{\fmsl@sh}{\fmsl@sh[0mu]}} 
\def\fmsl@sh[#1]#2{%
  \mathchoice 
    {\@fmsl@sh\displaystyle{#1}{#2}}%
    {\@fmsl@sh\textstyle{#1}{#2}}%
    {\@fmsl@sh\scriptstyle{#1}{#2}}%
    {\@fmsl@sh\scriptscriptstyle{#1}{#2}}} 
\def\@fmsl@sh#1#2#3{\m@th\ooalign{$\hfil#1\mkern#2/\hfil$\crcr$#1#3$}} 
\makeatother 

\usepackage{lscape}

\def\beq{\begin{equation}}
\def\eeq{\end{equation}}

\begin{document}\begin{titlepage}
\pagestyle{empty}
\begin{flushright}
{\tt KCL-PH-TH/2022-52}, {\tt CERN-TH-2022-172}  \\
{\tt UMN-TH-4204/22, FTPI-MINN-22/29} \\
\end{flushright}

\begin{center}{\bf \large{The CMSSM Survives {\it Planck}, 
the LHC, LUX-ZEPLIN, \\
\vspace{2mm} {\it Fermi}\,-LAT, H.E.S.S. and IceCube} }\\
\vskip 0.2in
{\bf John~Ellis}$^{a}$,
{\bf Keith~A.~Olive}$^{b}$,
{\bf Vassilis~C.~Spanos}$^{c}$  and
{\bf Ioanna~D.~Stamou}$^{d}$
\vskip 0.2in
{\small
{\em $^a$Theoretical Particle Physics and Cosmology Group, Department of
  Physics, King's~College~London, London WC2R 2LS, United Kingdom;\\
Theoretical Physics Department, CERN, CH-1211 Geneva 23,
  Switzerland;\\
National Institute of Chemical Physics and Biophysics, R\"{a}vala 10, 10143 Tallinn, Estonia}\\[0.2cm]
  {\em $^b$William I. Fine Theoretical Physics Institute, School of
 Physics and Astronomy,\\ University of Minnesota, Minneapolis, MN 55455,
 USA}\\[0.2cm]
{\em $^c$Section of Nuclear and Particle Physics, Department of Physics, \\
National and Kapodistrian University of Athens, 
   GR-157 84 Athens, Greece}\\[0.2cm] 
 {\em $^d$Service de Physique Th{\'e}orique, Universit\'e Libre de Bruxelles,
Boulevard du Triomphe CP225, B-1050 Brussels, Belgium}
}

\vspace{1.5cm}
{\bf Abstract}
\end{center}
\baselineskip=18pt \noindent
{\small
We revisit the viability of the CMSSM, searching for regions of parameter space that
yield a neutralino dark matter density compatible 
with {\it Planck} measurements, as well as LHC constraints including sparticle searches and the
mass of the Higgs boson, recent direct limits on spin-independent and -dependent
dark matter scattering from the LUX-ZEPLIN (LZ) experiment, the indirect constraints from
{\it Fermi}-LAT and H.E.S.S. on dark matter annihilations
to photons in dwarf spheroidal galaxies and the Galactic Centre,
and the IceCube limits on muons from annihilations to neutrinos in the Sun.
For representative values of $\tan \beta$ and $A_0$
we map in detail the {\it Planck}-compatible strips in CMSSM parameter
planes, which exhibit multiple distinctive features for large $\tan \beta$, $A_0 = 0$ and $\mu > 0$,
and identify portions of the strips that survive all the phenomenological constraints. 
We find that the most powerful
constraint is that from $m_h$, followed by the LZ limit on spin-independent scattering, whereas
sparticle searches at the LHC and indirect dark matter
searches are less restrictive. Most of the surviving CMSSM parameter space 
features a Higgsino-like dark matter particle with a mass $\sim 1000 - 1100$~GeV, 
which could best be probed with future direct searches for dark matter
scattering.
}


\vfill
\leftline{October   2022}
\end{titlepage}

\section{Introduction}
\label{sec:intro}

The Constrained Supersymmetric  extension of the Standard Model (CMSSM)
is a commonly-used template for supersymmetry phenomenology \cite{DN,cmssm,interplay,Ellis:2015rya,Ellis:2018jyl,Ellis:2019fwf}, which incorporates
the (over-?) simplified assumption that the soft supersymmetry-breaking
parameters are universal at the gauge coupling unification scale $\sim 10^{16}$~GeV.
The CMSSM has often been used, for example, to interpret supersymmetry searches
at the LHC and both direct and indirect searches for astrophysical dark matter.

The continuing lack of success in accelerator searches for supersymmetric
particles \cite{nosusy} as well as searches for massive astrophysical dark matter particles
\cite{XENON,LUX,PANDAX,LZ}
has been exerting ever-increasing pressure on supersymmetric models in general
and the CMSSM in particular. The low dimensionality of the CMSSM parameter space
enforces many links between supersymmetric observables and has relatively few
unexplored corners. Under these circumstances, an ever more pressing question is
whether the CMSSM survives all the phenomenological constraints. If it does, there
is likely to be an ever-decreasing region of CMSSM parameter space where
experimental searches should focus, which may be interesting also for more
general supersymmetric models. If the CMSSM does not survive, the time has come 
to move on, either to more general supersymmetric models or to other scenarios 
for dark matter and complementary possibilities for physics beyond the Standard Model.

Our philosophy in this paper is to regard the CMSSM as the canary in the
supersymmetric coalmine, whose survival is indicative of the general state of
health of supersymmetric phenomenology. As we shall see, the CMSSM continues
to survive the experimental onslaught in a restricted range of parameters that
is quite vulnerable to upcoming direct searches for astrophysical dark matter.

The following are the phenomenological constraints that we consider in this paper.

The most restrictive is the constraint on the density of dark matter provided by the
{\it Planck} satellite and other measurements, which has percent-level precision \cite{Planck}. Our calculation of the relic density assumes standard thermal freeze-out.  However,
it is subject to at least two important caveats. One is that the supersymmetric
relic density will in general be changed if the expansion history is modified or if the late decay of a massive particle (which comes to dominate the energy density) adds entropy to the radiation bath after freeze-out \cite{Gelmini:2006pw}
and the other is that there may be other sources of dark matter, in which case the
supersymmetric relic density may be less than the total dark matter density
indicated by {\it Planck} and other experiments. Nevertheless, we take the {\it Planck} dark matter density
constraint at face value, commenting later on the 
extension of our results to the possibility that the LSP
contributes only a fraction of the cold dark matter density.

The second most stringent constraint is provided by the mass of the Higgs boson, 
which has been measured at the permille level \cite{Aad:2012tfa}. Theoretical calculations of $m_h$
within supersymmetric models have estimated accuracies at the percent level.
In our analysis we use calculations with the {\tt FeynHiggs~2.18.1} code \cite{FH},
taking into account their estimated uncertainties. 
As we shall see, the $m_h$ constraint
is quite complementary to the {\it Planck} constraint.

Another set of constraints is provided by LHC searches for supersymmetric particles,
which yield (so far) only lower limits on sparticle masses \cite{nosusy} 
and hence the CMSSM
supersymm- etry-breaking parameters $m_0$ (the supposedly universal scalar mass at the
unification scale) and $m_{1/2}$ (the supposedly universal gaugino mass). In principle,
these limits depend somewhat on the values of $\tan \beta$ (the ratio of supersymmetric
Higgs vev's) and $A_0$ (the supposedly universal soft trilinear supersymmetry-breaking 
parameter). However, we find that the LHC constraints are not relevant in the regions
of CMSSM parameter space favoured by the measurement of $m_h$, implying that a detailed
implementation is not necessary.

We stress that our motivation for studying the CMSSM is the simplicity embodied in the assumptions on scalar and gaugino mass universality. Of course there are many more complicated extensions of this paradigm, such as models in which the Higgs masses differ from the scalar masses at the unification scale, which are commonly known as non-universal Higgs mass models (NUHM) \cite{nuhm}. 
Additional parameters will clearly relax the impact on the model imposed by experimental constraints. To the extent that the CMSSM survives them, other models will survive them more easily. As we will see, the CMSSM survives
these constraints with a supersymmetry breaking mass scale that is $\mathcal{O}(10)$ TeV \cite{Craig:2013cxa,mc12}, larger than the scale originally associated with supersymmetry when it was proposed as a solution to the hierarchy problem \cite{Maiani:1979cx}. However, supersymmetry at any scale below the GUT or Planck scale alleviates the hierarchy problem to some extent.

We implement several astrophysical constraints on dark matter interactions.
The first are the direct upper limits on spin-independent and -dependent astrophysical dark matter scattering
set recently by the LUX-ZEPLIN (LZ) experiment \cite{LZ}, which are more stringent than those set
previously by the LUX \cite{LUX}, PANDA-X \cite{PANDAX} and XENON \cite{XENON} experiments. As we show later, 
the spin-independent scattering constraint excludes parts of the dark matter strips that are compatible with $m_h$, the other
LHC and indirect dark matter constraints.

We also consider several indirect constraints on astrophysical dark matter annihilations~\cite{Strigari:2012acq}.
The first is provided by upper limits from {\it Fermi}-LAT on $\gamma$ fluxes from 
dwarf spheroidal satellite galaxies \cite{Fermi-LAT:2013sme,Fermi-LAT:2015ycq,Fermi-LAT:2015att,Boddy:2018qur,Hoof:2018hyn,Alvarez:2020cmw}. 
As discussed in \cite{Ellis:2011du} and in the text, we combine the $\gamma$
flux limits from all the available dwarf data sets. As in the case of the LHC constraints,
we find that the combined $\gamma$ limit does not impact the region of parameter space
favoured by $m_h$ and the dark matter density. The same is true for the upper limits from 
the $\gamma$ flux from the Galactic Centre~\cite{HESS:2022ygk,Johnson:2019hsm,Abazajian:2020tww}. The final
indirect limit we consider is that
on dark matter annihilation in the Sun \cite{indirectdet:solar} set by the IceCube search for energetic
neutrinos from the Sun \cite{IceCube:2016yoy,IceCube:2016dgk,IceCube:2021xzo}.

In order to set the stage for the applications of the different constraints,
we first discuss in detail in Section~2 the dark matter strips for different values of $\tan \beta$, $A_0 = 0$ and $\mu > 0$
and the underlying mechanisms that play dominant roles in bringing the cosmological dark matter density into the 
range allowed by {\it Planck} and other measurements. These include the focus-point 
and well-tempered neutralino mechanisms, coannihilation
between the LSP and other sparticles such as heavier neutralinos, charginos and stop squarks, and annihilations via
s-channel heavy Higgs bosons ($H/A$). As we discuss in Section~2, 
these different mechanisms may combine 
constructively in non-trivial ways. The pattern of dark matter strips is richest when $\tan \beta \ge 40$,
$A_0 = 0$ and $\mu >0$, simplifying significantly to the focus-point strip for smaller $\tan \beta$ when $A_0 = 0$ and
$\mu > 0$, to the stop coannihilation strip when $A_0 = 3 \, m_0$ and $\mu > 0$, and again to the focus-point strip
when $\mu < 0$.

Section~3 of this paper discuses the LHC constraints on the CMSSM when $A_0 = 0$ and $\mu >0$, with particular focus
on the $m_h$ constraint, which is generally much more important than the direct searches for supersymmetric particles.
We then discuss the astrophysical constraints on dark matter interactions when $A_0 = 0$ and $\mu >0$ in Section~4,
and combine these constraints in Section~5.
As we discuss in more detail below, for most of the values of $\tan \beta$ that we study
when $A_0 = 0$ and $\mu > 0$, there is a small region of the focus-point strip \cite{fp} that is allowed by all
the above constraints, where the LSP is mainly Higgsino-like with a mass 
$\sim 1000 - 1100$~GeV. 
However, there are some cases in which a lower neutralino mass $\gtrsim 450$~GeV
may be allowed, and other cases with an allowed strip at masses $\gtrsim 1200$~GeV. In all
these cases the spin-independent dark matter scattering cross section may be within reach of upcoming
experiments. On the other hand, when $A_0 = 3 m_0$ only LSP masses $\gtrsim 6$~TeV are
allowed, for which the dark matter scattering cross section is well below the neutrino ``floor" \cite{floor}.
Section~6 discusses the case when $A_0 = 3 \ m_0$ and $\mu > 0$, when there is a single dark matter 
strip where stop coannihilation is dominant, the uncertainty in the $m_h$ calculation is greater,
and dark matter interactions are well below the experimental limits. We find that when $\tan \beta \gtrsim 20$
the portion of the stop coannihilation strip that is allowed by $m_h$ within the calculational uncertainty extends 
from high masses down to the range of stop masses excluded by LHC searches.
Finally, Section~7 discusses the relatively simple case when $A_0 = 0$ and $\mu < 0$, and Section~8 summarizes
our conclusions.

\section{Dark Matter Strips in the CMSSM Parameter Space}
\label{sec:CMSSM}

\begin{figure}[t!]
\centering
\includegraphics[width=0.45\textwidth]{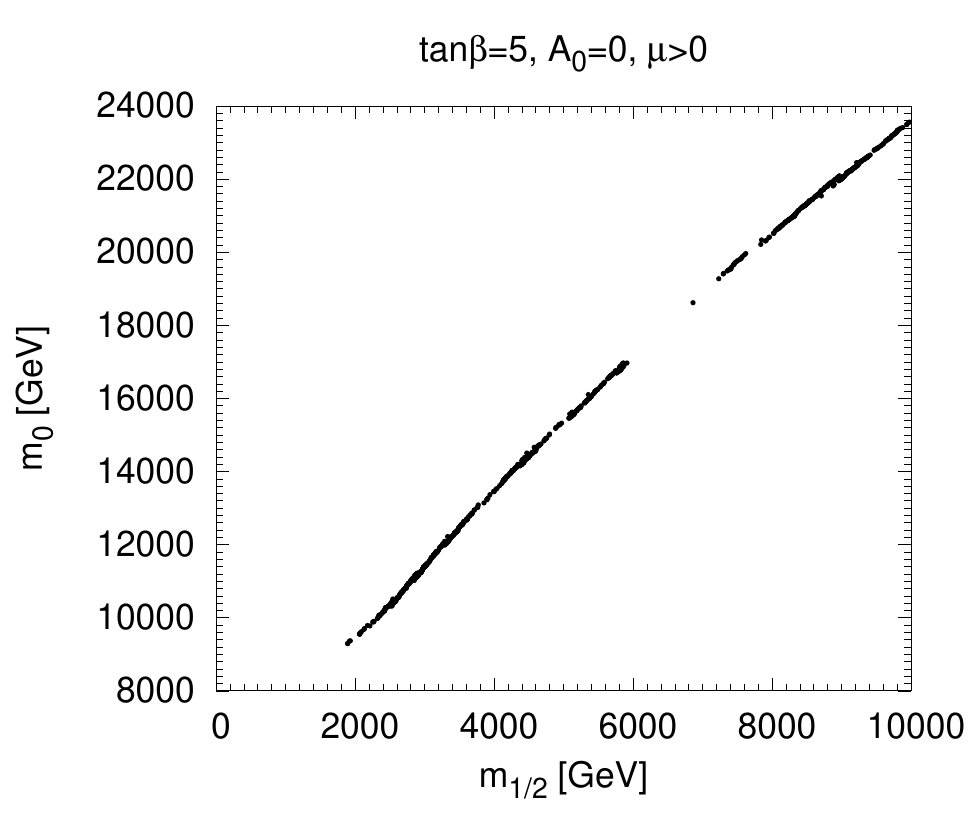}
\includegraphics[width=0.45\textwidth]{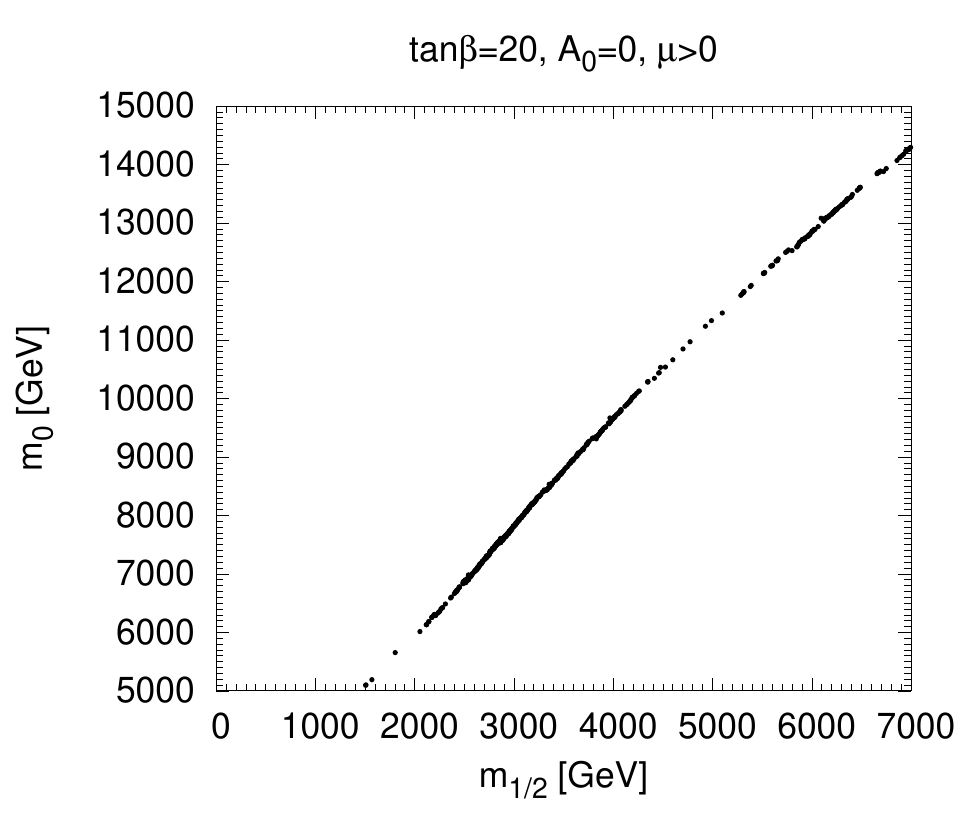}\\
\vspace{-2mm}
\includegraphics[width=0.45\textwidth]{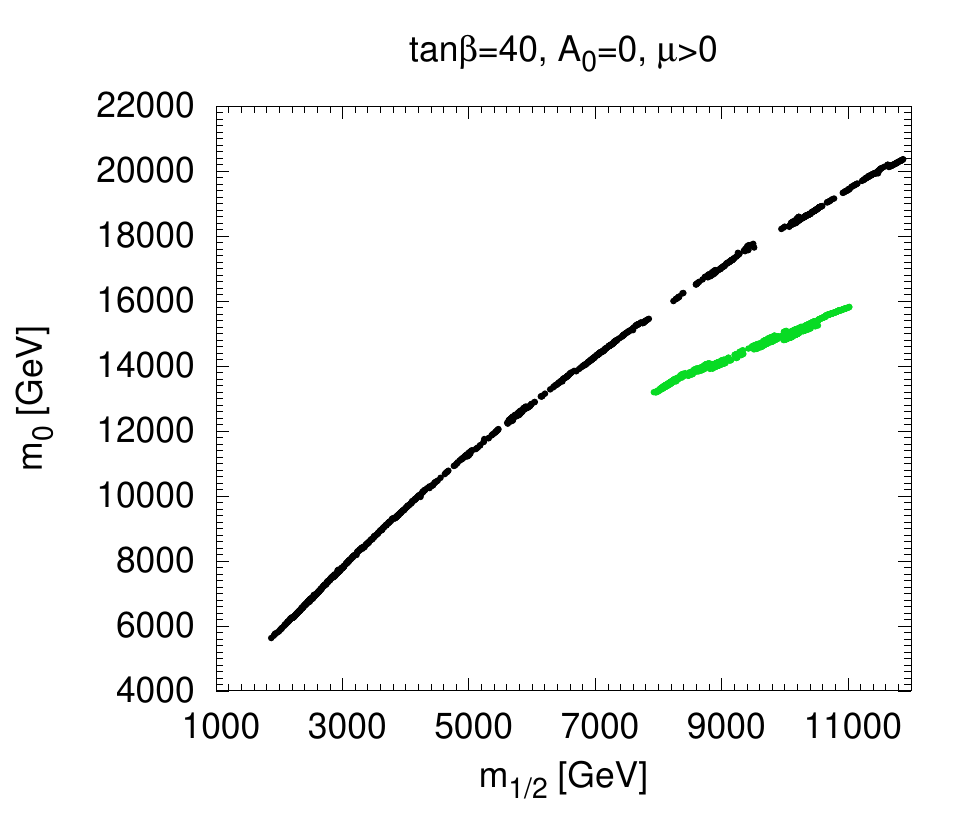}
\includegraphics[width=0.45\textwidth]{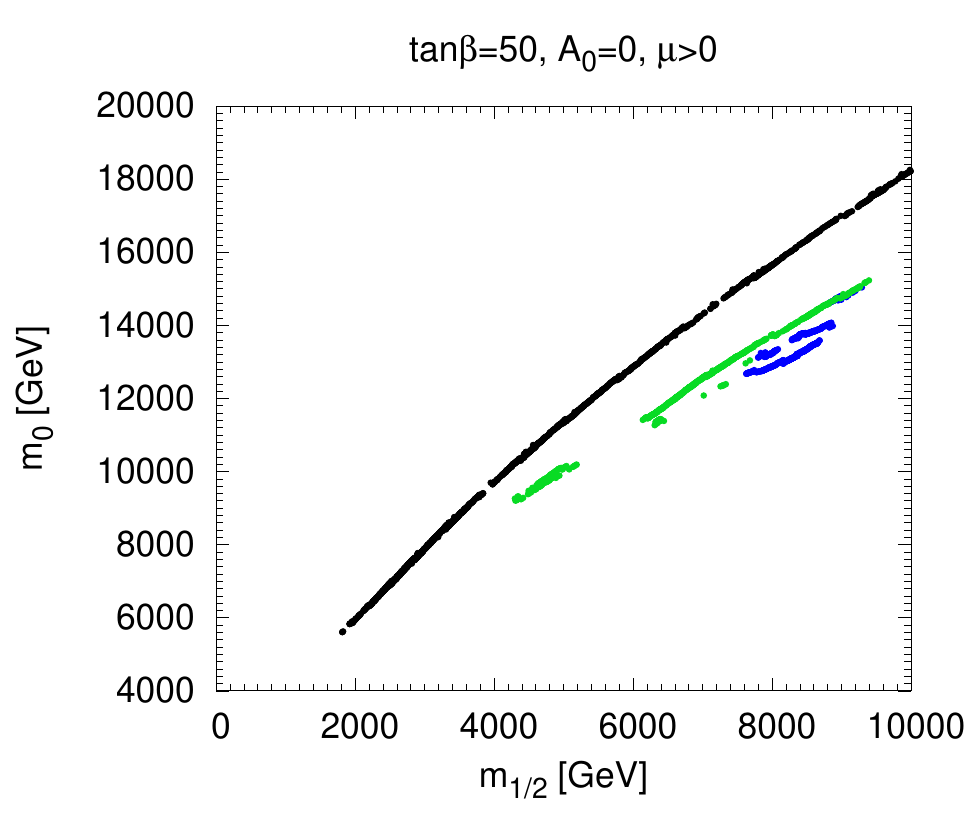} \\
\vspace{-2mm}
\includegraphics[width=0.45\textwidth]{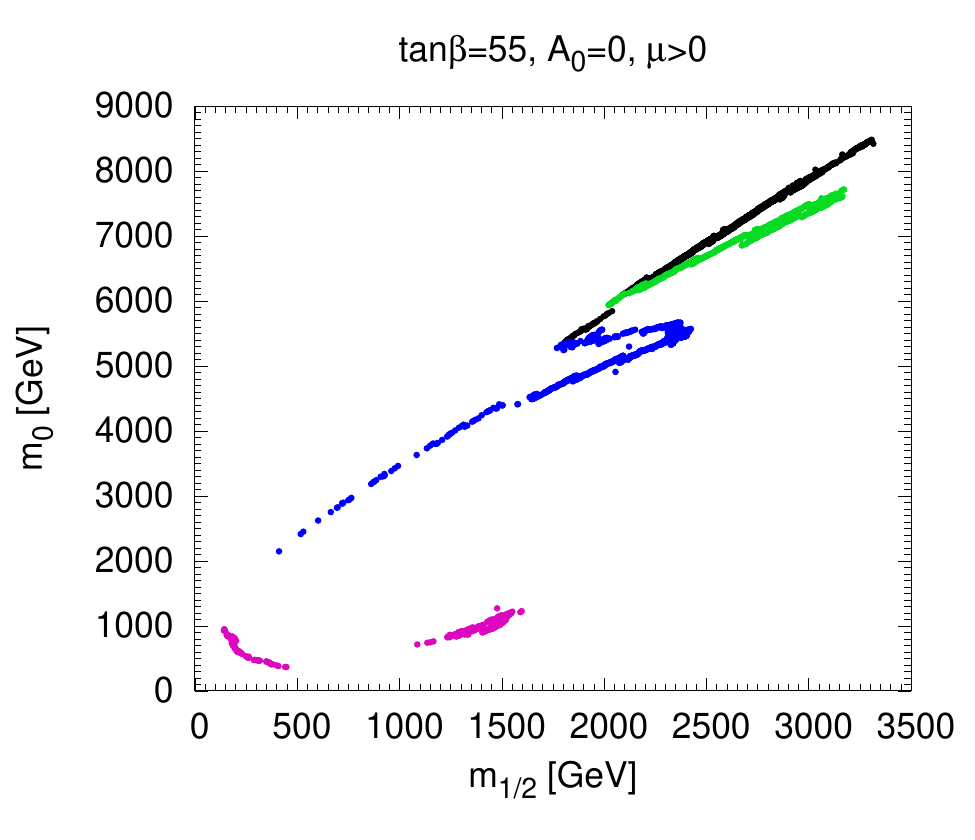} 
\includegraphics[width=0.45\textwidth]{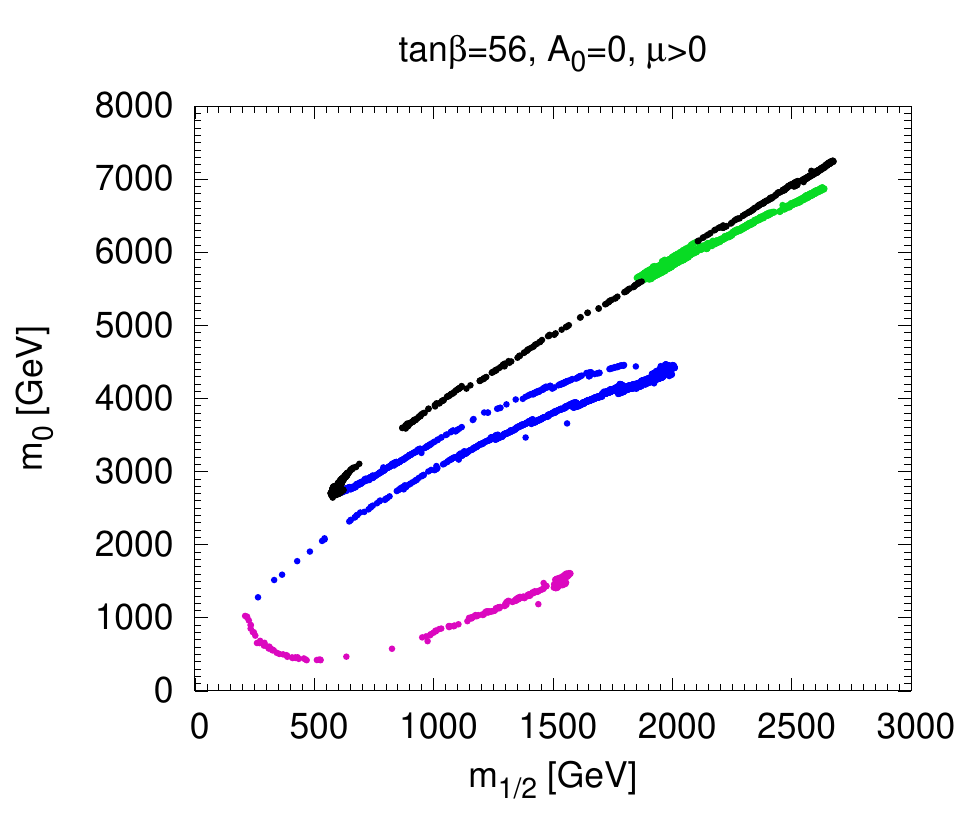}
\vspace{-4mm}
\caption{\it The dark matter strips for $\tan \beta = 5, 20,
40, 50, 55 \; and \; 56$, all assuming $A_0 = 0$ and $\mu > 0$.
}
\label{fig:m12_m0_planes}
\end{figure}

We first summarize the essential features of the CMSSM that are relevant for our
study. As usual, the assumption of universality for each of the various classes of 
soft supersymmetry-breaking parameters at the scale $\sim 10^{16}$~GeV where the SM
couplings unify leaves us with the following 4 continuous parameters: the
scalar mass, $m_0$, the gaugino mass, $m_{1/2}$, the trilinear soft 
supersymmetry-breaking parameter, $A_0$, and the ratio of supersymmetric Higgs vev's,
$\tan \beta$. In addition there is the unknown sign of the Higgsino mixing parameter,
$\mu$, which we take to be positive in this section. Historically, this choice has been motivated
by the sign of the difference between the SM calculation of $g_\mu -2$ \cite{Theory} and its
experimental value \cite{BNL1,BNL2,FNAL}. However, this choice can now be questioned on two grounds. 
One is that the supersymmetric contribution to $g_\mu - 2$ is negligible for the
relatively large supersymmetry-breaking scales found in our CMSSM analysis \cite{gm2,otherCMSSM}, and
the other is that recent lattice calculations \cite{bmw} cast doubt on the magnitude of the
significance between the SM calculation of $g_\mu -2$ and its experimental value.
Accordingly, we also comment on the status of the CMSSM for negative $\mu$ in Section~7.

The high precision of the determination of the cold dark matter density by
{\it Planck} and other experiments \cite{Planck} constrains the parameters of any supersymmetric
model to thin hypersurfaces in parameter space, which appear as narrow strips
in any planar slice through parameter space. Along these strips, a single
mechanism usually controls the dynamics of freeze-out, and obtaining
the appropriate density generally requires some specific mixing or relationship
between the masses of the sparticles controlling this mechanism. When the
LSP mass is relatively large, as suggested by the non-detection of
sparticles at the LHC \cite{nosusy} and value of $m_h$ \cite{Aad:2012tfa}, sometimes more than one mechanism plays a role. These relationships frequently
take the form of a near-degeneracy between the masses of the LSP and the
next-to-lightest supersymmetric particle (NLSP), so as to suppress the LSP
density via coannihilation \cite{gs,stauco,stopco,chaco,esug,Ellis:2018jyl}, a direct-channel resonance with mass 
$\sim 2 m_{\rm LSP}$ \cite{DN,funnel}, or some  other specific mass spectrum as in the 
focus-point region \cite{fp}.

In this section, we have imposed $A_0 = 0$. Generally, one might expect that the $A$-terms are comparable in magnitude to the other soft supersymmetry-breaking parameters, $m_{1/2}$ and $m_0$, at the GUT scale~\footnote{At low energies, due to the RGE running of all of the mass terms, the $A$-terms are of the same order as the other supersymmetry-breaking masses.}. As we  see below, large values of $A_0$ tend to increase the calculated Higgs mass, and thus would require a somewhat less massive supersymmetry spectrum. While one can motivate $A_0$ from, e.g., anomaly mediation \cite{anom}, our primary motivation lies in studying the CMSSM focus-point region, which contains much of the phenomenologically viable parameter space but disappears when $A_0 \gtrsim m_0$. 

The principal dark matter strips  in 
the CMSSM are those where the focus-point mechanism \cite{fp}
is operative, or where stop \cite{stopco,esug,Ellis:2018jyl} and/or neutralino/chargino coannihilation \cite{chaco,esug} is dominant, 
or where resonant exchanges of the neutral heavy Higgs bosons $H/A$ in the 
$s$-channel \cite{DN,funnel} suppress the relic LSP density. 
In some parameter planes there are also 
important regions of the dark matter strips where a single mechanism 
is insufficient to lower the relic density to its cosmological value, 
but this is made possible by combinations of these
mechanisms that are in play simultaneously.

The panels of Fig.~\ref{fig:m12_m0_planes} display the results of
performing a Markov chain Monte Carlo (MCMC) scan of 
$m_{1/2}, m_0$  parameter space with $A_0 = 0$ and $\mu > 0$
for the indicated fixed values of $\tan \beta$.
The MCMC scan determines the coordinates ($m_{1/2}, m_0$) for which
$\Omega_\chi h^2 = 0.12 \pm 0.0036$, which is the 3$\sigma$ range as determined by {\it Planck} \cite{Planck}.
The MCMC code  we run employs mainly a Metropolis–Hastings algorithm~\cite{mcmc}.
The basic criterion we use to run the MCMC code is to find points
in the parameter space that are 
compatible with the {\it Planck} $3\sigma$ range for $\Omega_\chi h^2$. 
This is an efficient way to delineate the dark matter strips in the 
parameter space. We stress that the MCMC method we employ does 
carry any statistical weight, but is merely a scanning code. 
Some of the gaps seen in the strips 
are due to incompleteness of the scanning process.

Information on the mechanism (or mechanisms) responsible for sufficient
annihilation prior to freeze-out is coded by the color of the point. 
Black points lie along the focus-point strip where $\mu \sim 1$ TeV 
and the LSP is predominantly a Higgsino of mass 1.1 TeV. At slightly
larger values of $m_0$ for each point on this strip, 
$\mu^2$ is driven to negative values and radiative electroweak symmetry
breaking is no longer possible. Below the strip, the LSP is predominantly
a bino, with some exceptions noted below.
In some cases, at values of $m_0$
below the focus-point region the bino LSP acquires non-negligible
Higgsino components, in which case the neutralino becomes 
`well-tempered'~\cite{wt}. These points are coloured green.

The purple points at lower
$m_0$ and large $\tan \beta$ lie along the funnel where
the LSP mass is close to half the heavy Higgs masses
and the dominant dark matter mechanism  is 
rapid LSP annihilation via these s-channel resonances.
However, as $m_{1/2}$ increases, this  $s$-channel annihilation is
suppressed by the heavy Higgs widths 
and the the annihilation cross section is
insufficient to reduce the relic density to the required value,
which is why the purple points terminate at an endpoint. At larger $m_0$, in addition to s-channel Higgs pole contributions, the LSP becomes well-tempered and the combination of these
contributions leads to a low relic density. These points are colored blue.
Coannihilations of the bino-like state, the two Higgsinos and chargino also play a role in determining the final relic density. In addition, as we will see, there is level crossing between the neutralino states in some areas of parameter space. Here the neutralino may be well-tempered and coannihilations may also be important. These points will be colored green. We describe in the following how each of these possibilities
arises for different values of $\tan \beta$.

The top left panel of Fig.~\ref{fig:m12_m0_planes} shows the
$(m_{1/2}, m_0)$ plane for $\tan \beta = 5, A_0 = 0$ and $\mu > 0$.
It features a single dark matter strip where the focus-point
mechanism is operative, which extends beyond the range of $m_{1/2} \le 10$~TeV
that we display. The $(m_{1/2}, m_0)$ 
plane for $\tan \beta \lesssim 20$, $A_0 = 0$ and $\mu > 0$ is similar,
as seen in the top right panel of 
Fig.~\ref{fig:m12_m0_planes}. This one is truncated at $m_{1/2} = 8$~TeV, because
the extensions of this strip to larger $m_{1/2}$ is not compatible with the LHC
measurement of $m_h$, as we see later. 

In the $\tan \beta = 40$ case as seen in the middle left panel of Fig.~\ref{fig:m12_m0_planes}, in addition to the black focus point strip, 
a closely-spaced pair of green strips
appears when $m_{1/2} > 7$~TeV. 
Here, the neutralino (mostly bino) is well-tempered
and the non-negligible Higgsino components in the LSP lead to enhanced annihilations
that bring the relic density into the allowed range. 
On either side (higher and lower $m_0$) of the pair of strips, 
the relic density is too high, and between the strips the relic density is 
suppressed.
However, as we will see these points are excluded by both $m_h$ and the upper limit
on spin-independent dark matter scattering.

\begin{figure}[ht!]
\includegraphics[width=0.45\textwidth]{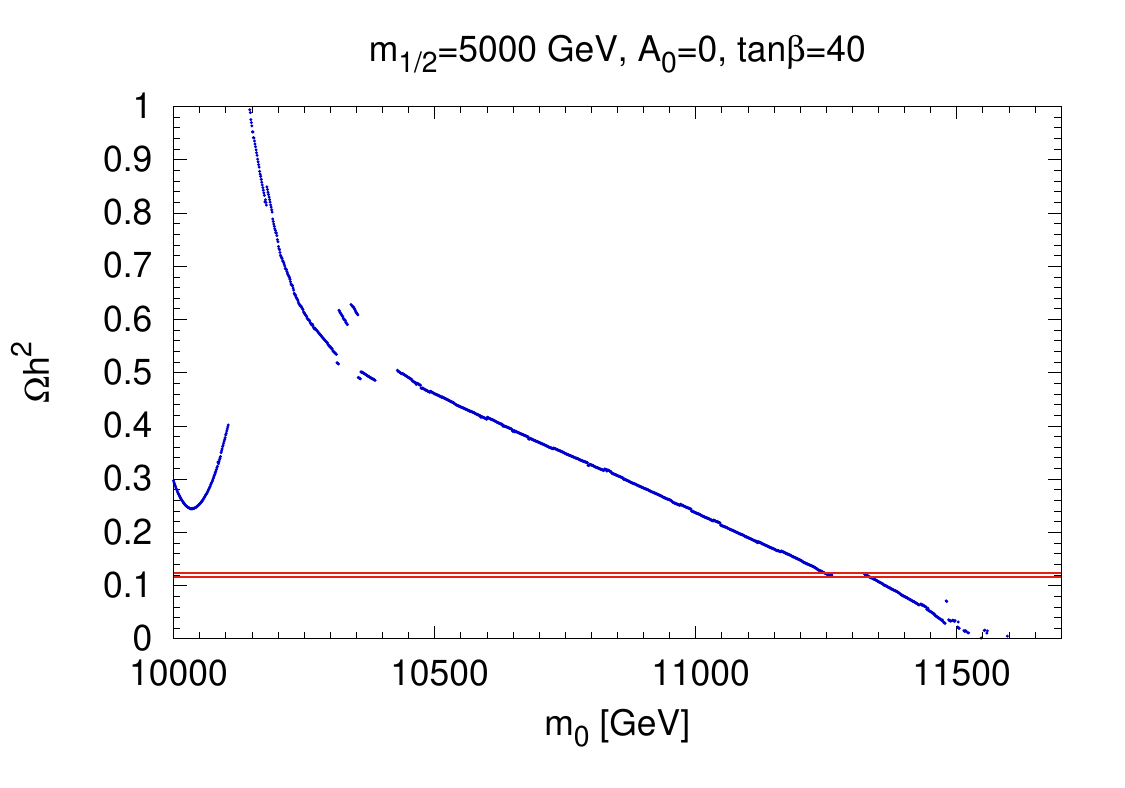} 
\includegraphics[width=0.45\textwidth]{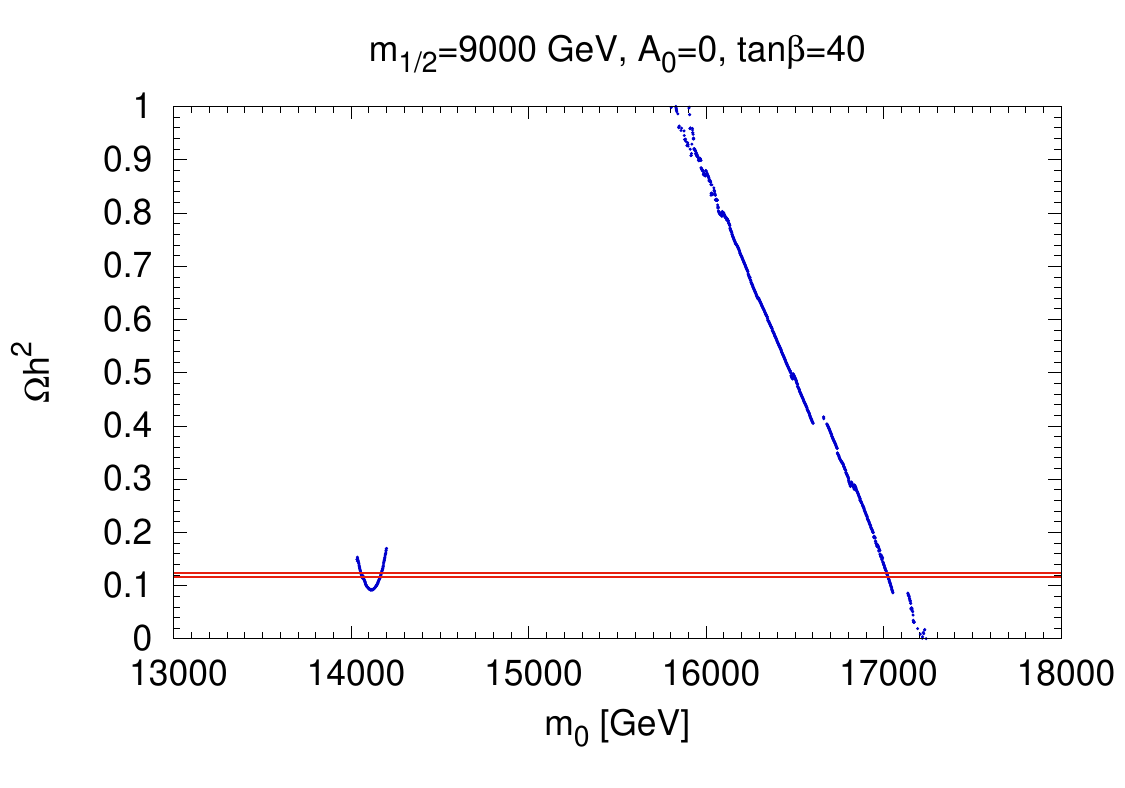} 
\vspace{-2mm}
\includegraphics[width=0.45\textwidth]{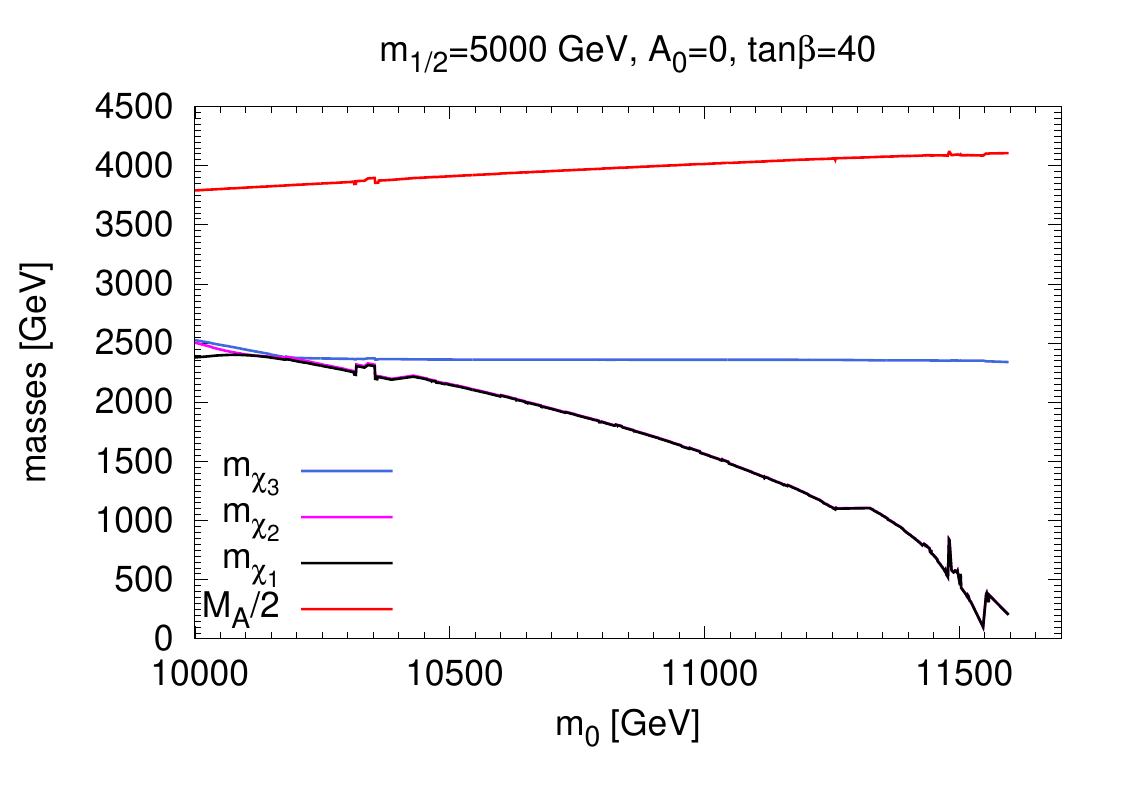} 
\includegraphics[width=0.45\textwidth]{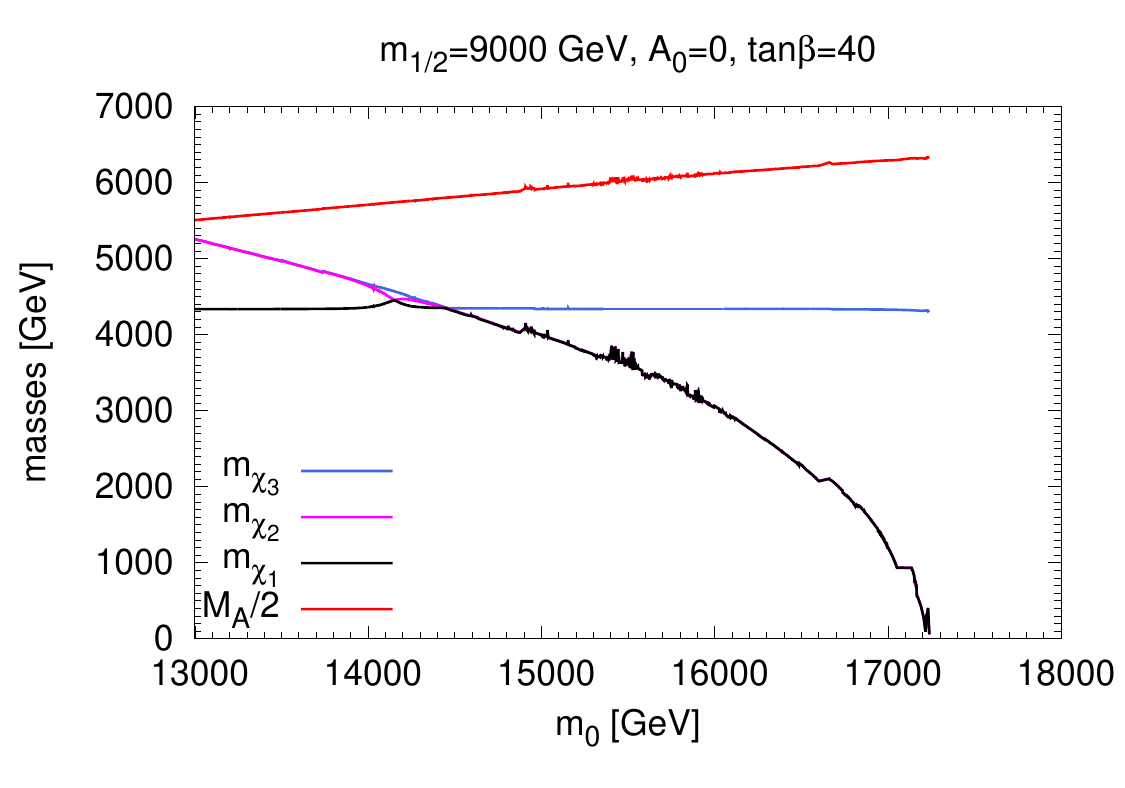} 
\vspace{-2mm}
\includegraphics[width=0.45\textwidth]{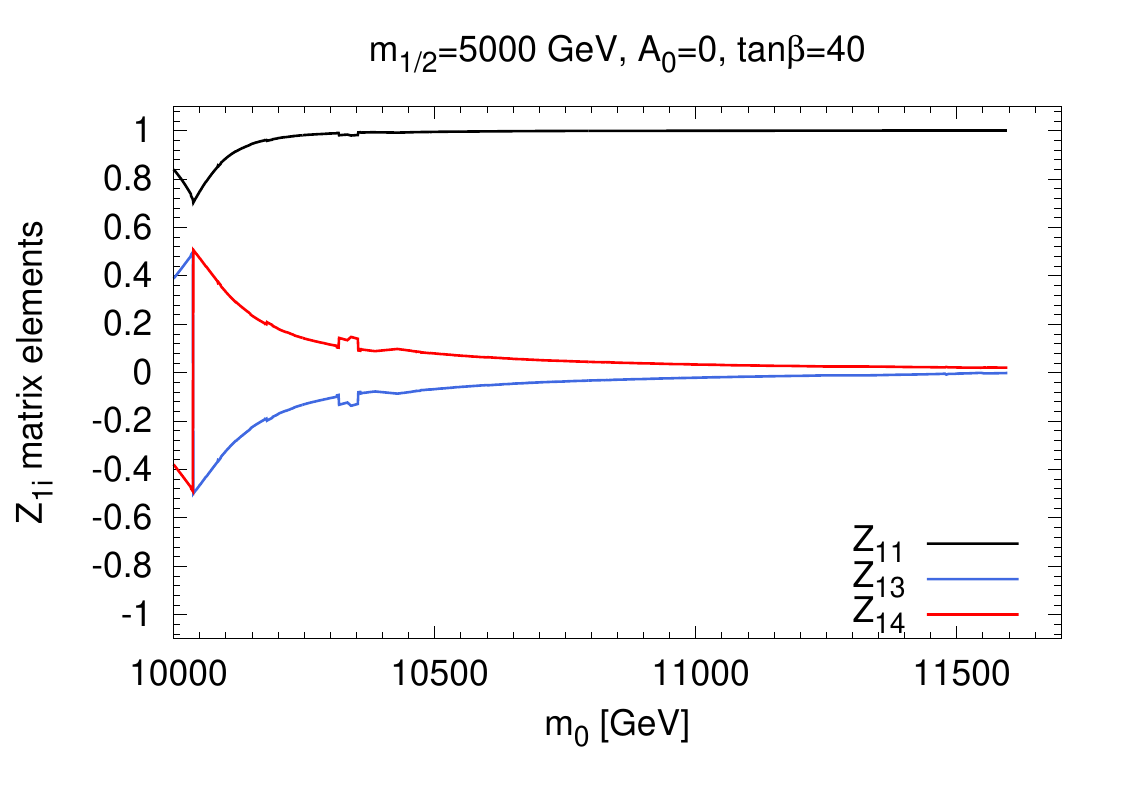} 
\hspace{5mm}
\includegraphics[width=0.45\textwidth]{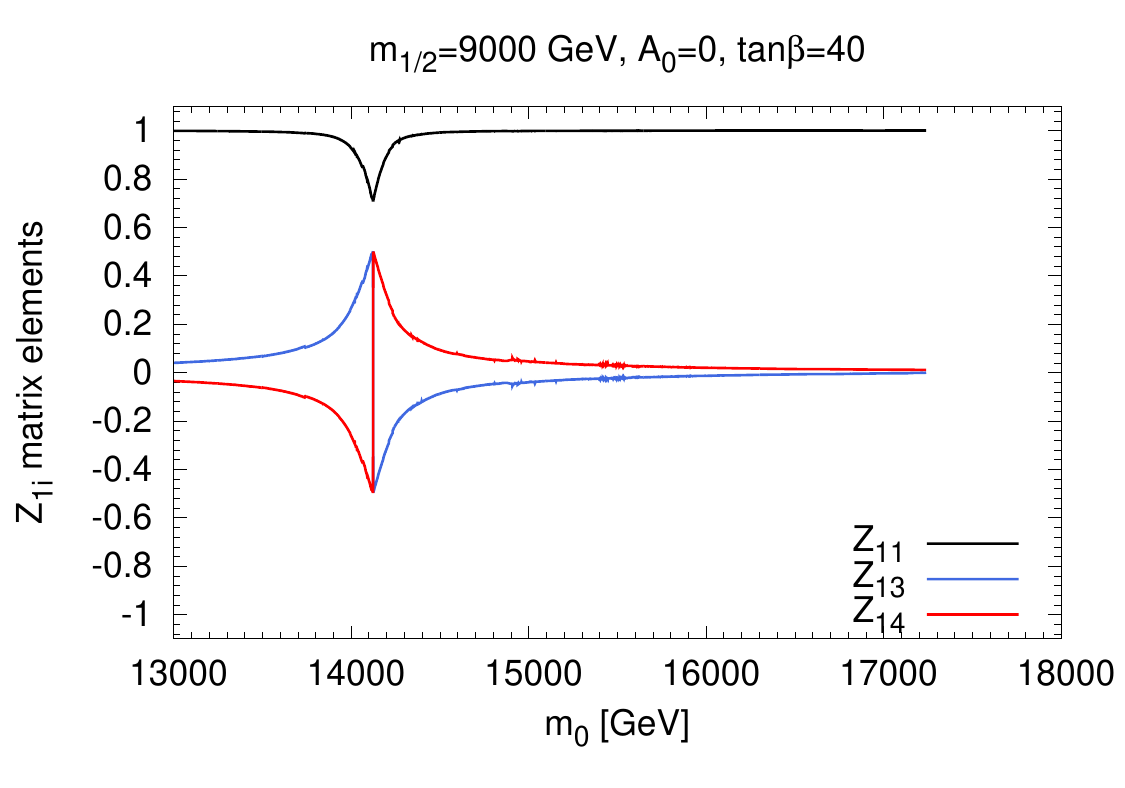} 
\vspace{-4mm}
\caption{\it 
Cuts across the $(m_{1/2}, m_0)$ planes for $\tan \beta = 40$, $A_0 = 0$ and $\mu > 0$  for  varying $m_0$ and fixed
$m_{1/2} = 5.0$ TeV (left panels) and $9.0$ TeV (right panels), showing the dark matter density (top panels), particle
masses (middle panels) and the composition of the
bino-like neutralino (bottom panels).}
\label{fig:Slices40}
\end{figure}

As aids to understanding the origins of the strips, we plot slices across
the $(m_{1/2}, m_0)$ plane showing the relic density, masses, and neutralino composition as functions of $m_0$
for fixed values of $m_{1/2}$. As a first example
for $\tan \beta = 40$, $A_0 = 0$ and $\mu > 0$, we fix
$m_{1/2} = 5$ TeV in the left panels of
Fig.~\ref{fig:Slices40} 
and in $m_{1/2} = 9$ TeV in the right panels. The top panels illustrate how the relic density varies with $m_0$. We see that along
both slices the relic density decreases as the boundary of electroweak symmetry breaking is approached
when $m_0 \gtrsim 11 \ (17)$~TeV, corresponding to the focus-point strip in the middle left panel
of Fig.~\ref{fig:m12_m0_planes}. The dark matter density is
generally larger than allowed by {\it Planck} and other measurements at lower $m_0$. However, there
are dips in the relic density at $m_0 \sim 10 \ (14)$~TeV for $m_{1/2} = 5.0 \ (9.0)$ TeV, 
including a pair of $m_0$ values where the {\it Planck} relic density is found for $m_{1/2} = 9$~TeV. 
These values lie on the pair of green lines seen in the middle left panel of Fig.~\ref{fig:m12_m0_planes}.

The middle panels of Fig.~\ref{fig:Slices40} display
the masses of the lightest neutralino, $m_{\chi_1}$ (black line), the next two neutralino masses, $m_{\chi_2}$ (purple line) and $m_{\chi_3}$ (blue line) and half the mass of the pseudoscalar Higgs boson, $m_A/2$ (red line). The three mass eigenstates are either mostly pure or mixed states of the bino and two Higgsinos, as the wino is always heavy and does not enter in the discussion
of the parameter ranges discussed here. The bino-like state
(whether the LSP or not) has a relatively constant mass, while the two Higgsino masses decrease with increasing $m_0$ as $\mu$ decreases over the displayed range of $m_0$, as a result of the Higgs potential minimization conditions. 
Coannihilation is the most important dark matter mechanism when two or more of the neutralinos are close in mass. Comparing the middle and top panels, one sees that the dips in the relic density coincide with the degeneracy in the neutralino masses leading to coannihilation, whereas $s$-channel annihilation via heavy Higgs poles is less important in this case because
$m_\chi$ is quite different from $m_A/2$.

The bottom panels of Fig.~\ref{fig:Slices40} show that the dips in the relic density 
at relatively low $m_0$ are correlated
with crossovers in the neutralino mass eigenstates and corresponding variations in the LSP composition, 
with the LSP acquiring a 
significant Higgsino component near the crossovers, leading to a well-tempered neutralino.
Around this cross-over the lightest eigenstate transitions from a bino-like state with Higgsino components $Z_{13} \simeq 1/2$ and $Z_{14} \simeq -1/2$ to a Higgsino-like state with  $Z_{33} \simeq 1/2$ and $Z_{34} \simeq -1/2$ (with $Z_{11} \simeq 1/
\sqrt{2}$ in both cases). 
As one can see, these states are maximally well-tempered. 
We display the bino component ($Z_{11}$) and the two Higgsino components ($Z_{13}$ and $Z_{14}$) of the {bino-like state, which is not necessarily the LSP}. This state is the LSP for $m_0$ below the dip. When the LSP eigenstate flips, the Higgsino components of the bino-like state are now $Z_{13} \simeq -1/2$ and $Z_{14} \simeq 1/2$ still with $Z_{11} \simeq 1/\sqrt{2}$. At larger $m_0$, this state is again mostly a pure bino, though it is no longer the LSP. 
As seen more clearly in the middle right panel of Fig.~\ref{fig:Slices40}, 
the crossover is reflected in a bump
in the LSP mass where coannihilation with $\chi_2$ and the lighter chargino is enhanced,
 suppressing the relic density around the local
minimum seen in the top right panel of Fig.~\ref{fig:Slices40}.~\footnote{In these and subsequent
analogous figures we see other glitches in the
composition of the LSP and particle masses, and consequently the relic density, which are numerical artifacts.}

The middle right panel of Fig.~\ref{fig:m12_m0_planes} shows the
$(m_{1/2}, m_0)$ plane for $\tan \beta = 50$, where the focus-point strip (black points) is again
supplemented by additional features at lower $m_0$ that resemble the second feature
that appeared for $\tan \beta = 40$, but extend to lower $m_{1/2}$ and exhibit a richer structure.  As in the case of $\tan \beta = 40$,
there are a pair of strips where the neutralino is well-tempered. These are represented by the green points which
extend down to $m_{1/2} \approx 4$ TeV.  At larger $m_{1/2}$, an additional pair of strips appear where the LSP is somewhat less well-tempered, but sits near the heavy Higgs pole. For these points, it is the combination of the $s$-channel annihilations with the bino-Higgsino mixing contributions that lead to a cross section that is large enough to obtain the correct relic density.

\begin{figure}[ht!]
\includegraphics[width=0.45\textwidth]{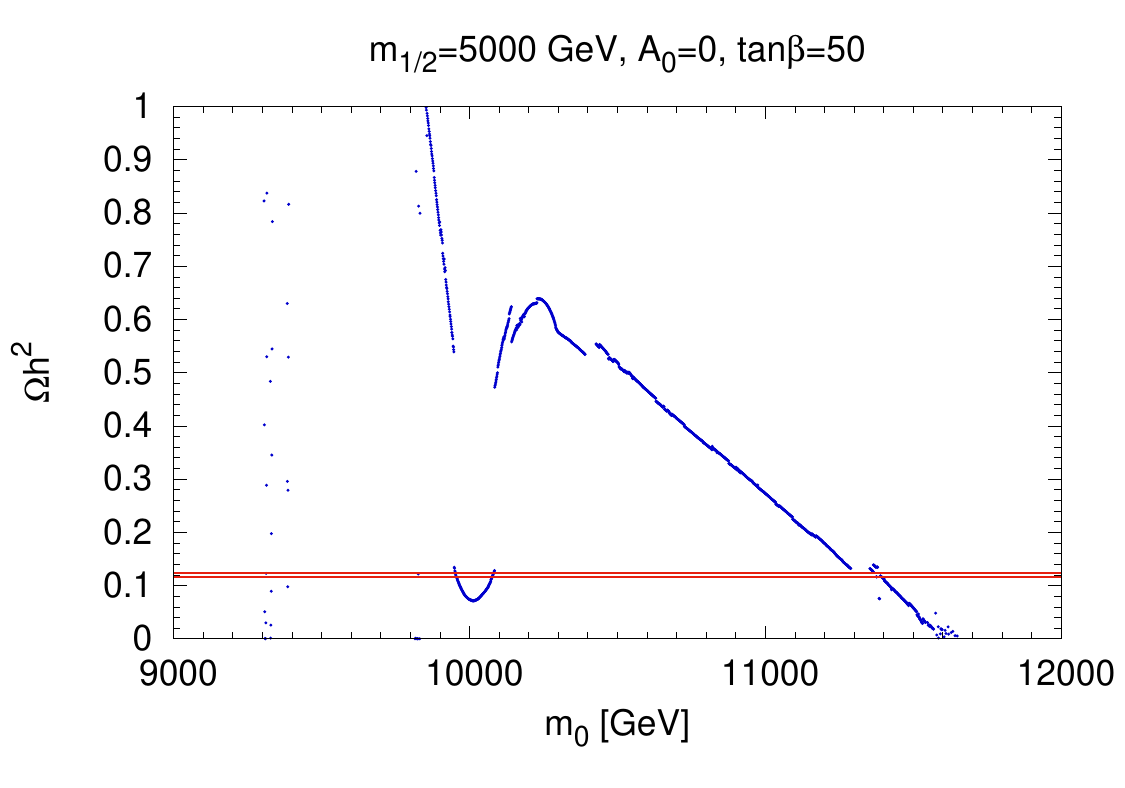} 
\includegraphics[width=0.45\textwidth]{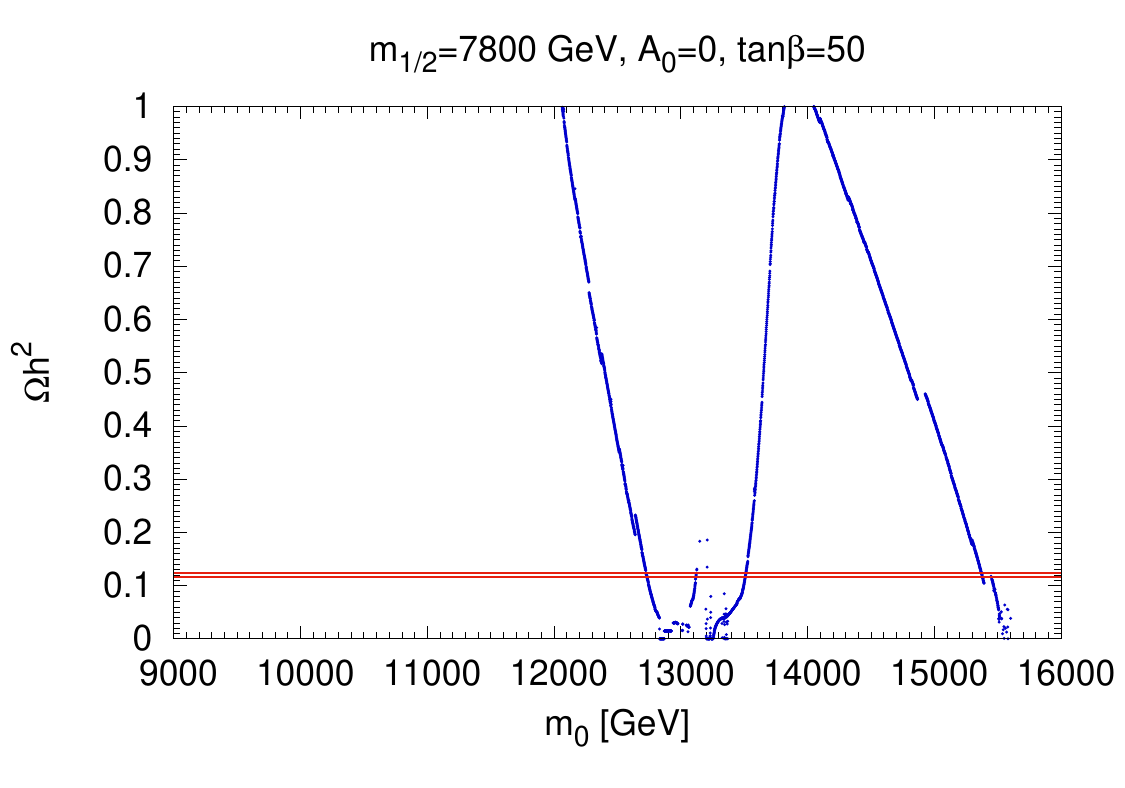} \\
\vspace{-2mm}
\includegraphics[width=0.45\textwidth]{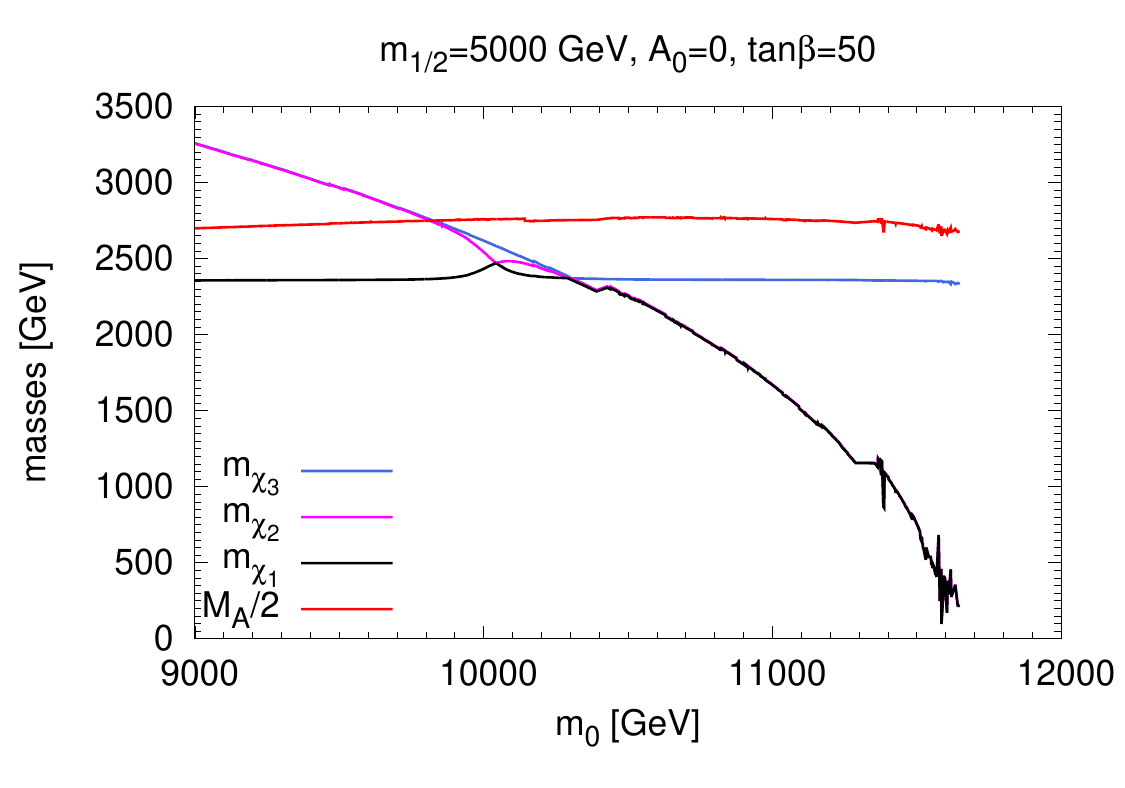} 
\includegraphics[width=0.45\textwidth]{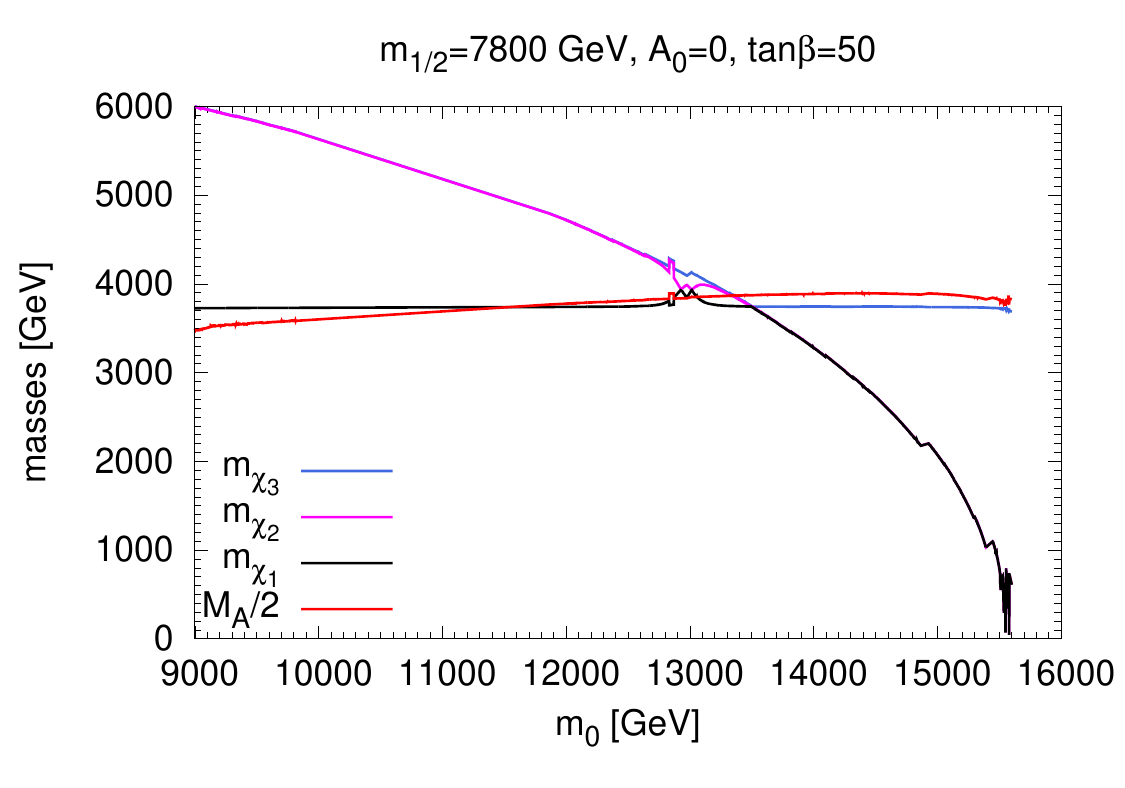} \\
\vspace{-2mm}
\includegraphics[width=0.45\textwidth]{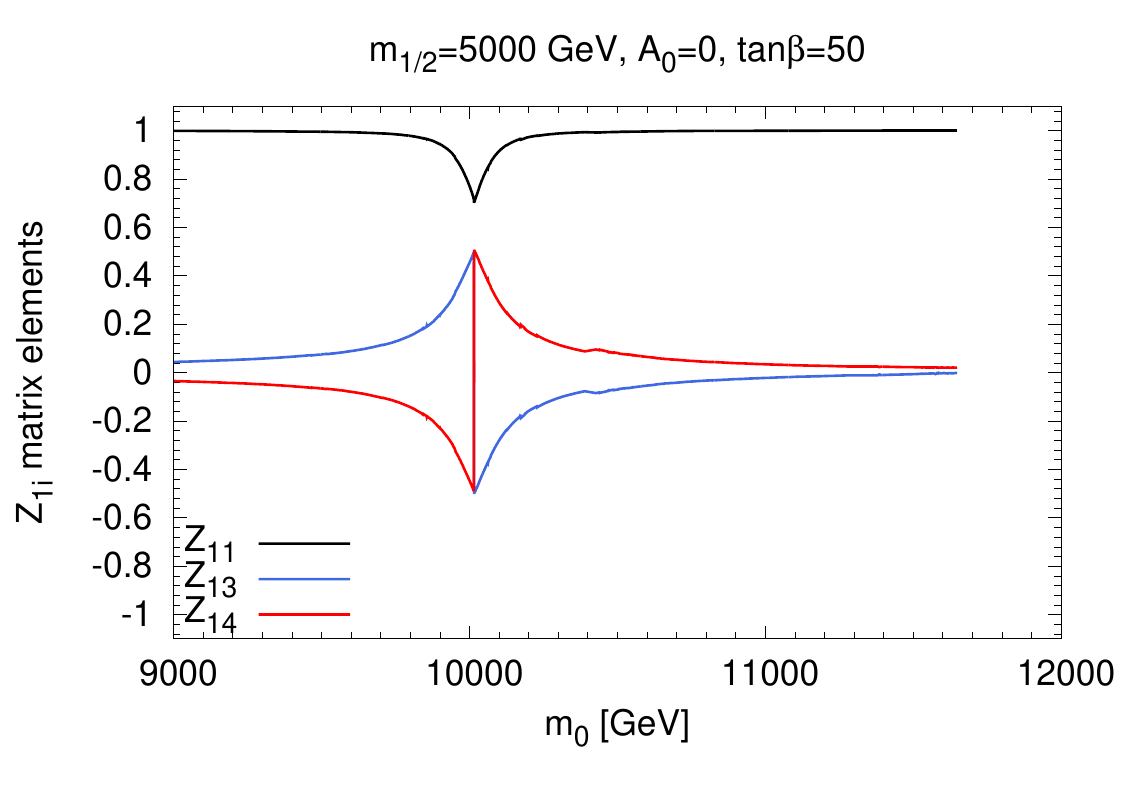} 
\includegraphics[width=0.45\textwidth]{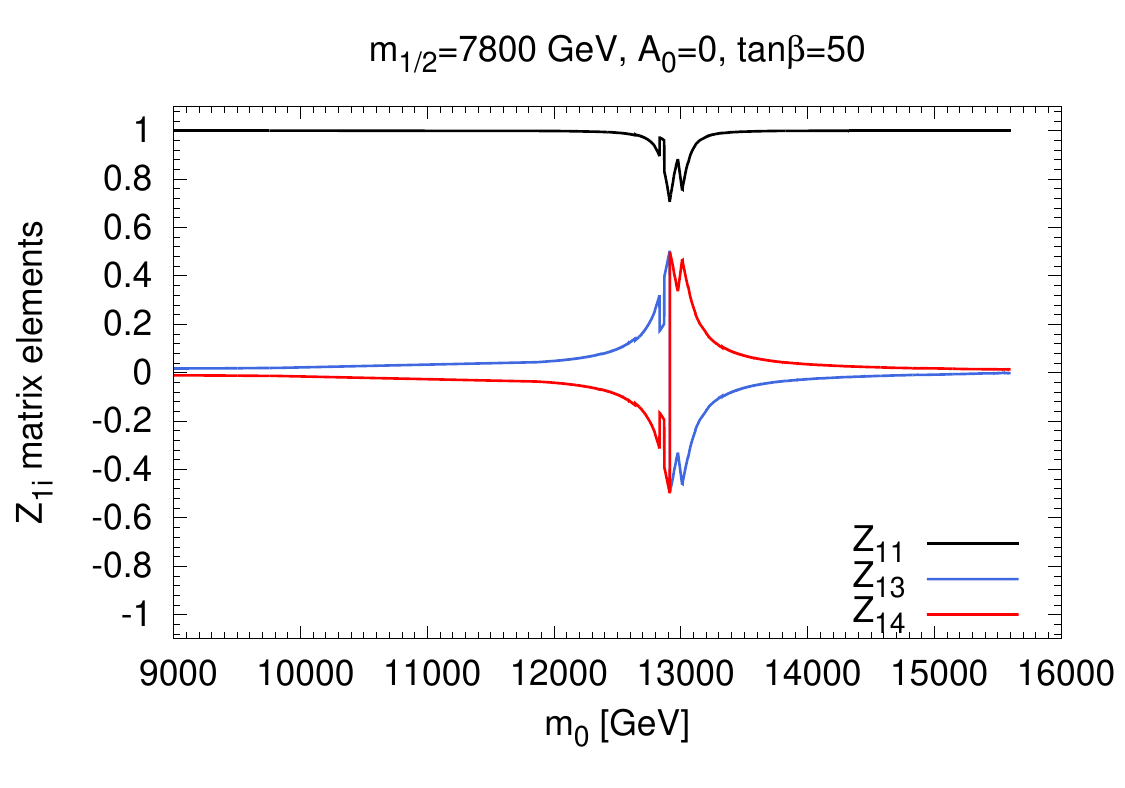} 
\vspace{-4mm}
\caption{\it 
Cuts across the $(m_{1/2}, m_0)$ planes for $\tan \beta = 50$, $A_0 = 0$ and $\mu > 0$  for  varying $m_0$ and fixed
$m_{1/2} = 5$ TeV (left panels) and $7.8$ TeV (right panels), showing the dark matter density
(top panels), particle masses  (middle panels) and the composition of the LSP (bottom panels).}
\label{fig:Slices50}
\end{figure}

The interplay of different dark matter mechanisms for $\tan \beta = 50$
is illustrated in Fig.~\ref{fig:Slices50}, where $m_0$ 
varies along slices across the $(m_{1/2}, m_0)$ plane
at fixed $m_{1/2} = 5$ TeV (left panels) and 7.8~TeV (right
panels).  We see in the top left panel that the {\it Planck} value
of the relic density is attained for three values of $m_0$~\footnote{There are also a handful of points seen at $m_0 \approx 9.3$ TeV. These points are almost exactly on a coannihilation pole where $(m_{\chi_1} + m_{\chi_2})/m_H - 1 <  10^{-6} $, where $H$ is the heavy Higgs scalar. We do not discuss these points further.}.
These correspond to the focus-point strip at $m_0 \approx 11.3$ TeV and a pair of values of $m_0$ around 10~TeV. The latter correspond to the 
green strips seen in Fig.~\ref{fig:m12_m0_planes} where the
well-tempered nature of the LSP is sufficient for reducing the
relic density into the cosmological range. In contrast, at
higher $m_{1/2}$, as seen in the right panels, there are five values of $m_0$ where the desired relic density is attained. Once again the highest value at $m_0 \approx 15.4$ TeV corresponds to the focus-point strip. The cluster of points
around $m_0 \simeq 13$ TeV
contains 4 values of $m_0$ where the cosmological value $\Omega h^2 = 0.12$
is crossed.
These 4 values arise from the interplay of rapid $s$-channel annihilation via the $A/H$ resonances
(which is most important where the red and black lines cross in the middle panels, which does not occur for $\tan \beta = 40$) 
and LSP coannihilation with the Higgsinos and lighter chargino (which is most important where the blue and black lines approach each other in the lower panels).
Within this cluster of points, those
with lower $m_0$ require both s-channel annihilations and a well-tempered neutralino, whereas at higher $m_0$,  the neutralino level crossing is also sufficient to suppress the relic density. These correspond to the blue and green strips seen in Fig.~\ref{fig:m12_m0_planes}.
At higher $m_0$, the Higgsino becomes the LSP, and the annihilation cross section enhancements disappear until $m_0$ is sufficiently large that $m_\chi \approx \mu \approx 1.1$ TeV, and the relic density falls back to $0.12$ due to neutralino/chargino coannihilations among the Higgsinos. Finally, the boundary of the electroweak symmetry breaking region is reached when $m_0 \approx 15.5$ TeV.

The $(m_{1/2}, m_0)$ plane for $\tan \beta = 55$
shown in the bottom left panel of Fig.~\ref{fig:m12_m0_planes} exhibits
additional features. We note that the focus-point strip terminates because the renormalization-group running of the top Yukawa coupling diverges at larger $m_{1/2}$.
The purple strip corresponds to a funnel region 
starting at low $m_0$ where s-channel annihilation through the heavy Higgs bosons dominates. 
This terminates at about $m_{1/2} \sim 1.6$ TeV when, due to the increasing heavy-Higgs width, the annihilation cross section becomes too small 
to reduce the relic density into the range allowed by {\it Planck} and other measurements. 
In the region above and to the right of this funnel, the relic density is too high.  For fixed $m_{1/2}$, e.g., $m_{1/2} = 2$~TeV, 
as $m_0$ is increased, the heavy Higgs masses increase and $2 m_\chi < m_A$, until at still higher $m_0$,
$\mu$ begins to decrease. This has the effect of decreasing $m_A$ and moving the LSP back on the Higgs pole, lowering the 
Higgsino/chargino masses and increasing the Higgsino component of the LSP. This results in the pair of 
blue strips that form a second funnel region,
where the annihilation cross section is augmented by an increasing neutralino coannihilation contribution 
sufficiently to lower the relic density 
to an acceptable value. Further increasing $m_0$ moves the LSP off the pole and the relic density is again too big until $m_0$
is sufficiently large that $\mu$ has dropped enough for the LSP to become mostly Higgsino and the focus-point strip is reached. 
At higher $m_{1/2}$, e.g., $m_{1/2} = 2.8$ TeV, the funnel is not able to bring the relic density sufficiently low, 
even though enhanced by coannihilation.
However, at larger $m_0 \sim 7$ TeV a large Higgsino component in the LSP increases the cross section producing the 
green pair of strips below the black focus-point strip.

\begin{figure}[ht!]
\includegraphics[width=0.45\textwidth]{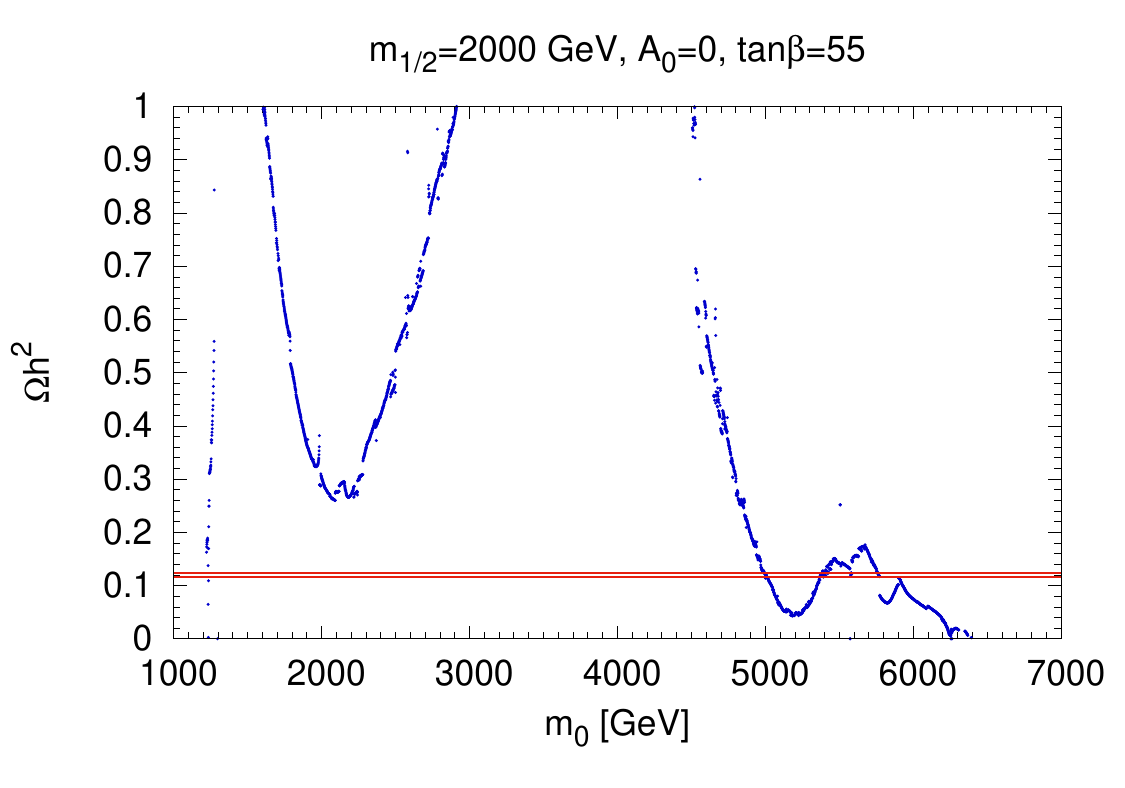}
\includegraphics[width=0.45\textwidth]{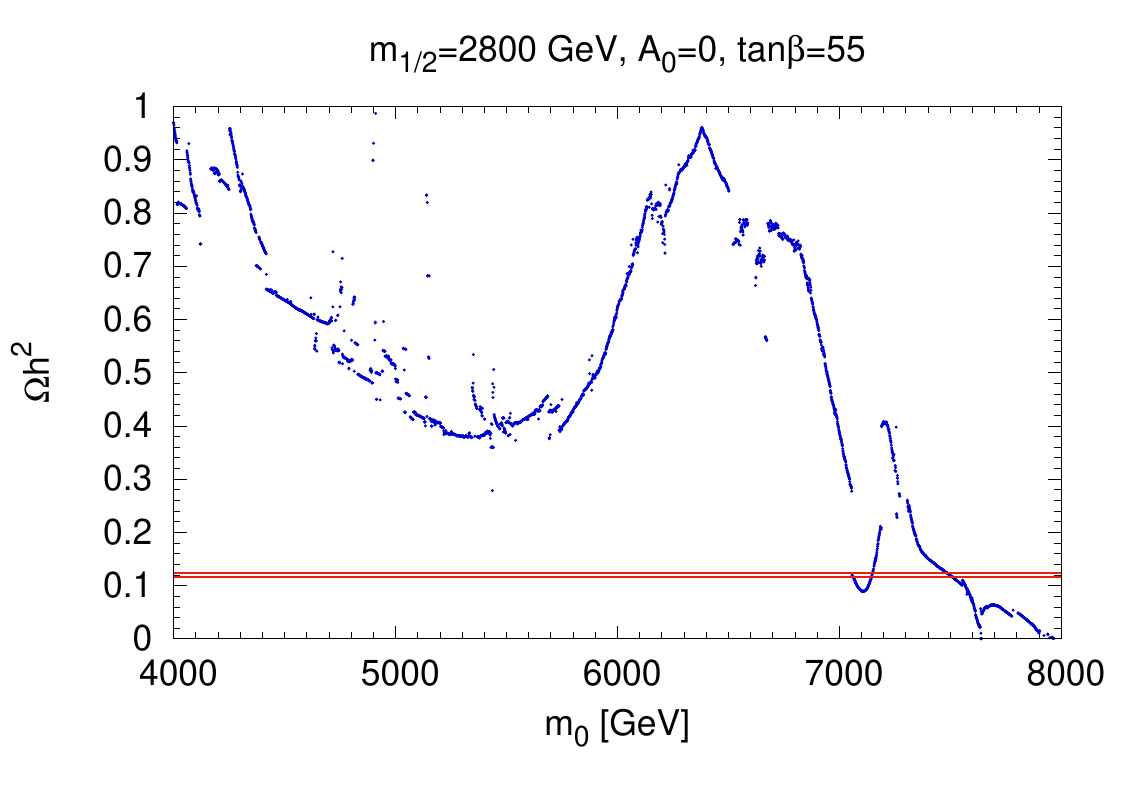} \\
\vspace{-2mm}
\includegraphics[width=0.45\textwidth]{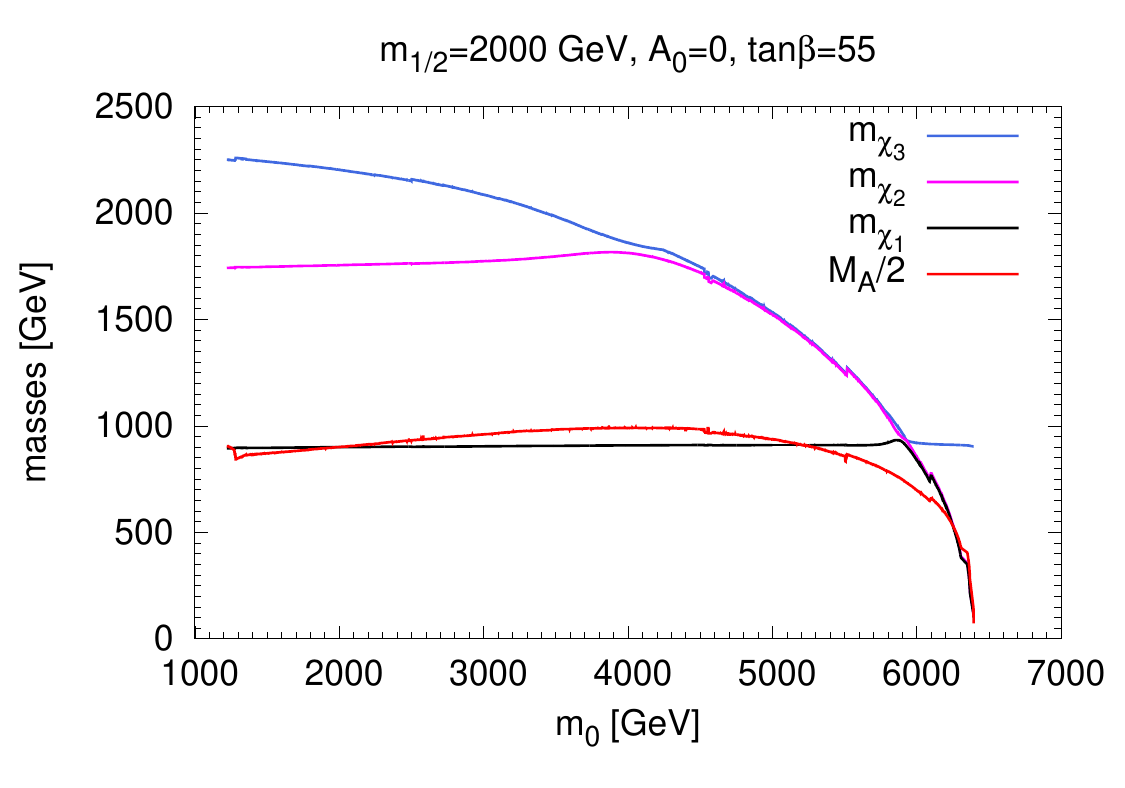} 
\includegraphics[width=0.45\textwidth]{ 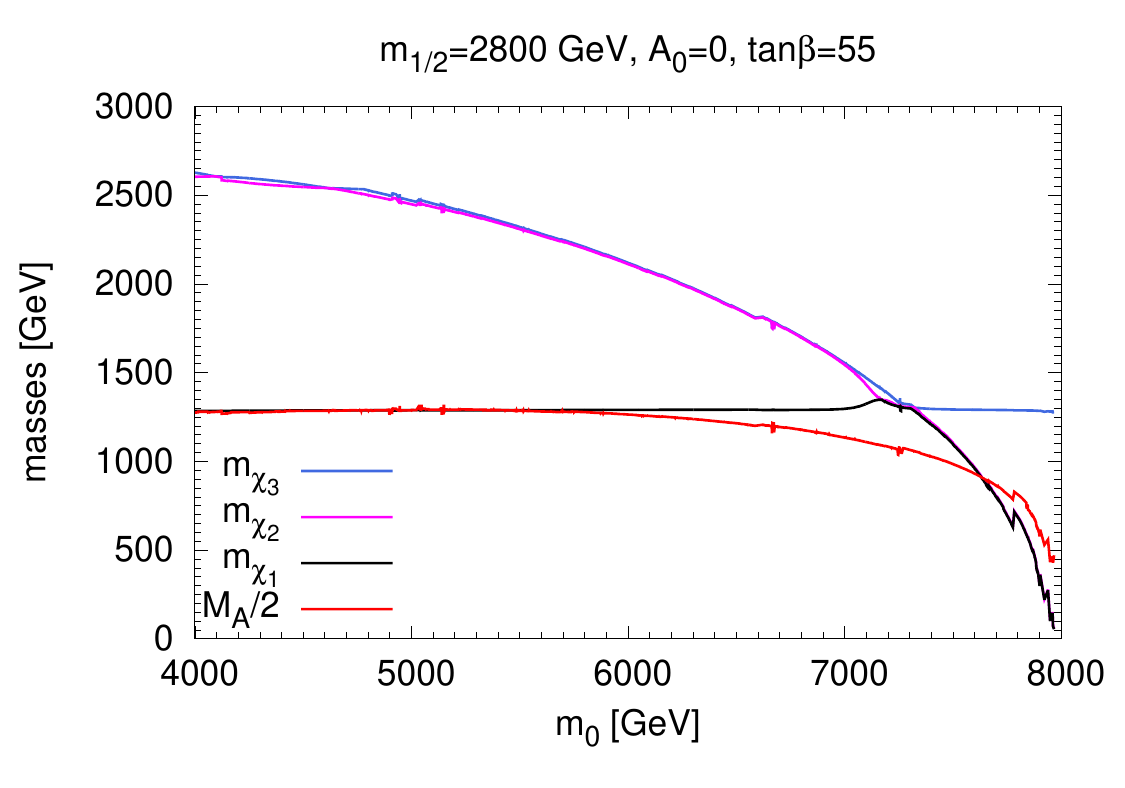} \\
\vspace{-2mm}
\includegraphics[width=0.45\textwidth]{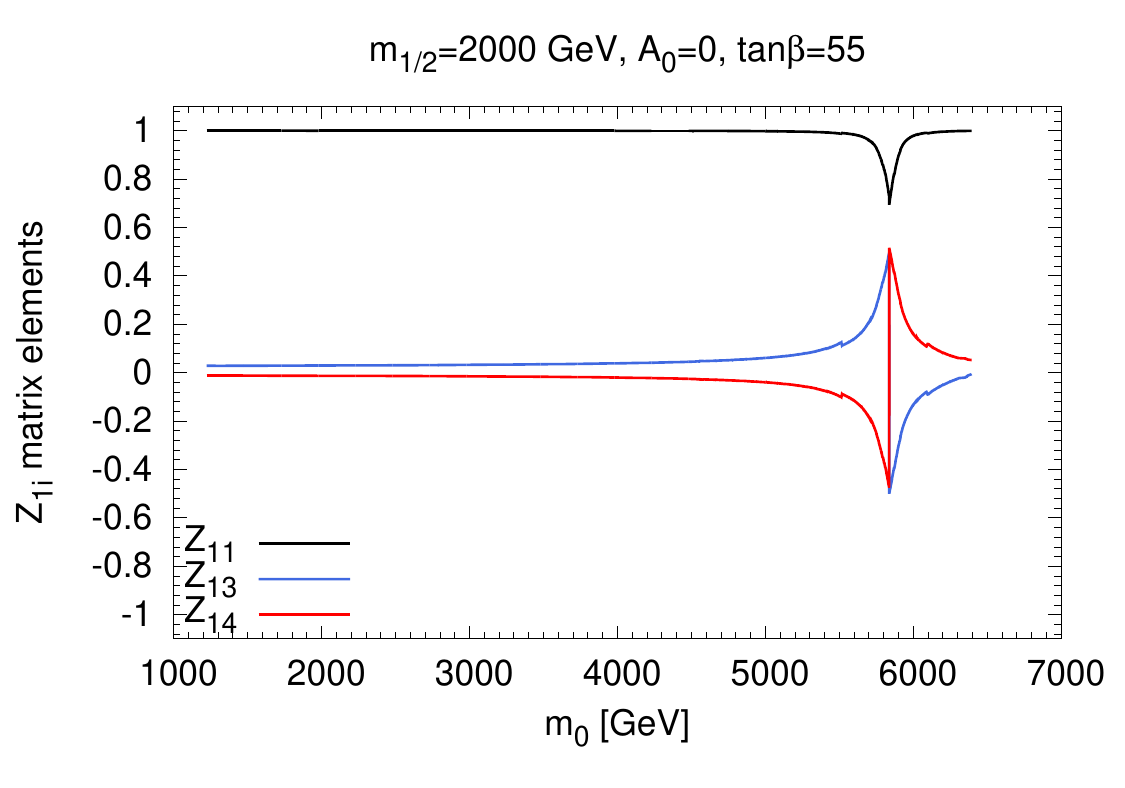} 
\includegraphics[width=0.45\textwidth]{ 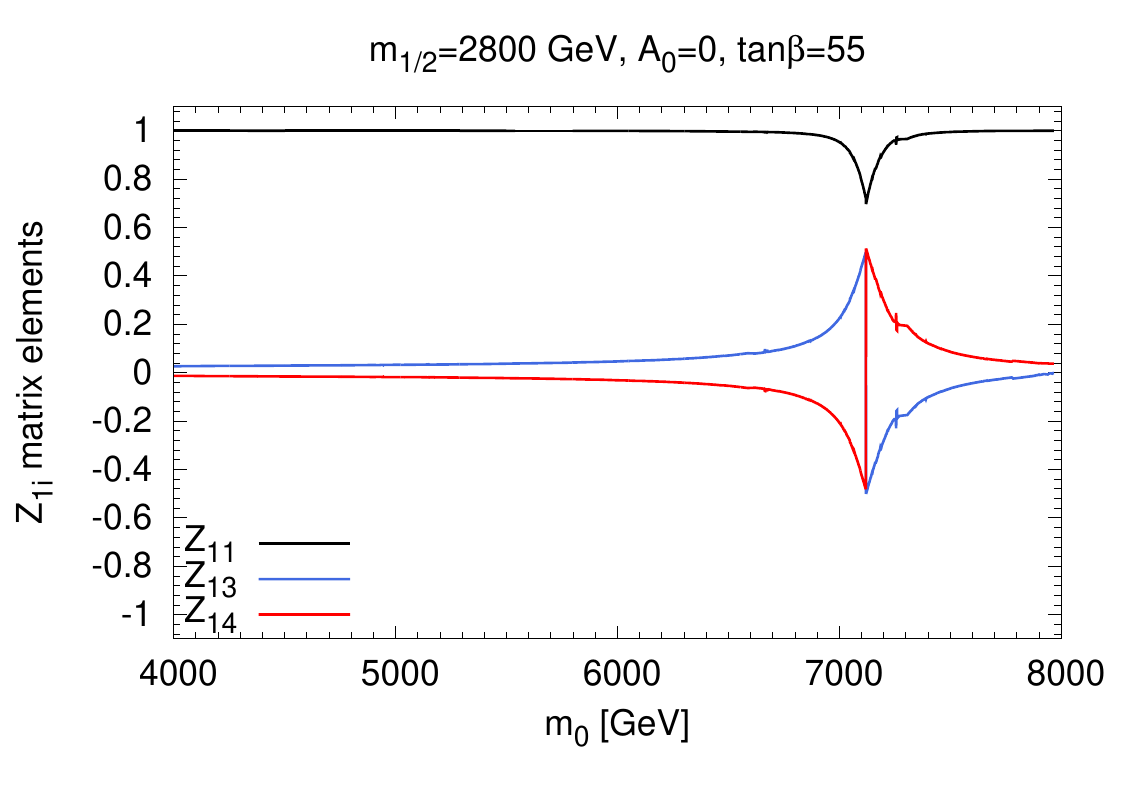} 
\vspace{-4mm}
\caption{\it Cuts across the $(m_{1/2}, m_0)$ planes for $\tan \beta = 55$, $A_0 = 0$ and $\mu > 0$  for  varying $m_0$ and fixed
$m_{1/2} = 2.0$ TeV (left panels) and $2.8$ TeV (right panels), showing the dark matter density
(top panels), particle masses  (middle panels) and the composition of the LSP (bottom panels).}
\label{fig:Slices55}
\end{figure}

Fig.~\ref{fig:Slices55} shows slices through the $(m_{1/2}, m_0)$ plane for $\tan \beta = 55$ with 
$m_{1/2} = 2$ TeV (left panels) and 2.8 TeV (right panels). 
As we see in the top panels, for both values of $m_{1/2}$ the relic density exceeds the {\it Planck} range 
at low $m_{0}$, namely for $m_0 < 5000$ and 7000~GeV, respectively.
For $m_{1/2} = 2000$~GeV, when $m_0 < 1.22$ TeV, the lighter stau is the LSP.
At higher $m_0$, 
the bino becomes the LSP and its mass is close to half the heavy Higgs mass, but it is past the endpoint of the funnel region and, although the relic density is low, it is still above the range allowed by cosmology~\footnote{An exception is presented by a handful of points where the relic density is suppressed at $m_{1/2}\simeq 1.23$~TeV where LSP coannihilations with the lighter stau are important {and} the stau mass is very close to one half of the heavy Higgs mass, leading to enhanced $s$-channel stau annihilations.}. At slightly higher $m_0$, $m_{A/H}$ drops, LSP annihilation via the heavy Higgs poles is suppressed and the relic density is high. However, as one sees in the middle left panel of Fig.~\ref{fig:Slices55}, $m_{H/A}$ rises, LSP annihilation via the heavy Higgs poles becomes possible, 
and the relic density decreases leading to the dip seen in the upper left panel at $m_0 \sim 2$ TeV. However, it does not fall sufficiently before the LSP mass moves away from half the pole mass, 
and begins to rise for $m_0 \gtrsim 2$ TeV. At $m_0 \gtrsim 4$ TeV, $\mu$ and the Higgsino masses begin to drop. 
The Higgs pseudoscalar mass also drops and heavy-Higgs pole annihilation becomes possible again, 
though this time with an increased Higgsino component and a smaller mass splitting with the chargino. 
Then, at $m_0 \approx 5$ TeV the relic density is sufficiently low to become compatible with {\it Planck} and other data.
The ``valley" around 5200~GeV is characteristic of s-channel annihilation via the
$A/H$ bosons and explains the funnel-like feature of the blue strips  in the bottom left panel
of Fig.~\ref{fig:m12_m0_planes}. As seen there, this strip
terminates when $m_{1/2} \sim 2400$~GeV.
Above $m_0 \simeq 5.2$ TeV, the LSP moves off the pole again and the relic density begins to rise again until $m_0 \approx 5.7$ TeV, where the Higgsino becomes the LSP and the focus-point strip is encountered.

Similar behaviour is seen in the right panels of Fig.~\ref{fig:Slices55} for $m_{1/2} = 2.8$ TeV. While the relic density shows a minimum
around $m_0 \approx 5.5$ TeV, 
we see that the funnel alone is not sufficient in obtain the 
{\it Planck} value for relic density. 
As the LSP moves off the pole 
at higher $m_0$, the relic density begins to rise until
the Higgsino component of the LSP increases sufficiently to raise the cross section and hence reduce the relic density. The relic density hits a local minimum and then rises
as the LSP becomes a pure Higgsino, 
subsequently falling through the {\it Planck} value when the Higgsino mass falls to $\sim$1 TeV.

All these features are present also in the bottom right panel of
Fig.~\ref{fig:m12_m0_planes}, which displays the corresponding
$(m_{1/2}, m_0)$ plane for $\tan \beta = 56$. 
As for $\tan \beta = 55$,
the focus-point strip terminates because the
renormalization-group running of the top Yukawa coupling diverges at larger $m_{1/2}$.

\begin{figure}[ht!]
\includegraphics[width=0.45\textwidth]{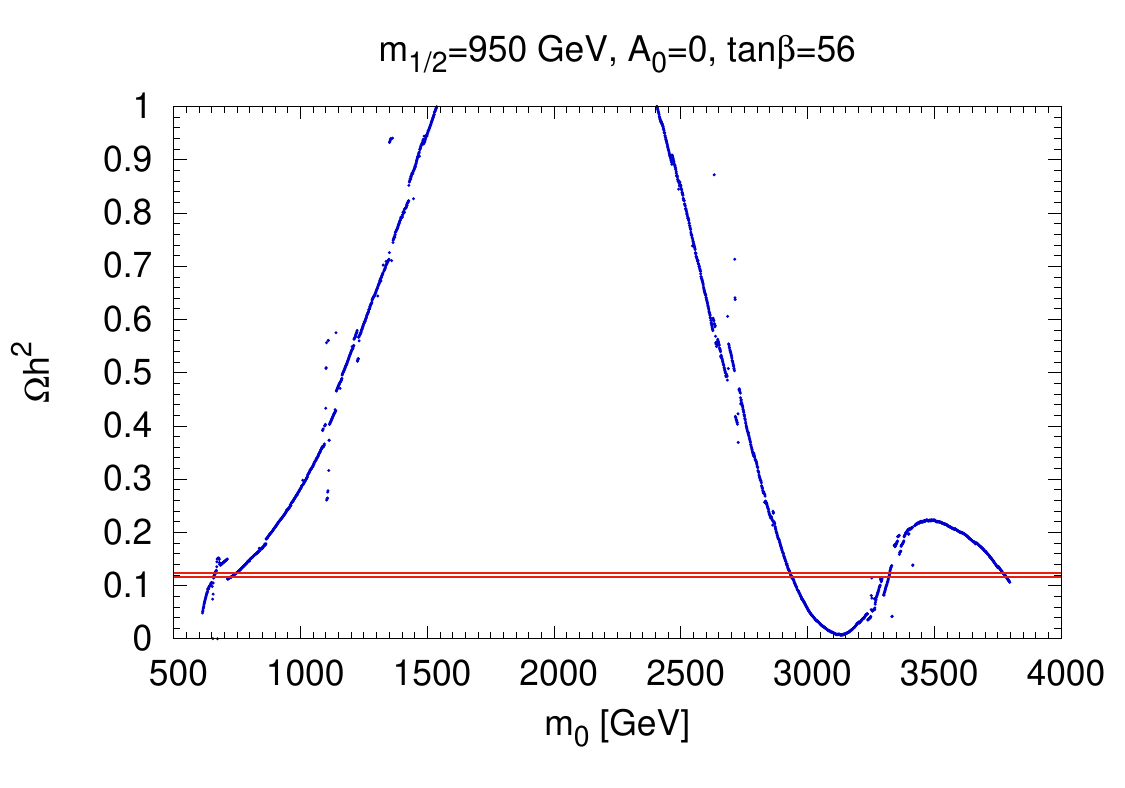} 
\includegraphics[width=0.45\textwidth]{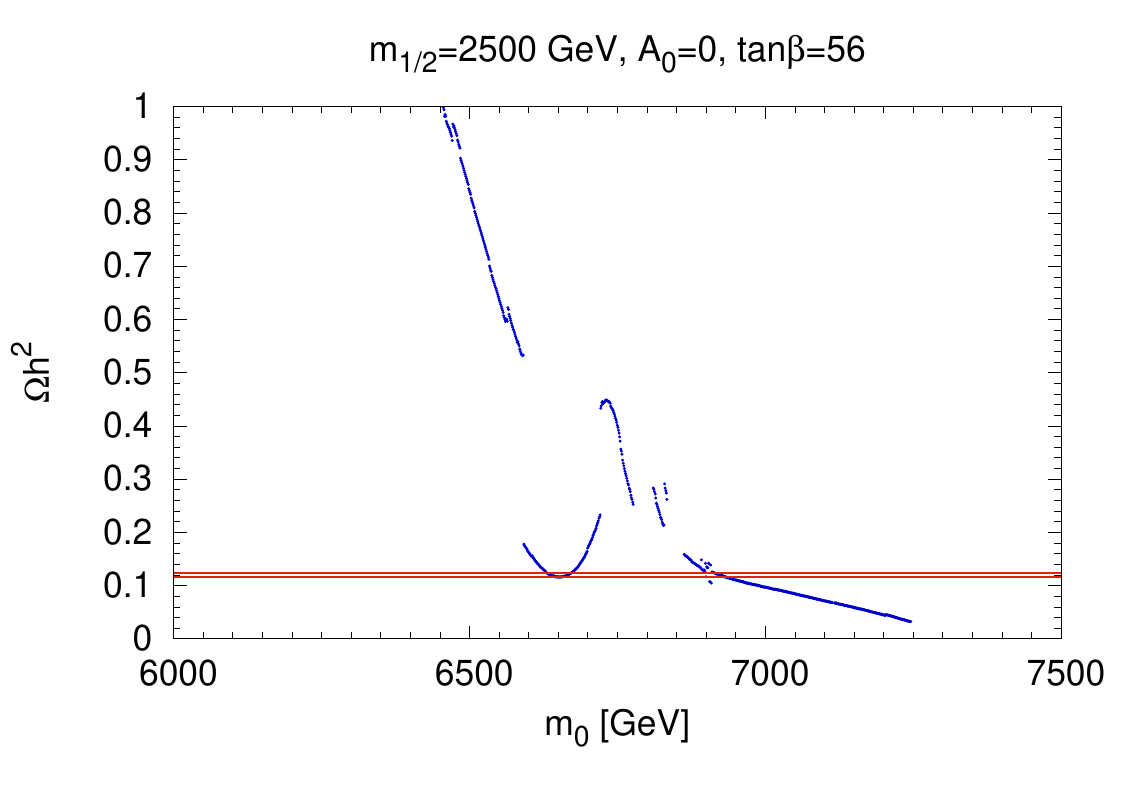} \\
\vspace{-2mm}
\includegraphics[width=0.45\textwidth]{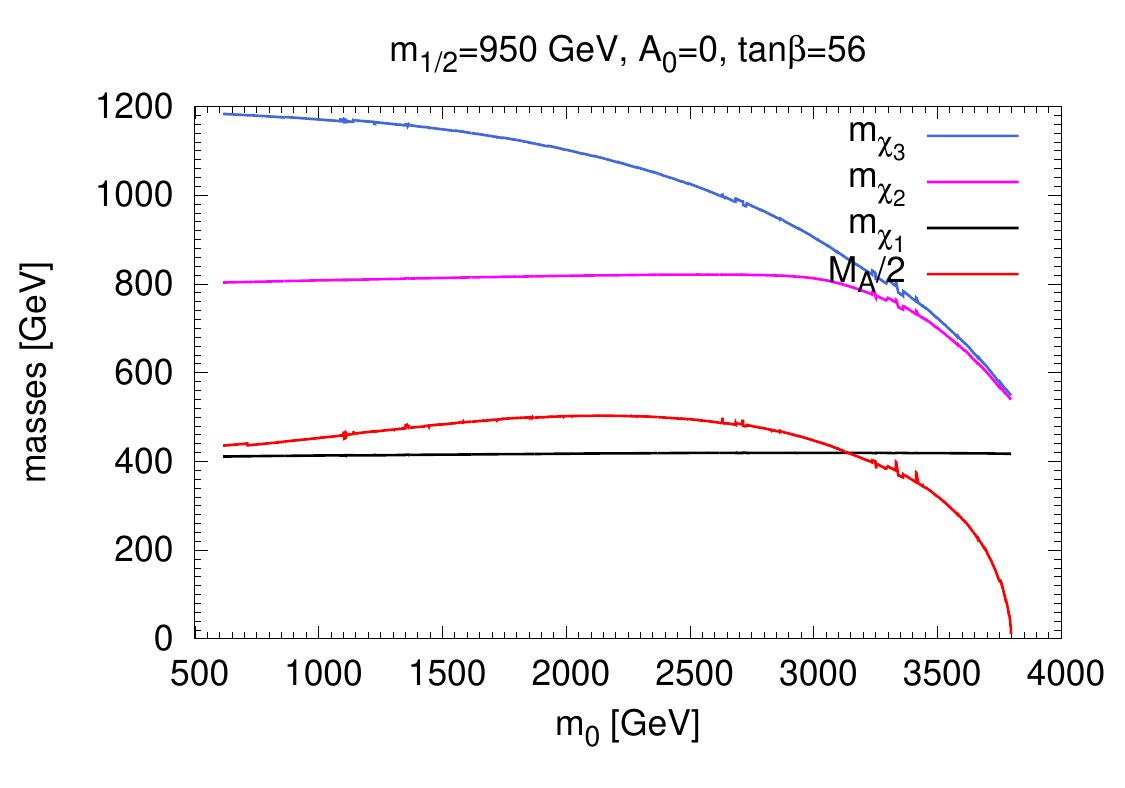} 
\includegraphics[width=0.45\textwidth]{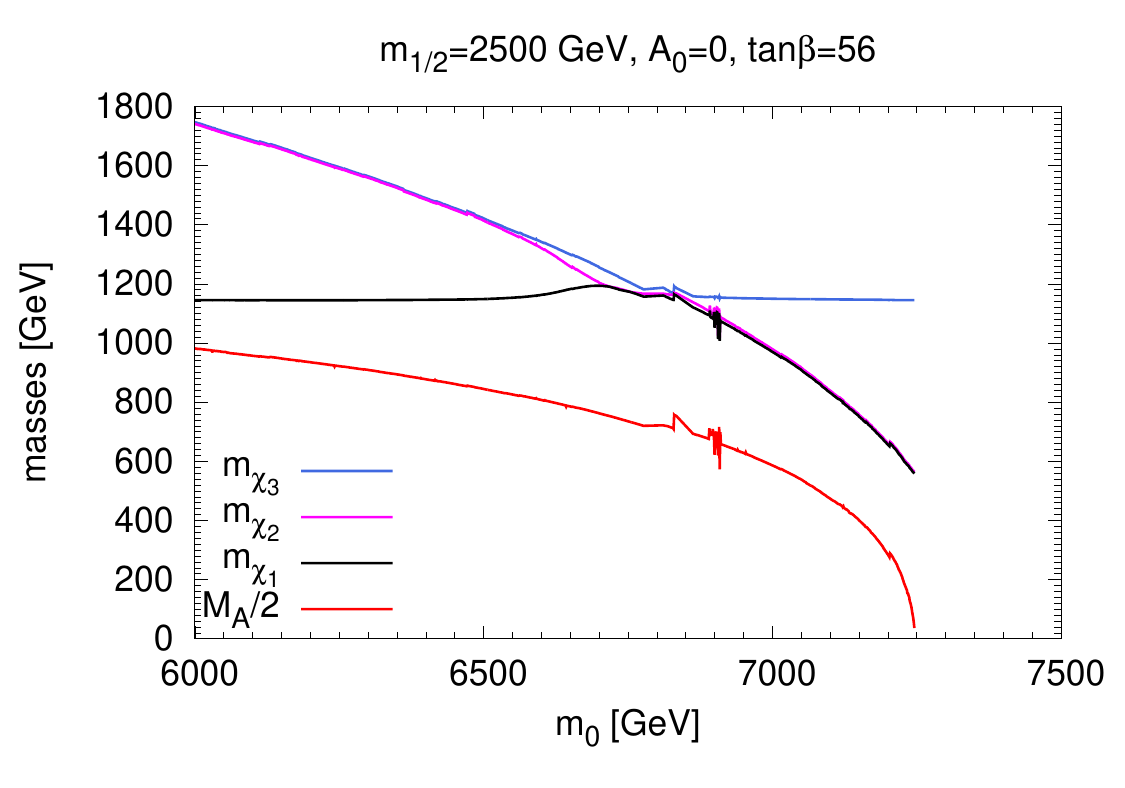} \\
\vspace{-2mm}
\includegraphics[width=0.45\textwidth]{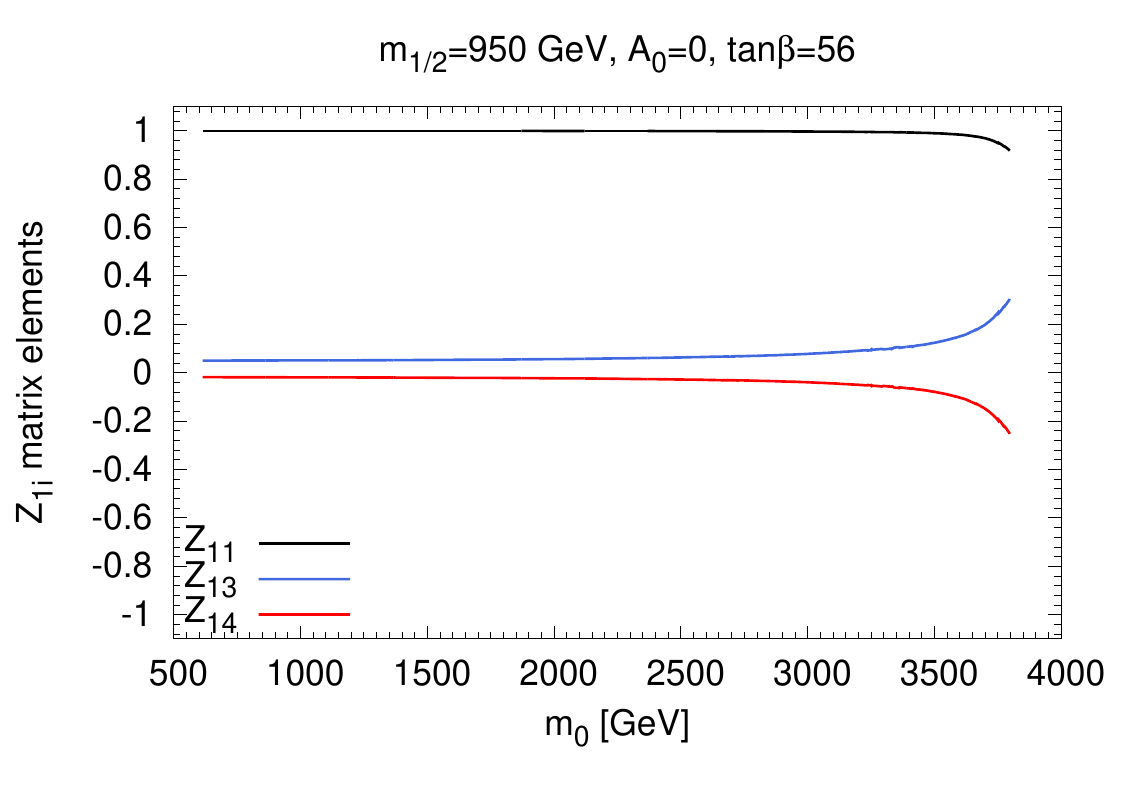} 
\includegraphics[width=0.45\textwidth]{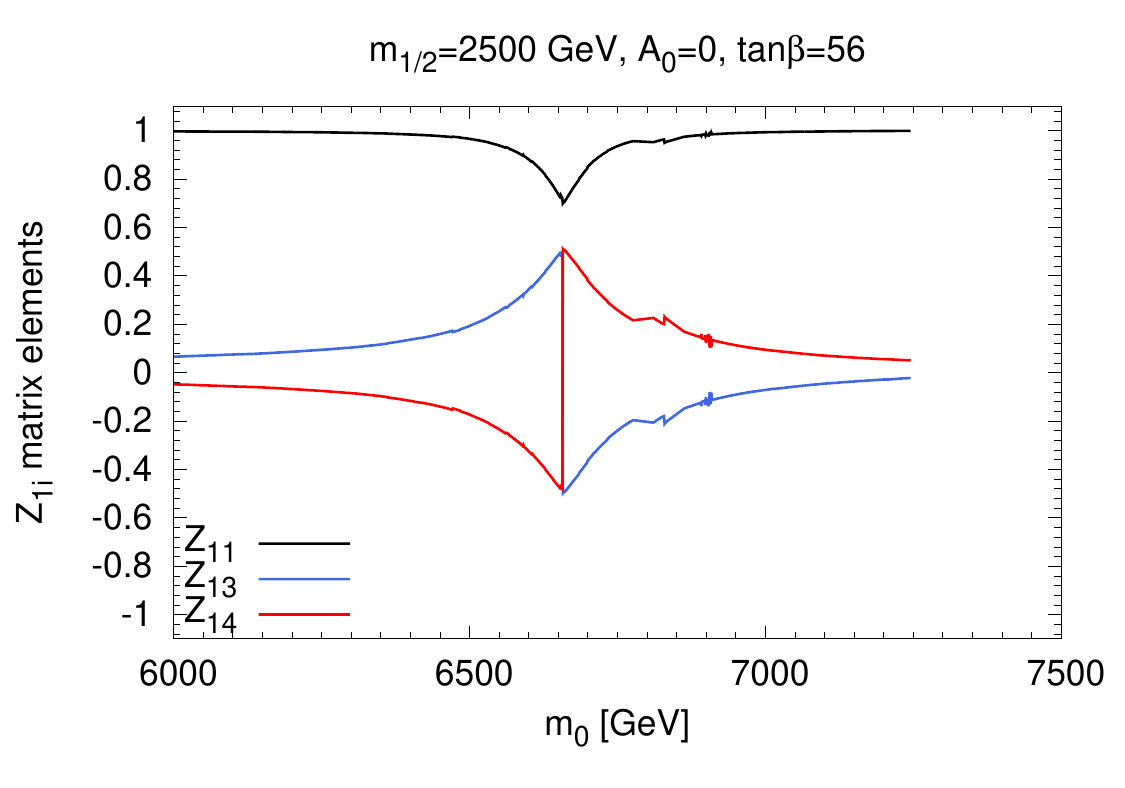} \\
\vspace{-4mm}
\caption{\it  Cuts across the $(m_{1/2}, m_0)$ planes for $\tan \beta = 56$, $A_0 = 0$ and $\mu > 0$  for  varying $m_0$ and fixed
$m_{1/2} = 950$ GeV (left panels) and $2.5$ TeV (right panels), showing the dark matter density
(top panels), particle masses  (middle panels) and the composition of the LSP (bottom panels).
 }
\label{fig:Slices56}
\end{figure}

Fig.~\ref{fig:Slices56}
shows slices across the $(m_{1/2}, m_0)$ plane 
for $\tan \beta = 56$ for the fixed values of $m_{1/2} = 950$ (upper panels) and 2500~GeV (lower panels).
In the first case we see clearly that the {\it Planck} value of the relic density is attained
at four different values of $m_0$, corresponding to the funnel, two strips corresponding to the coannihilation/funnel combination and finally
the focus-point region. 
In the left panel we see again that the LSP is near the heavy-Higgs pole at both low and high $m_0$. 
On the other hand, for $m_{1/2} = 2500$~GeV
(right panels of Fig.~\ref{fig:Slices56}) we see that the {\it Planck} density is attained only
for two very similar values of $m_0 \gtrsim 6700$~GeV that are in well-tempered region and at higher $m_0 \sim 6900$ GeV when the focus point strip is reached.  \\

\section{LHC Constraints}
\label{sec:LHC}

\subsection{Sparticle Searches}
\label{sec:sparticles}

\begin{figure}[ht!]
\includegraphics[width=0.45\textwidth]{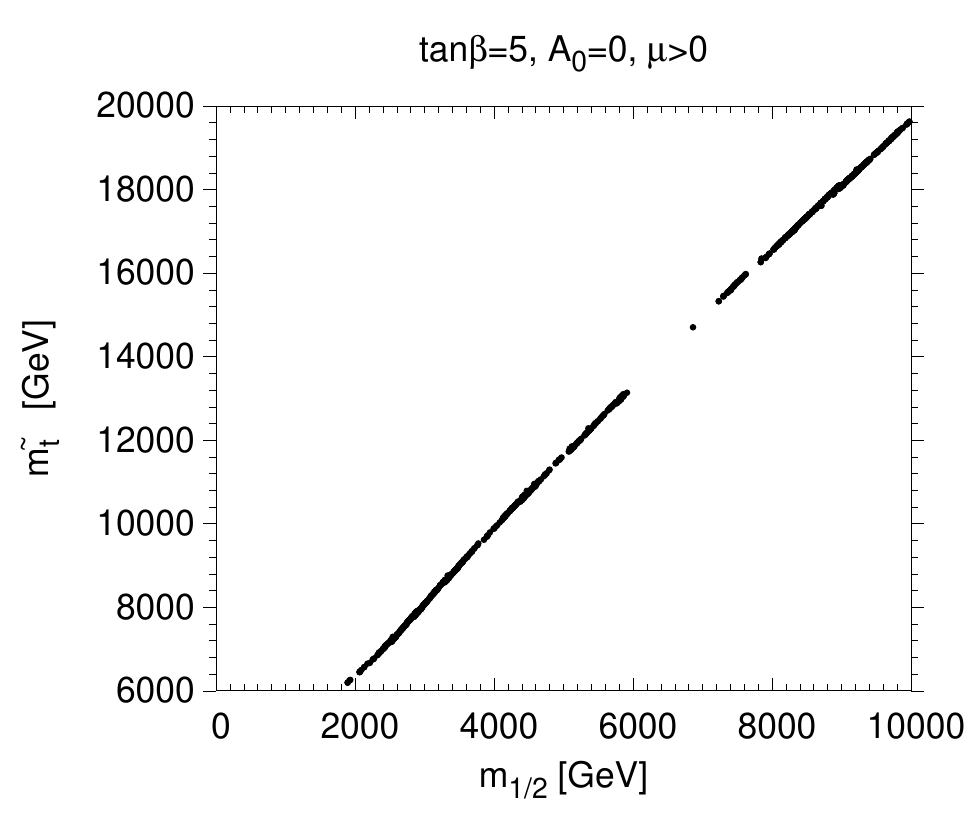} 
\includegraphics[width=0.45\textwidth]{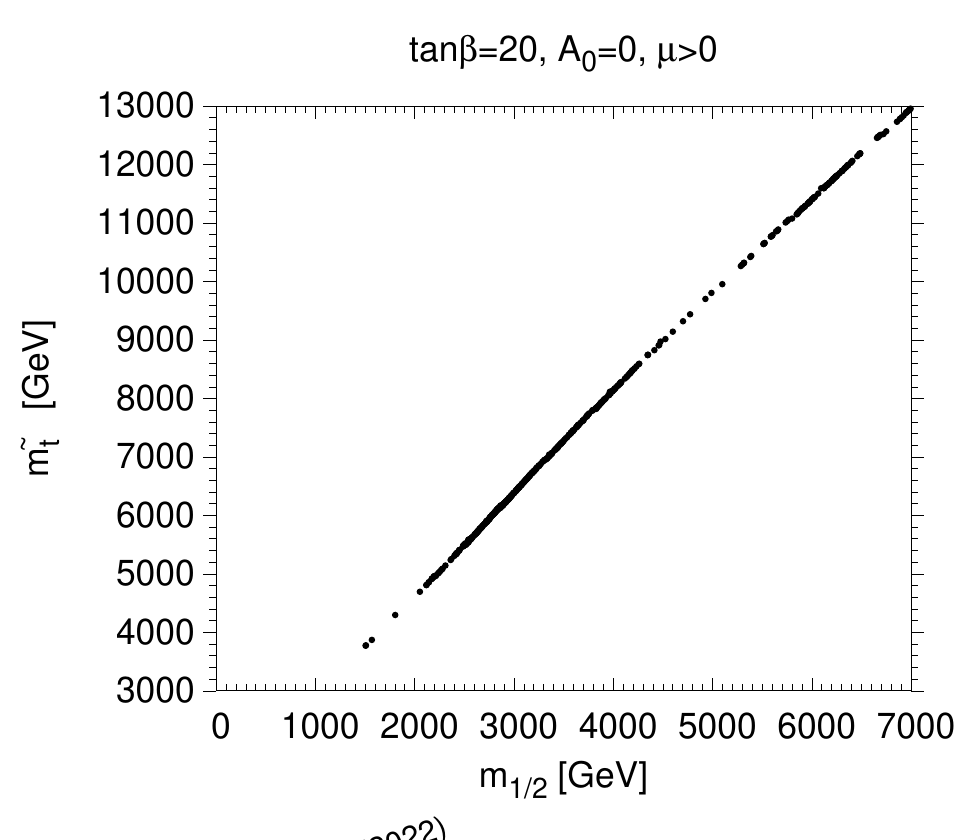} \\
\includegraphics[width=0.45\textwidth]{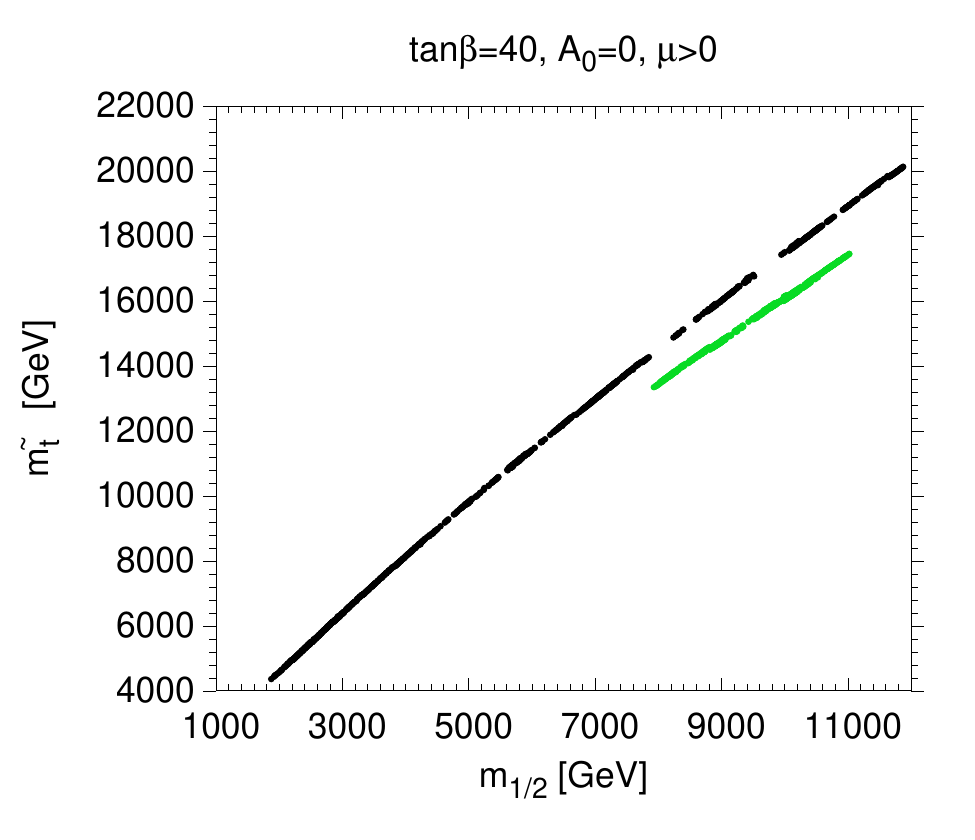}
\includegraphics[width=0.45\textwidth]{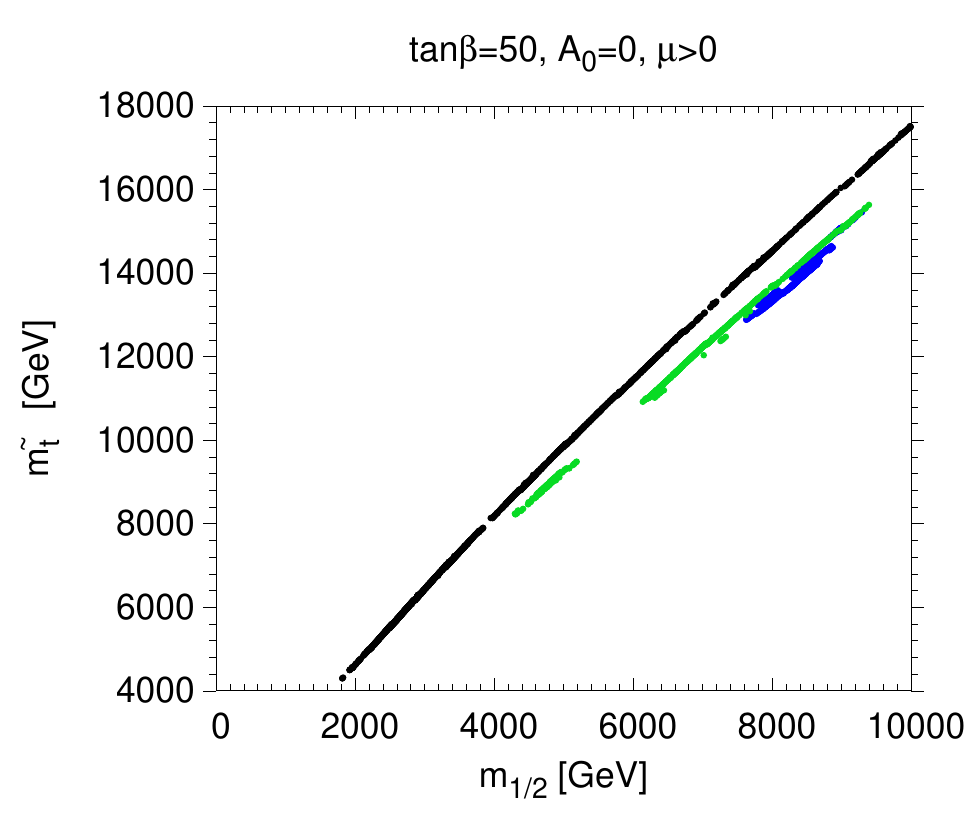} \\
\includegraphics[width=0.45\textwidth]{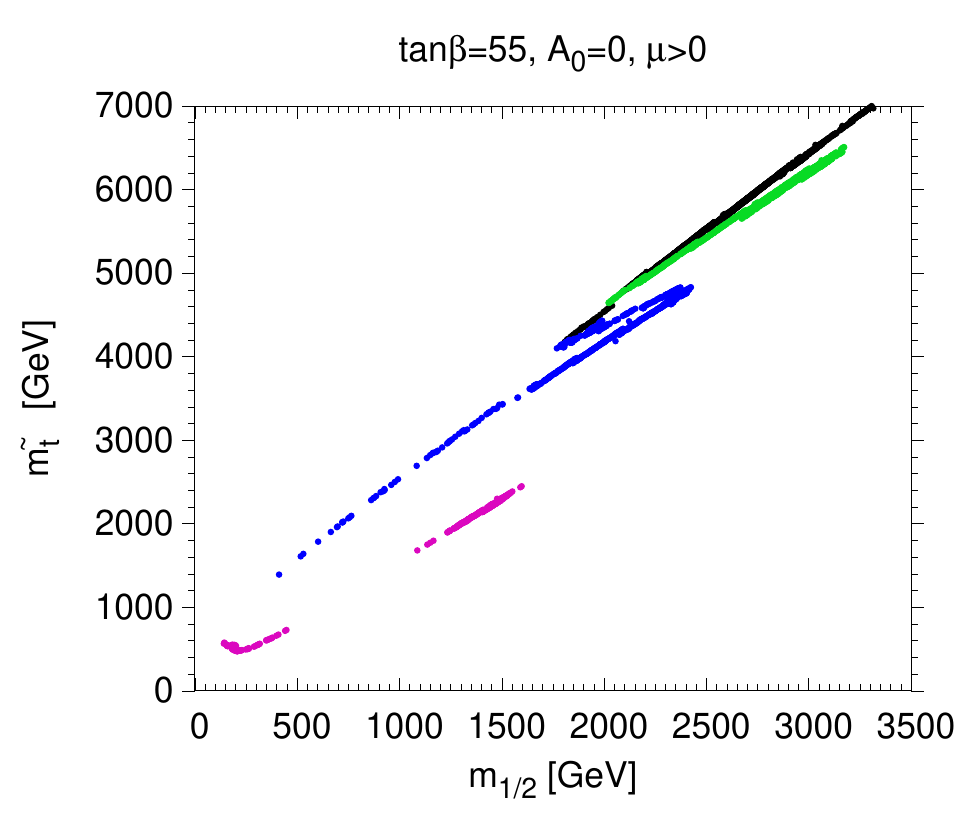} 
\includegraphics[width=0.45\textwidth]{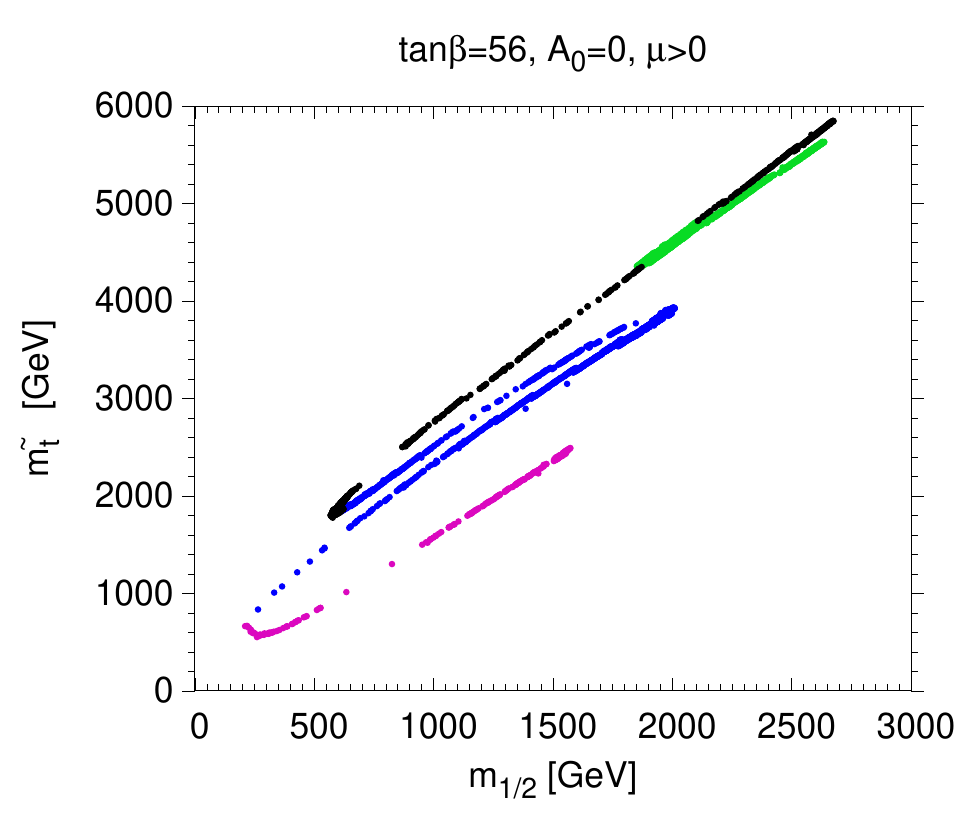} 
\caption{\it The lighter stop mass 
along the dark matter strips
for $\tan \beta = 5, 20, 40, 50, 55 \; and \; 56$, with $A_0 = 0$ and $\mu > 0$.}
\label{fig:mstop}
\end{figure}

\begin{figure}[ht!]
\includegraphics[width=0.45\textwidth]{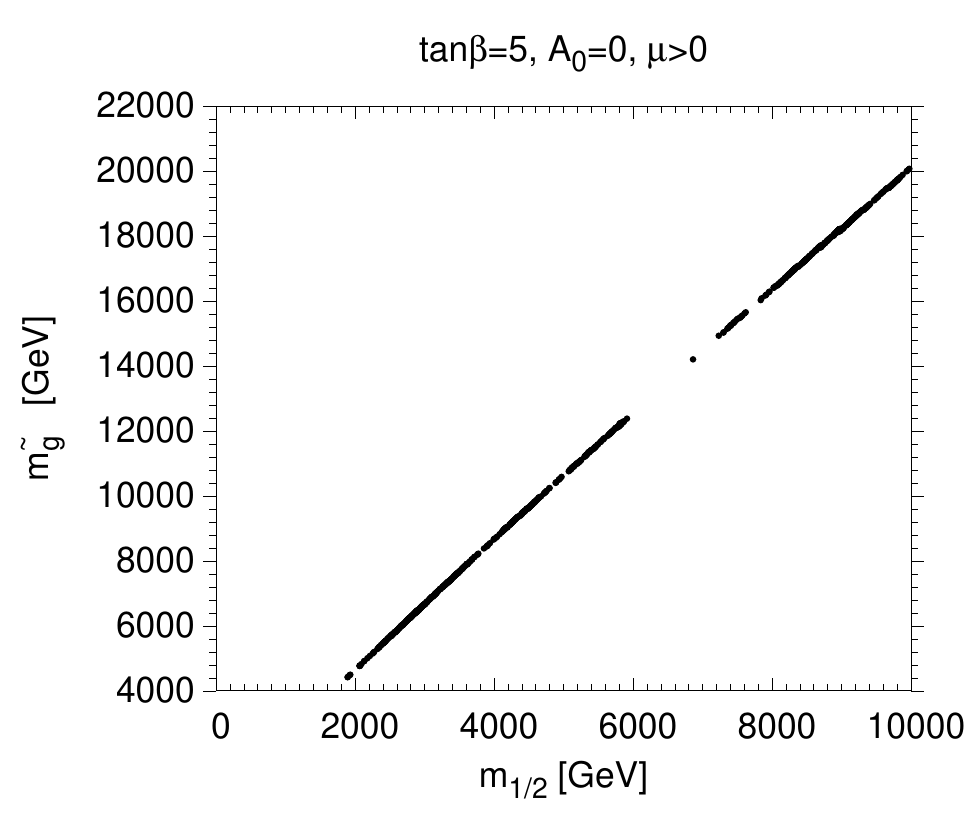} 
\includegraphics[width=0.45\textwidth]{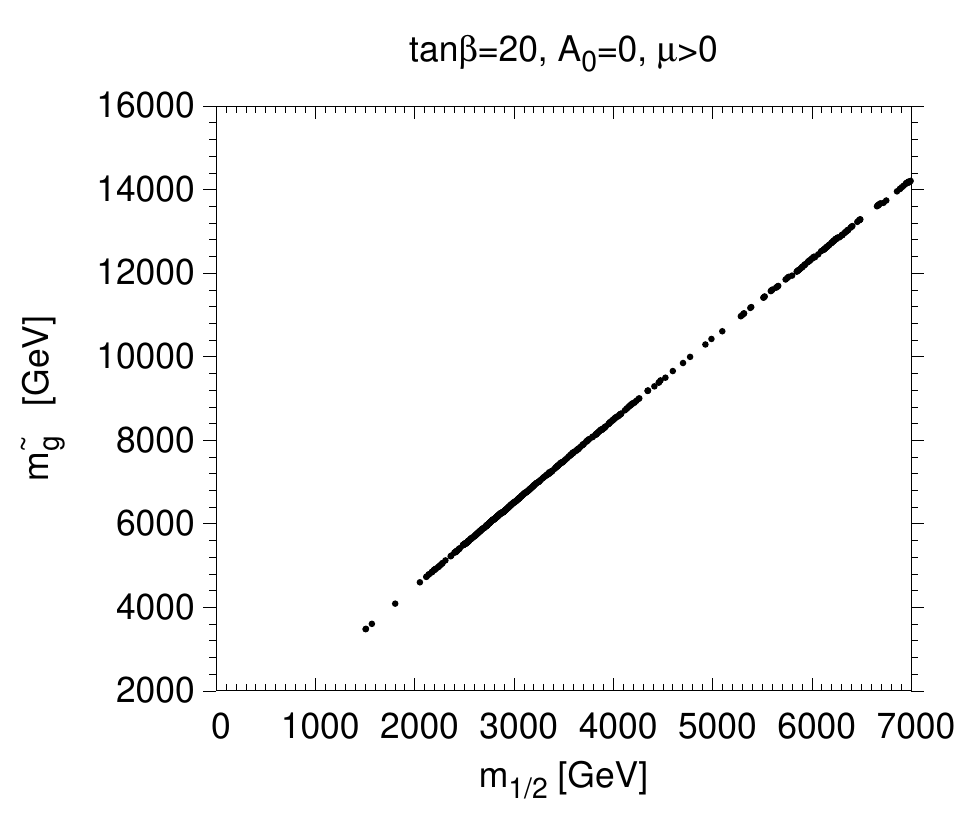} \\
\includegraphics[width=0.45\textwidth]{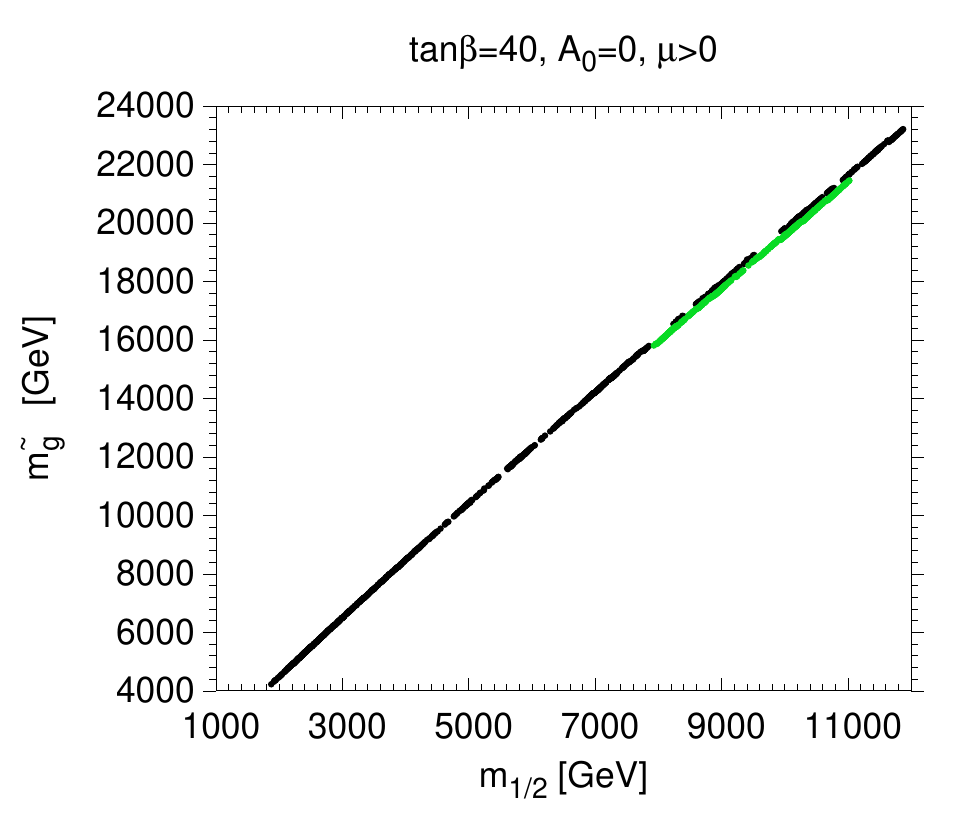}
\includegraphics[width=0.45\textwidth]{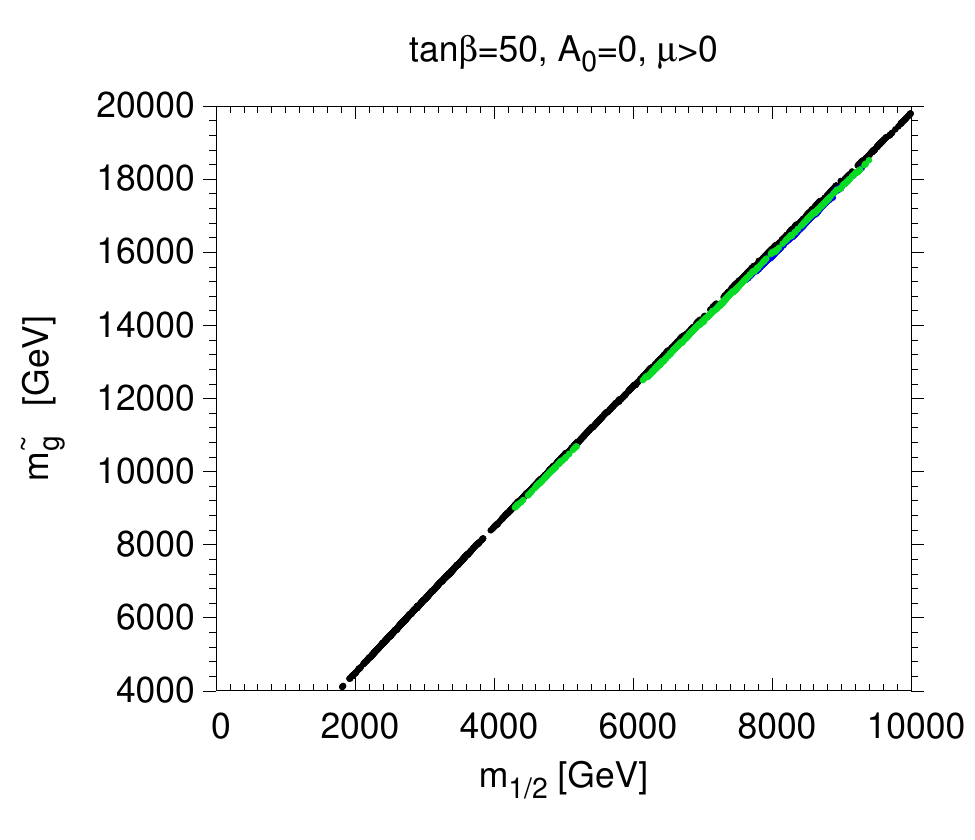} \\
\includegraphics[width=0.45\textwidth]{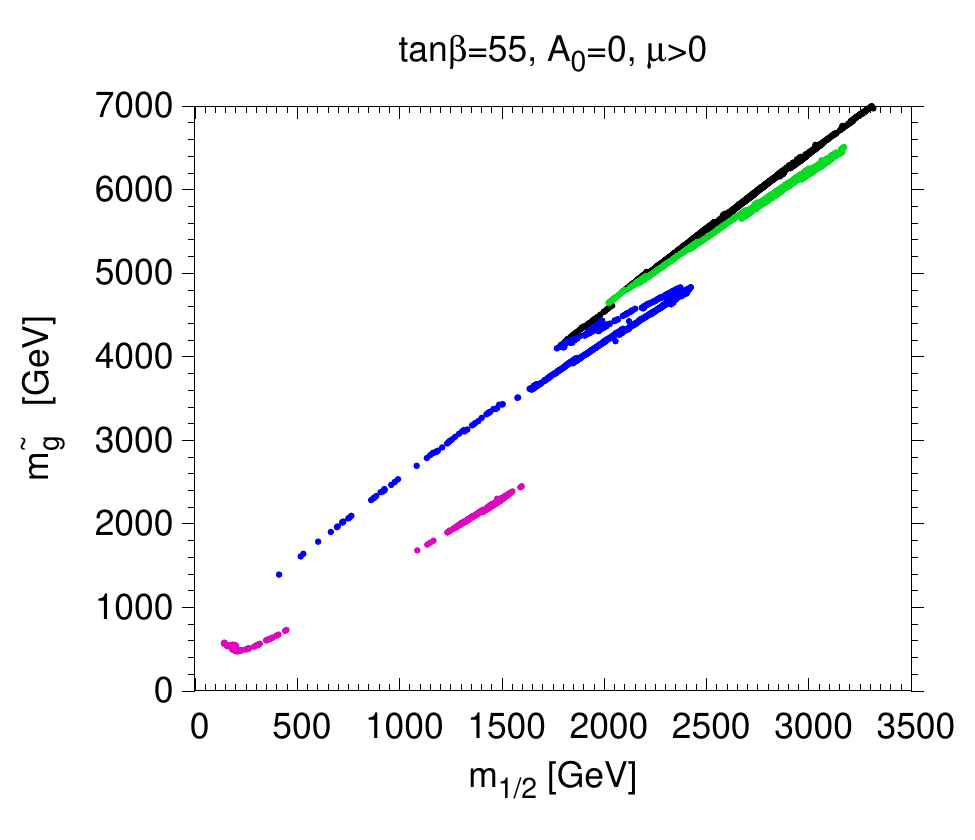} 
\includegraphics[width=0.45\textwidth]{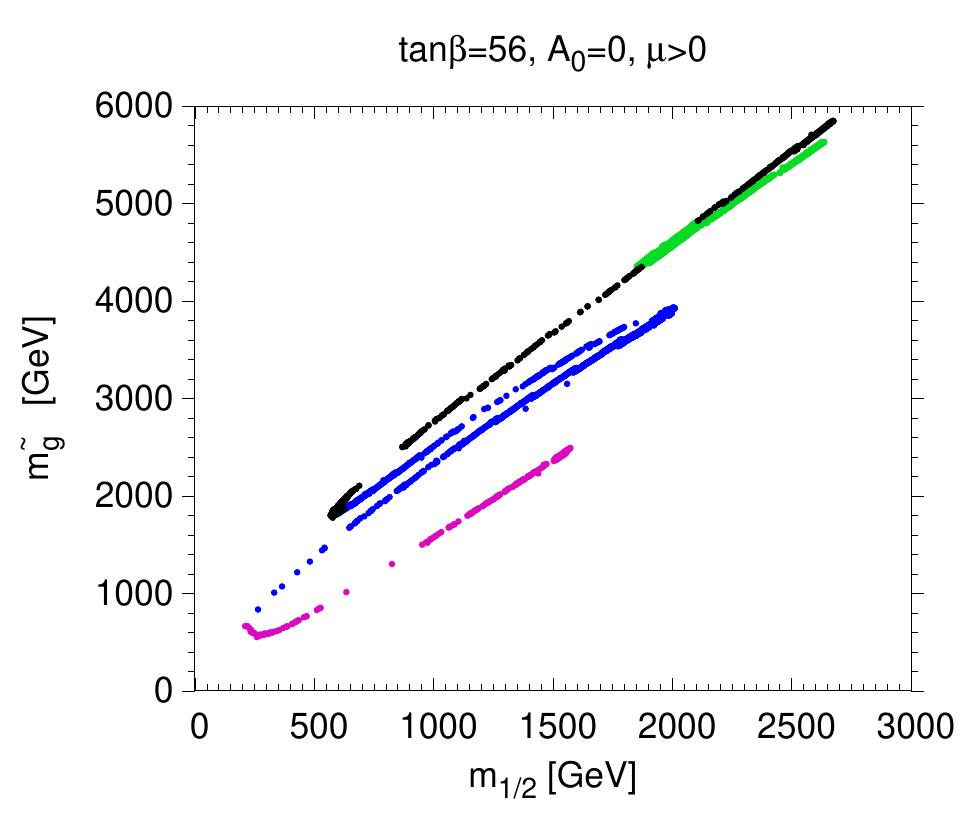} 
\caption{\it The gluino  mass 
along the dark matter strips
for $\tan \beta = 5, 20, 40, 50, 55 \; and \; 56$, with $A_0 = 0$ and $\mu > 0$.}
\label{fig:mgluino}
\end{figure}

The headline sparticle searches at the LHC are those for squarks and gluinos,
both of which are sensitive to masses $\lesssim 2$~TeV in simplified models, e.g.,
assuming that some specific decay mode has a branching ratio of 100\% and
the LSP mass is negligible. The squark limit would correspond to $m_0 \gtrsim 2$~TeV
for small $m_{1/2}$, and the gluino limit would correspond to $m_{1/2} \gtrsim 800$~GeV.
These constraints do not impinge on the $A_0 = 0$ regions of interest in Fig.~\ref{fig:m12_m0_planes}.  Indeed the mass spectrum we find is quite heavy. In Fig.~\ref{fig:mstop}, we show the lighter stop mass as a function of $m_{1/2}$ along the dark matter strips. As one can see, the stop masses are similar to (though slightly smaller than) $m_0$.
Similarly in Fig.~\ref{fig:mgluino}, we show the gluino mass along the strips. The gluino mass is typically $\simeq 2 - 3 m_{1/2}$. 
On the other hand, as we discuss in Section~6, for $A_0 = 3 \ m_0$ searches for light stops with
a compressed spectrum exclude masses just below the region of the stop coannihilation
strip allowed by the {\tt FeynHiggs~2.18.1} calculation of $m_h$ at the 1-$\sigma$ level.
The stop masses in this case are nearly degenerate with the LSP mass. The gluino mass is again roughly $2 - 3 m_{1/2}$. 

\subsection{Higgs Mass Constraint}
\label{sec:mH}

The world average of the Higgs boson mass is $125.25 \pm 0.17$~GeV~\cite{PDG}. In order
to calculate $m_h$ in the CMSSM we use the {\tt FeynHiggs~2.18.1} code \cite{FH}, which returns
an uncertainty estimate $\Delta m_h$ as well as the best estimate of $m_h$
for each choice of model parameters. The uncertainty 
estimate is quite variable being, e.g., significantly smaller for $A_0 = 0$ than for 
$A_0 = 3 \ m_0$. Conservatively, for $A_0 = 0$ we assume that $\Delta m_h = 1.5$~GeV, 
which corresponds to somewhat more than twice the typical numerical estimate of $\Delta m_h$, commenting in passing 
on the prospective implications for $\Delta m_h \sim 0.5$~GeV. On the other hand, the typical
{\tt FeynHiggs~2.18.1} uncertainty estimate for $A_0 = 3 \ m_0$ is much larger, $\Delta m_h \gtrsim 3$~GeV (and $\tan \beta$ dependent),
which does not exclude any portions of the stop coannihilation strips for $A_0 = 3 \ m_0$.

\begin{figure}[ht!]
\includegraphics[width=0.45\textwidth]{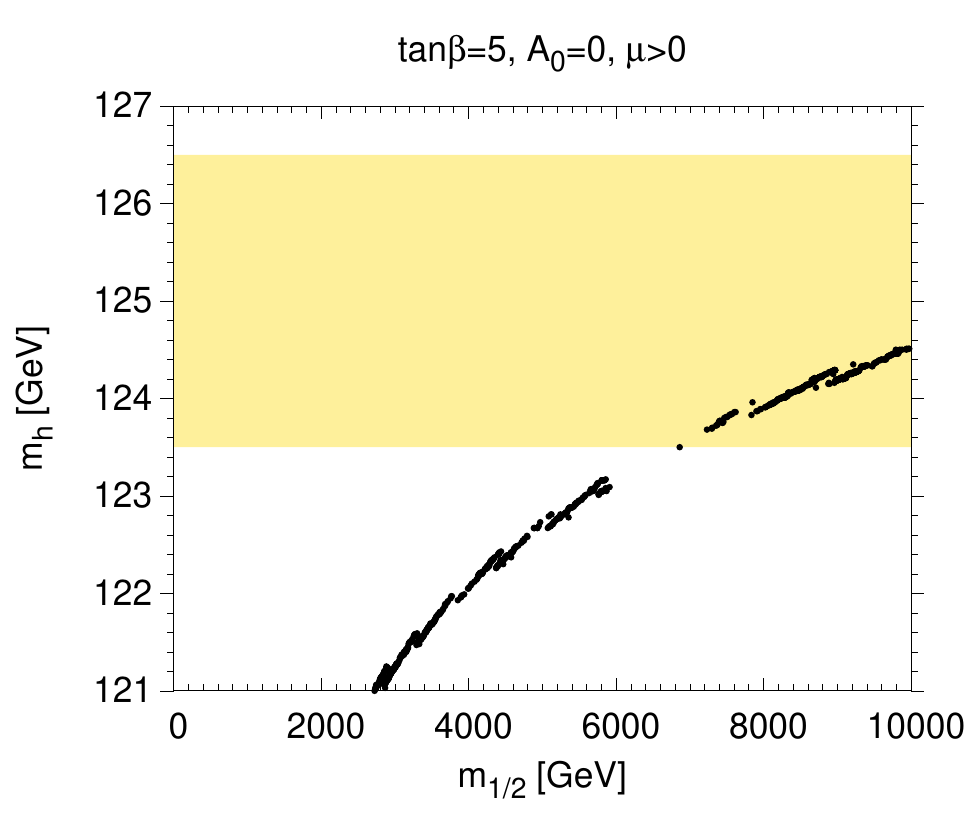} 
\includegraphics[width=0.45\textwidth]{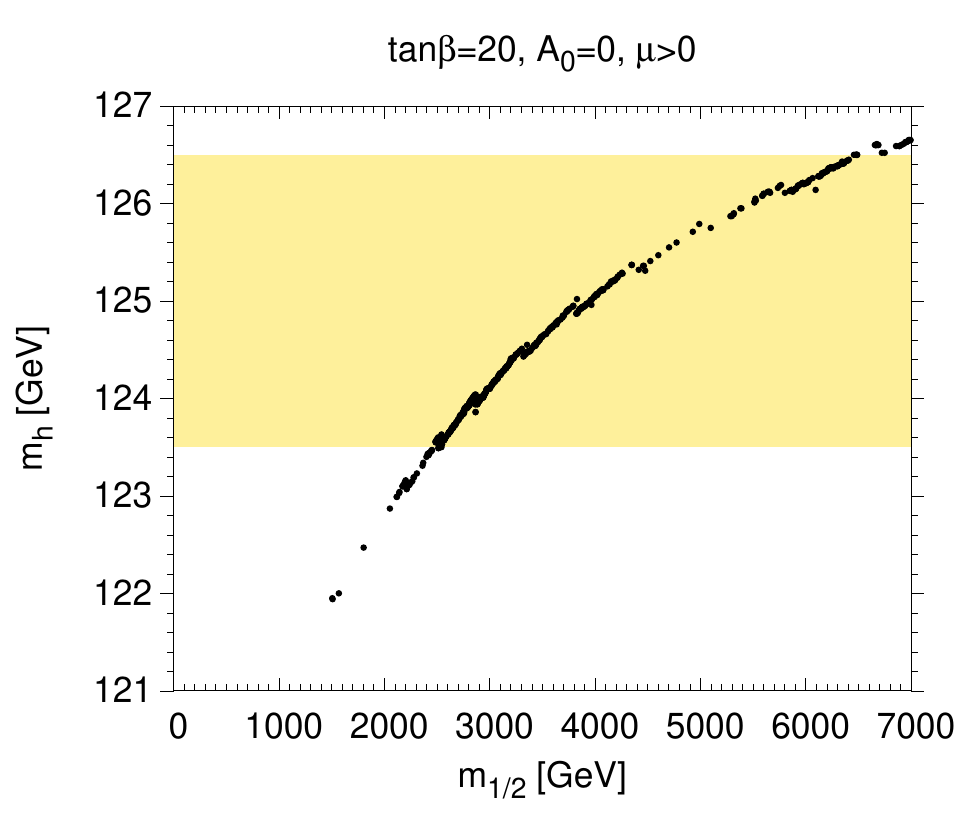} \\
\includegraphics[width=0.45\textwidth]{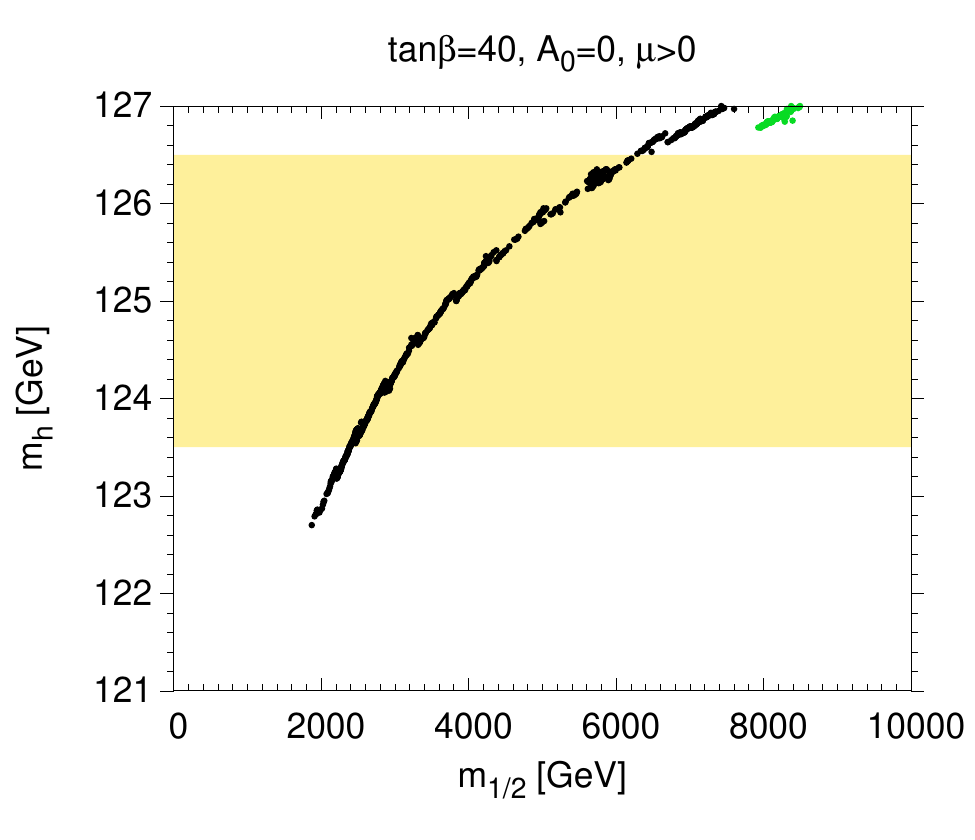}
\includegraphics[width=0.45\textwidth]{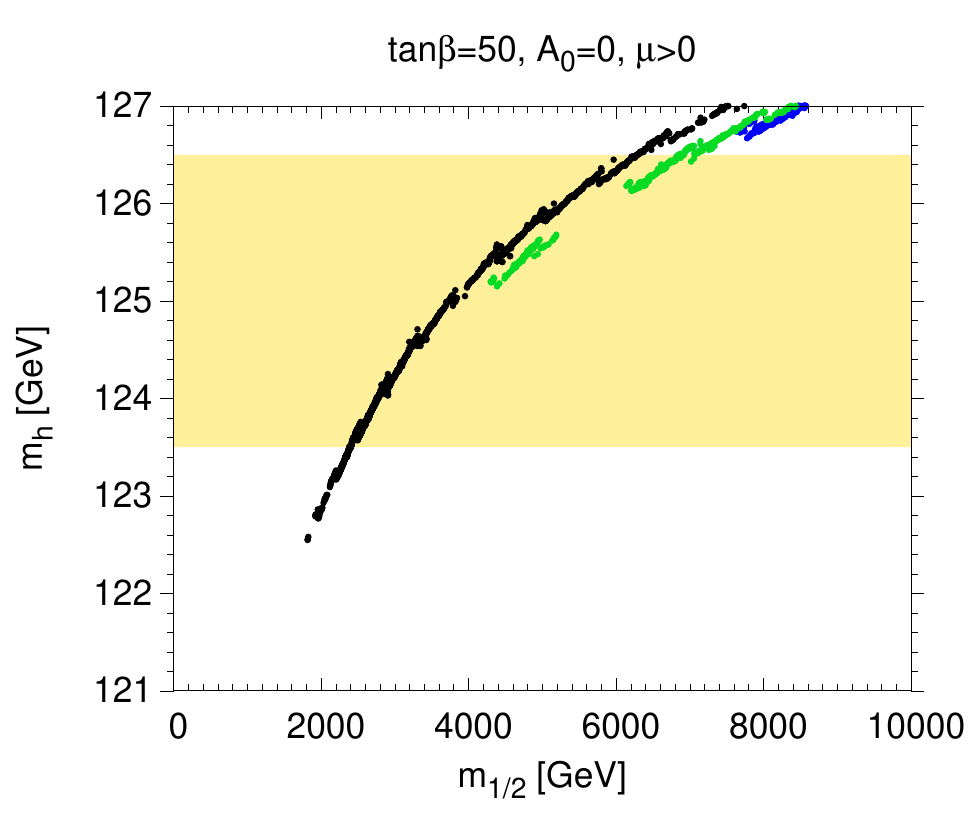} \\
\includegraphics[width=0.45\textwidth]{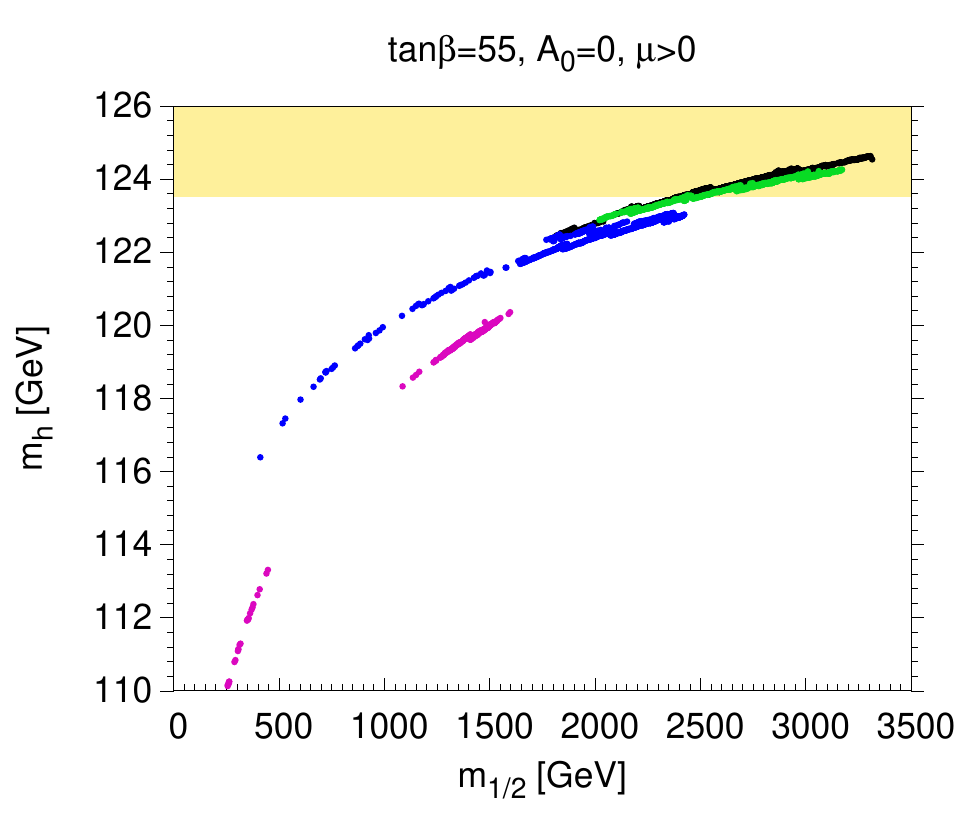} 
\includegraphics[width=0.45\textwidth]{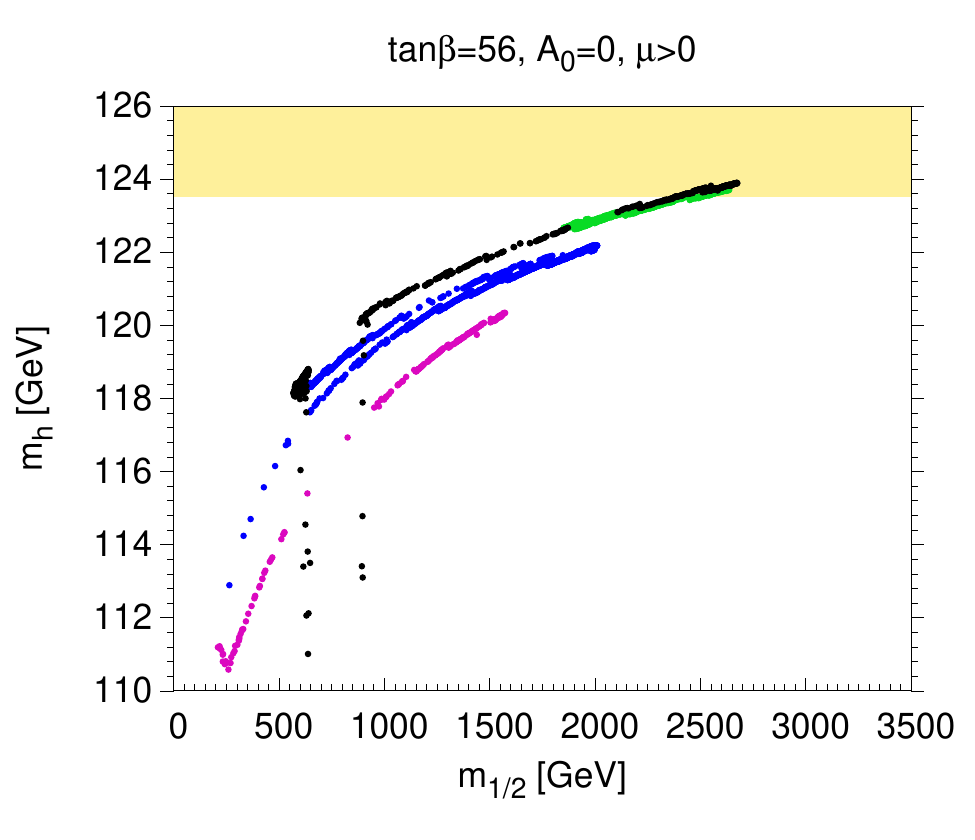} 
\caption{\it The yellow shading indicates the impacts 
of the LHC measurement of $m_h$,
as calculated using {\tt FeynHiggs~2.18.1} and allowing an uncertainty of 1.5~GeV,
on the dark matter strips
for $\tan \beta = 5, 20, 40, 50, 55 \; and \; 56$ with $A_0 = 0$ and $\mu > 0$.}
\label{fig:mh}
\end{figure}

The panels in Fig.~\ref{fig:mh} illustrate how the $m_h$ constraint restricts the allowed ranges
of the dark matter strips for  $\tan \beta = 5, 20, 40, 50, 55$ and 56 when $A_0 = 0$ 
and $\mu > 0$. For $\tan \beta = 5$ (top left panel), $m_{1/2} \gtrsim 7$~TeV is allowed for
$\Delta m_h = 1.5$~GeV, but only a limited range of $m_{1/2} \gtrsim 9$~TeV if
we assume 
$\Delta m_h \simeq 0.5$~GeV. (Note that the focus-point strip continues to larger values of $m_{1/2}$ than those displayed. The apparent endpoint is due to the finite range of our MCMC scan.)  
For $\tan \beta = 20$ (top right panel), the range $2.4 \; {\rm TeV} \lesssim m_{1/2} \lesssim 6.6 \; {\rm TeV}$
is allowed for $\Delta m_h = 1.5$~GeV, whereas only the range 
$3.6 \; {\rm TeV} \lesssim m_{1/2} \lesssim 5 \; {\rm TeV}$ would be favoured for $\Delta m_h \simeq 0.5$~GeV.
Similar ranges of $m_{1/2}$ are allowed (favoured) for $\tan \beta = 40$, as seen in the middle left panel of Fig.~\ref{fig:mh}. We see now that the well-tempered (green) strips at $m_{1/2} > 8$ TeV are excluded as the calculated Higgs mass is too large even when
conservative allowance is made for the uncertainty in the calculation.
When $\tan \beta = 50$ (middle right panel), a portion of the well-tempered strip is
allowed, extending to $m_{1/2} \simeq 7$~TeV for $\Delta m_h = 1.5$~GeV.
However, the additional strips at large $m_{1/2}$ where s-channel annihilations contribute (blue points) lead to values of $m_h$ that are too large. 
When $\tan \beta = 55$ (bottom left panel of Fig.~\ref{fig:mh}), only a portion
of the focus-point strip with $m_{1/2} \gtrsim 2.5$~TeV is allowed for $\Delta m_h = 1.5$~GeV, increasing to $m_{1/2} \gtrsim 3$~TeV
for $\Delta m_h \simeq 0.5$~GeV.
We see also that parts of the coannihilation/well-tempered (green) strips for $\tan \beta = 55$ also fall within the experimental range for the Higgs mass. In contrast, in the funnel/coannihilation/well-tempered strips (blue)
the Higgs mass is always too small, as it is in the funnel region (purple).
Finally, we see that for $\tan \beta = 56$ (bottom right panel) the correct value of $m_h$ is only attained 
when $m_{1/2} \gtrsim 2.5$ TeV at the tip of the black and green strips.~\footnote{We recall that these strips cannot be extended
for this value of $\tan \beta$ because of the divergence of the Yukawa couplings during renormalization group evolution.}

\section{Constraints on Dark Matter Interactions}
\label{sec:DM}

\subsection{Spin-Independent Dark Matter Scattering}
\label{sec:sidirect}

Many experiments have used massive nuclear targets to probe
coherent spin-independent LSP-nucleon scattering. The most
recent such experiment is LUX-ZEPLIN (LZ) \cite{LZ}, which uses a xenon target
to establish a stronger upper limit on the spin-independent LSP-nucleon
scattering cross section than the previous XENON1T \cite{XENON}, PandaX-4T \cite{PANDAX},
LUX \cite{LUX} and DEAP-3600 \cite{DEAP} experiments.

The computation of the spin-independent cross section 
that we use was described in detail in \cite{sospin}.
As is well known, the spin-independent cross section is very sensitive to the quark matrix elements,
$\langle N| {\bar q} q| N \rangle$. These can be expressed in terms of $\sigma$ terms, notably the $\pi-N$ $\sigma$ term $\Sigma_{\pi N}$ and $\sigma_s$ (see \cite{sospin} for more detail). Here
we have used $\Sigma_{\pi N} = 46 \pm 11$ MeV and $\sigma_s = 35 \pm 16$ MeV, respectively. The
corresponding typical uncertainties in the elastic cross sections are of order 10-15 \%.

\begin{figure}[ht!]
\includegraphics[width=0.45\textwidth]{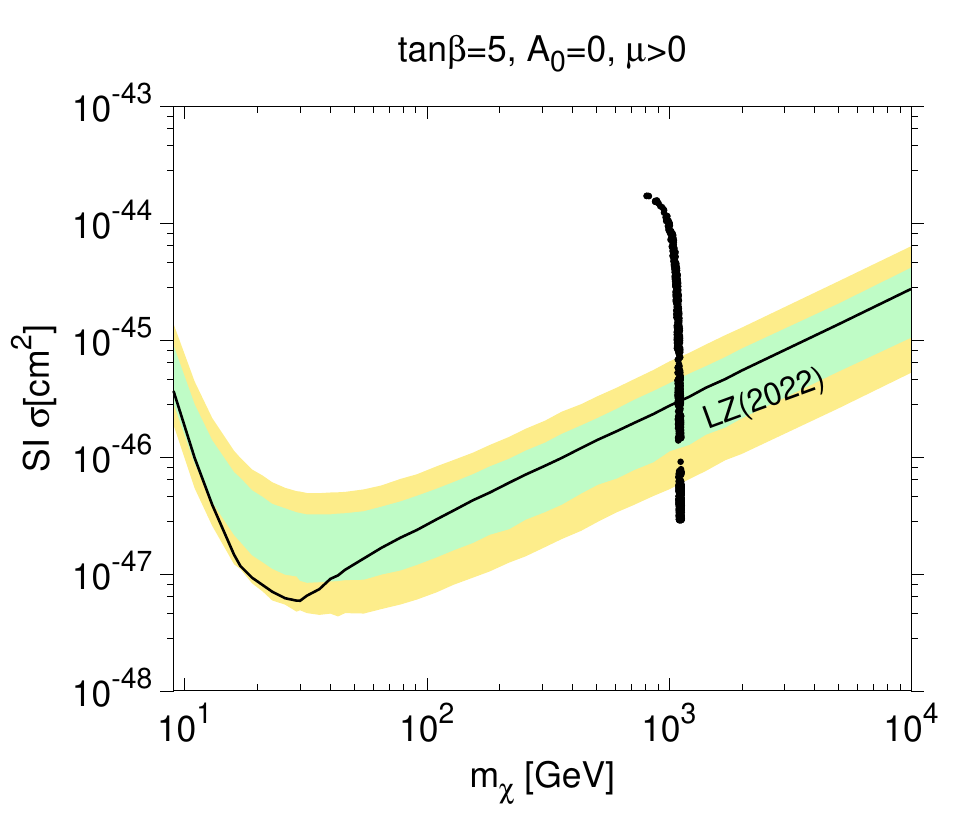} 
\includegraphics[width=0.45\textwidth]{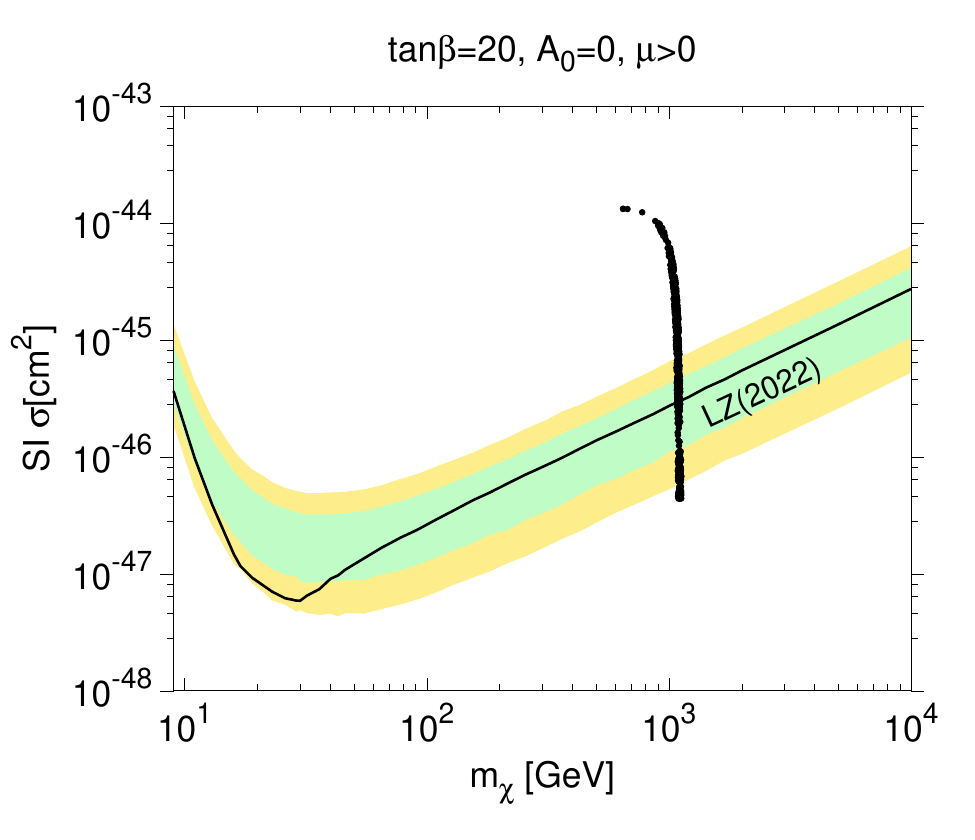} \\
\vspace{-2mm}
\includegraphics[width=0.45\textwidth]{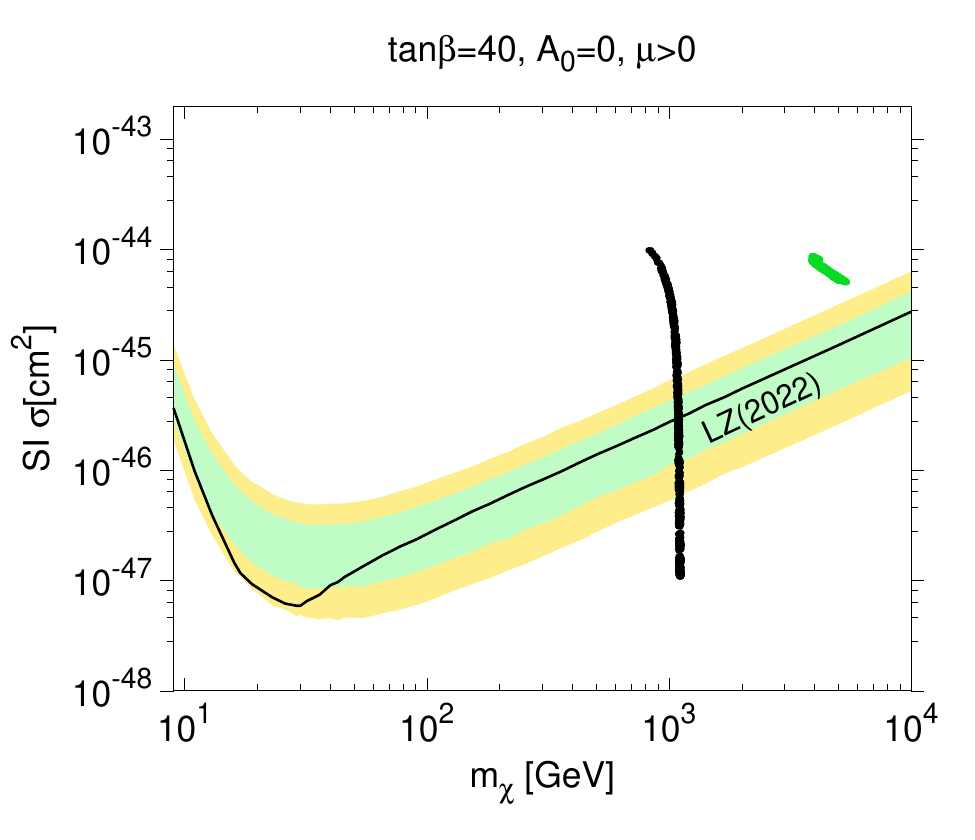}
\includegraphics[width=0.45\textwidth]{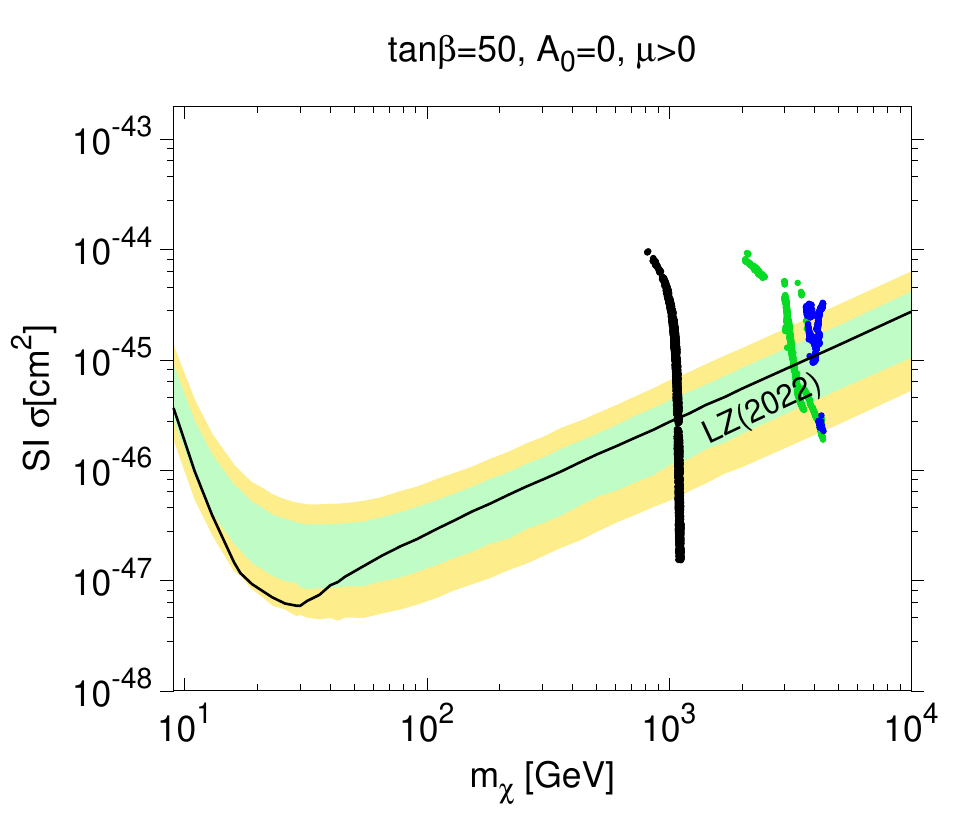} \\
\vspace{-2mm}
\includegraphics[width=0.45\textwidth]{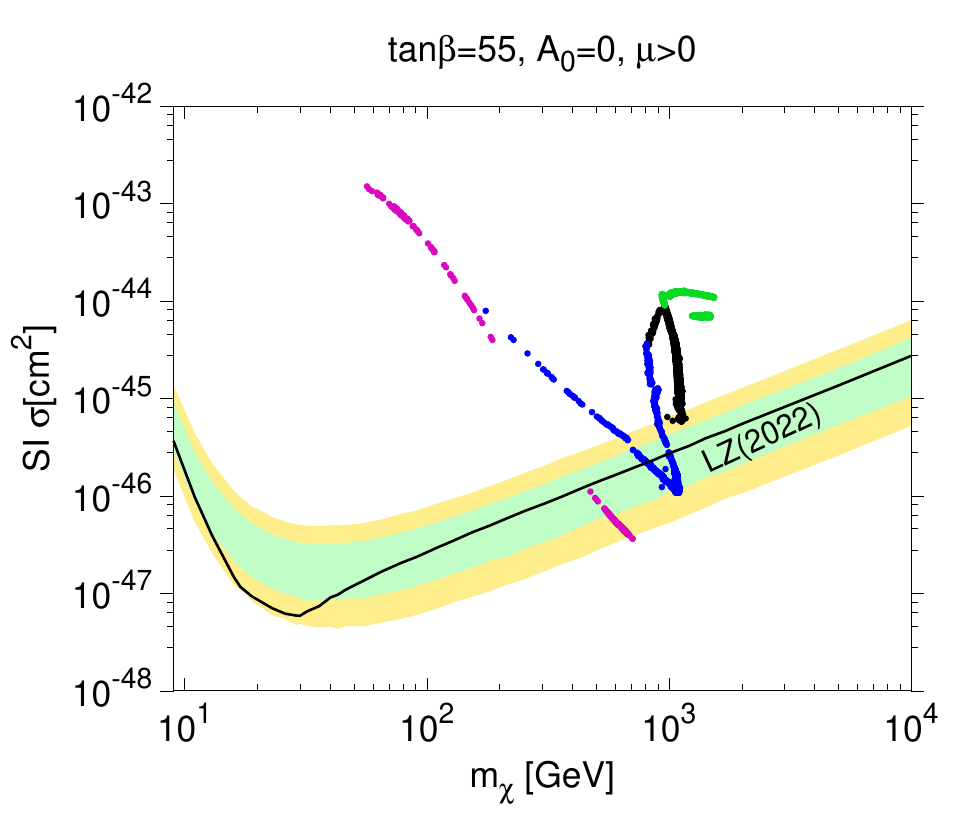} 
\includegraphics[width=0.45\textwidth]{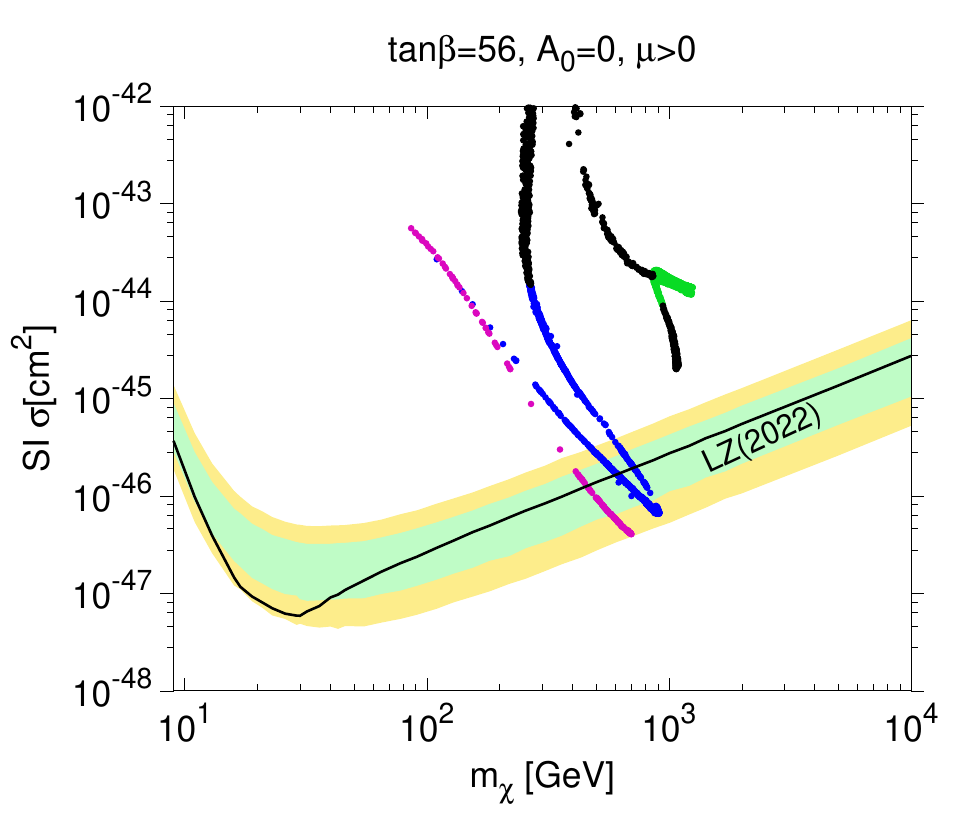} 
\vspace{-4mm}
\caption{\it Calculations of spin-independent
WIMP-nucleon scattering for points along the dark matter strips
for $\tan \beta = 5, 20, 40, 50, 55 \; and \; 56$ with $A_0 = 0$ and $\mu > 0$
compared with the LUX-ZEPLIN upper limit 
(solid black line). The green and yellow bands are the 1$\sigma$ and 2$\sigma$ LZ sensitivities.}
\label{fig:xsec}
\end{figure}

We see in the top and middle panels of Fig.~\ref{fig:xsec} that
the spin-independent LSP-nucleon scattering constraint allows
limited regions of the dark matter focus-point strips where $m_\chi \sim 1.0 - 1.1$~TeV
for $\tan \beta = 5, 20, 40$ and $50$. 
Along the near vertical strips shown here, although $m_{1/2}$ (and $m_0$) changes, the LSP mass is nearly constant. For $\tan \beta = 40$, we see that the direct detection experiments are (like the Higgs mass constraint) able to exclude the well-tempered LSP points (green). We also see in the middle right
panel of Fig.~\ref{fig:xsec} that only a small fraction of the
well-tempered and funnel points (green and blue) with larger values of $m_\chi$ are allowed.
We see in the bottom panels of Fig.~\ref{fig:xsec}
for $\tan \beta = 55$ and $56$ that points along the focus-point strips and
in the well-tempered regions (green) are excluded, and only a small fraction of well-tempered/funnel points (blue)
are still allowed. These points have $m_\chi \lesssim 1$~TeV and,
as we have discussed previously, correspond to  Higgs masses that are unacceptably small.
Some ranges of the funnel strips (purple) are also still  allowed 
by the spin-independent dark matter scattering
constraint, and have also shifted to lower values of $m_\chi \lesssim 800$~GeV. 
However, these points also correspond to unacceptably low Higgs masses.

These figures show the strong impact of the spin-independent scattering limits when $A_0$ and $\mu > 0$
and illustrate their potential for even more stringent constraints 
in the future as their sensitivities reach down towards the neutrino `floor'~\cite{floor}. We find spin-independent scattering
cross sections that lie above the `floor' for all values
of $\tan \beta$ when $A_0 = 0$, for both signs of $\mu$.

\subsection{Spin-Dependent Dark Matter Scattering}
\label{sec:sddirect}

Spin-dependent LSP-nucleon scattering on a nucleus is not
coherent, but is proportional to the nuclear spin, which
is mainly carried by an odd nucleon. The LZ experiment 
also set new upper limits on spin-dependent LSP scattering on both the neutron
and the proton. The LZ limit on spin-dependent LSP-neutron
scattering exploits the fact that two common xenon isotopes have non-zero nuclear spins
that are carried mainly by unpaired neutrons: $^{129}$Xe (abundance 26.4\%, spin 1/2) 
and $^{131}$Xe (21.2\% abundance, spin 3/2). The LZ limit on spin-dependent LSP-proton
scattering exploits configuration mixing between proton and neutron spin states in these isotopes,
but has larger uncertainties and is significantly weaker than their limit on spin-dependent LSP-neutron
scattering.~\footnote{There is also an upper limit on spin-dependent LSP-proton scattering from the
PICO-60 experiment using a $^{19}$F target~\cite{PICO}, with somewhat less sensitivity than the LZ limit.}
The spin-dependent LSP- neutron and -proton scattering cross sections are calculated to be quite similar,
so we use here only the LZ limit on spin-dependent LSP-neutron scattering, which provides the
most stringent constraint on spin-dependent LSP-nucleon scattering.

\begin{figure}[ht!]
\includegraphics[width=0.45\textwidth]{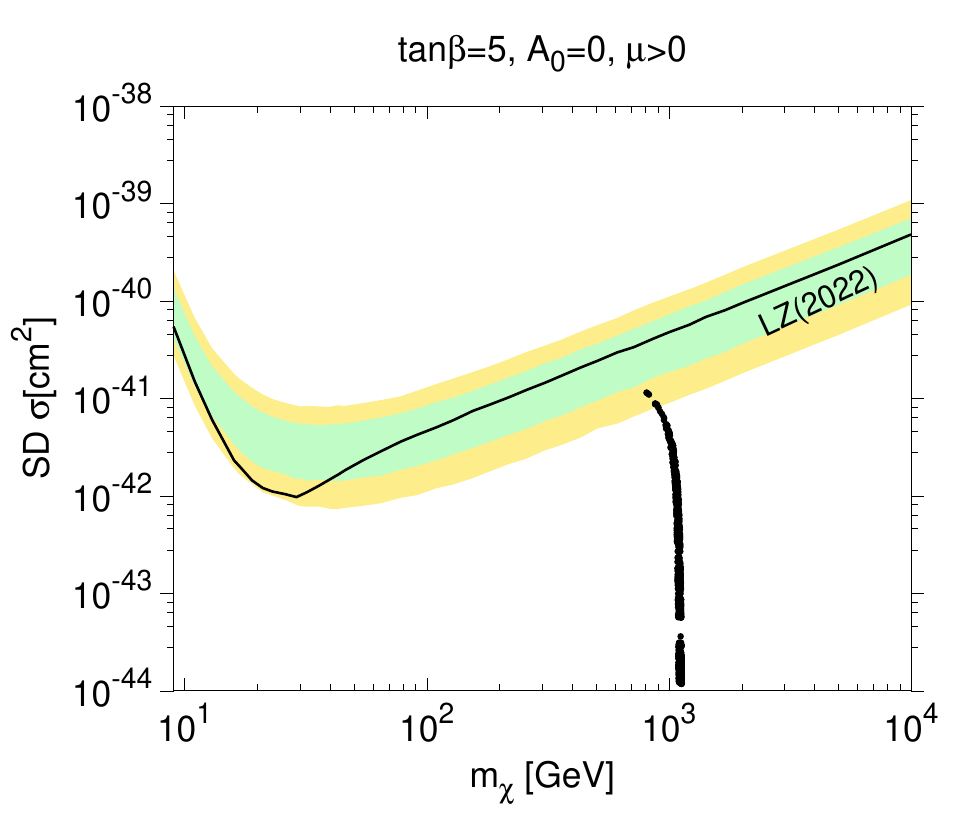} 
\includegraphics[width=0.45\textwidth]{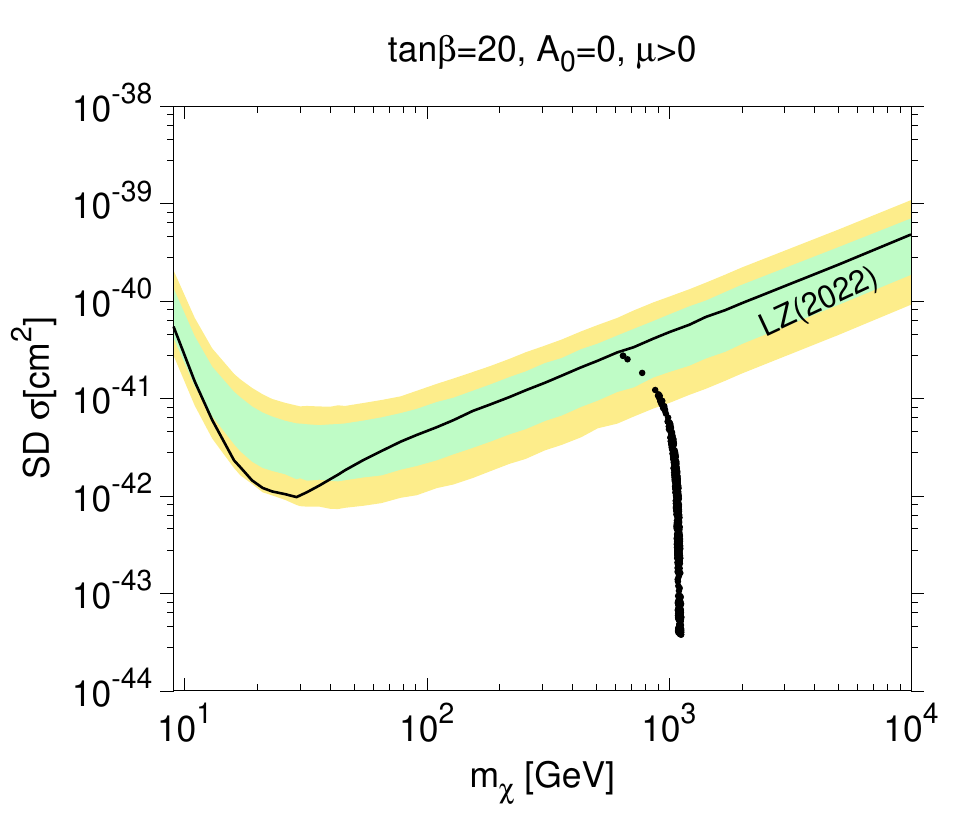} \\
\vspace{-2mm}
\includegraphics[width=0.45\textwidth]{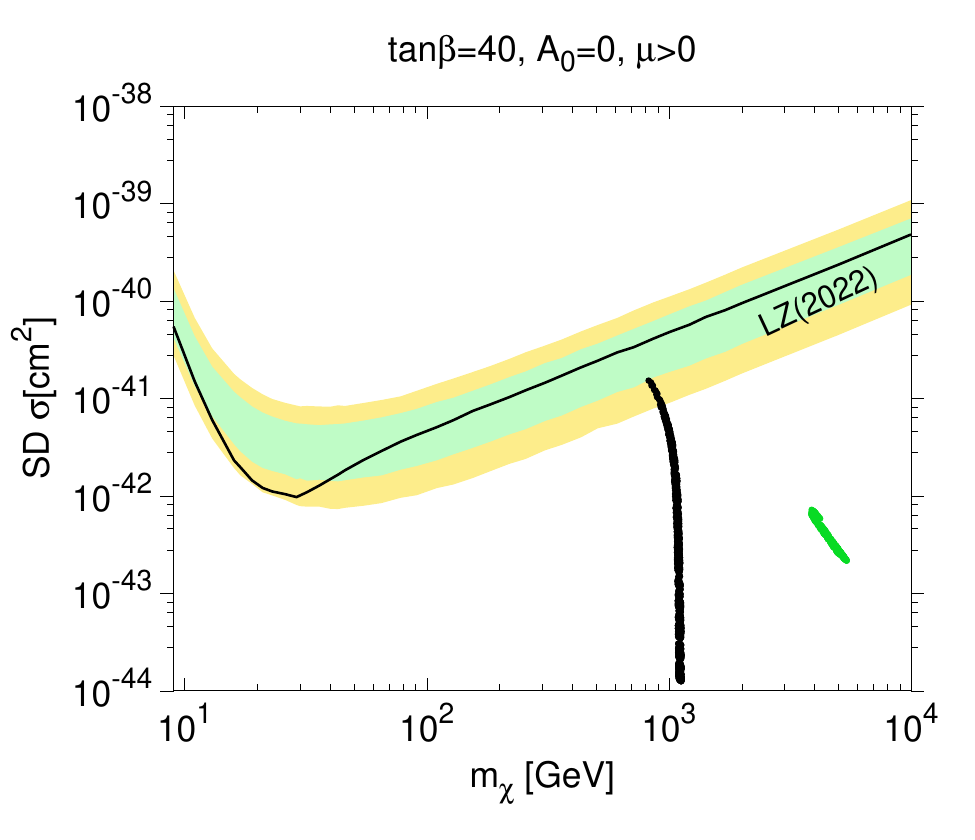}
\includegraphics[width=0.45\textwidth]{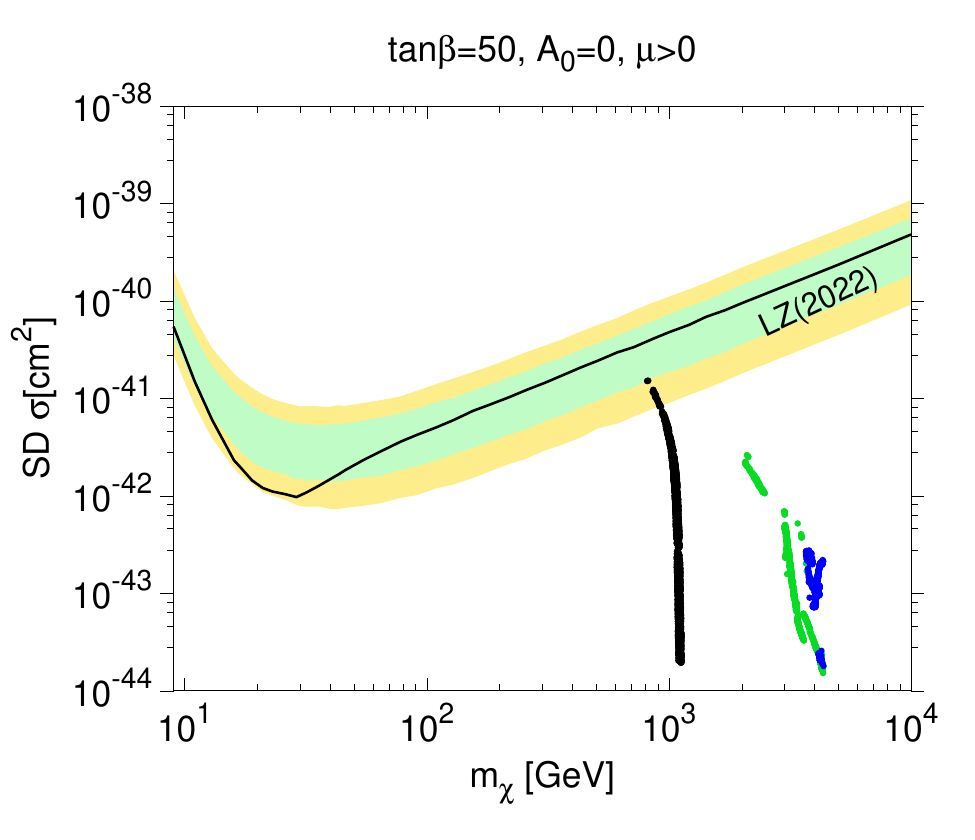} \\
\vspace{-2mm}
\includegraphics[width=0.45\textwidth]{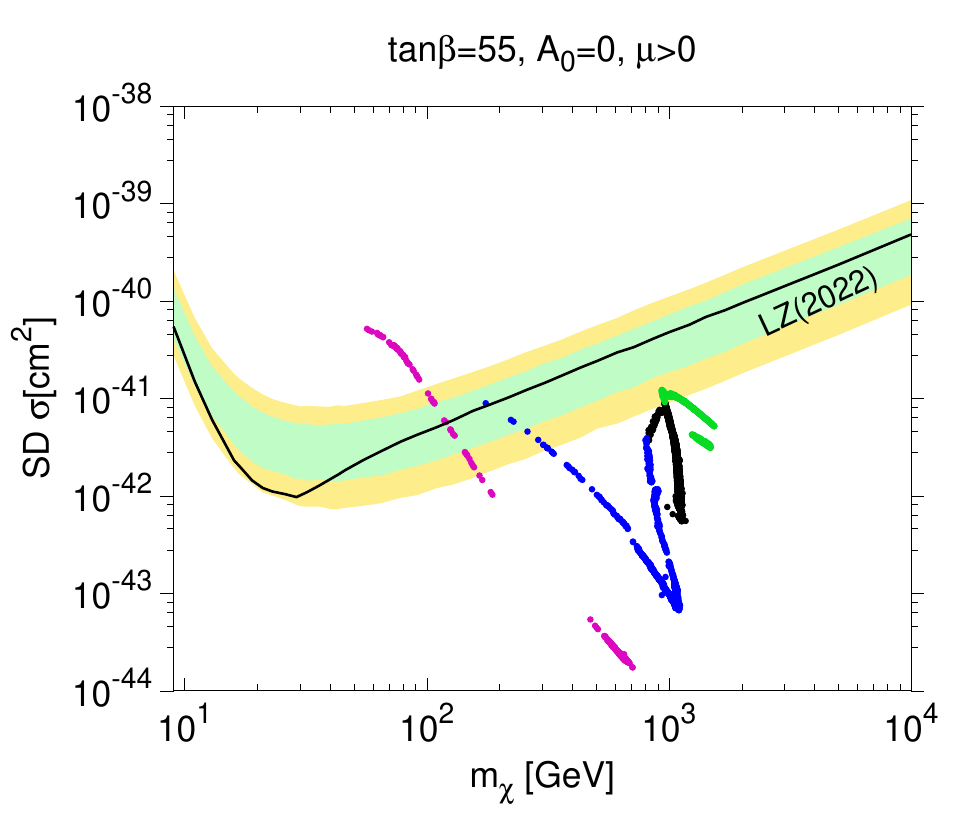} 
\includegraphics[width=0.45\textwidth]{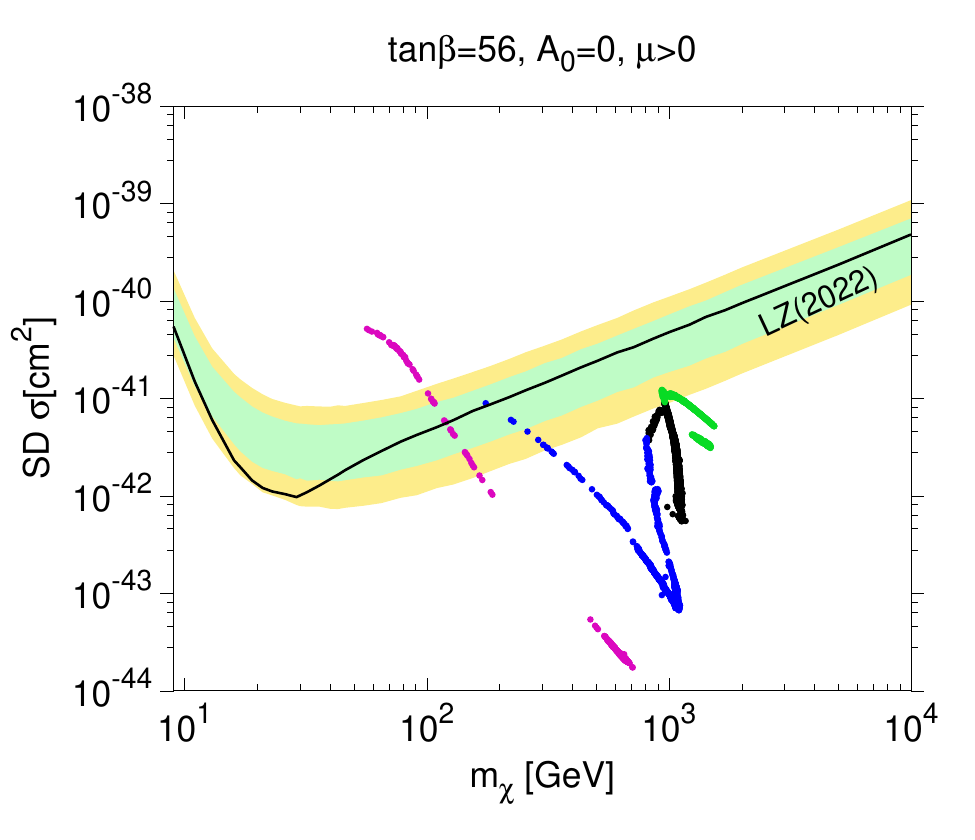} 
\vspace{-4mm}
\caption{\it Calculations of spin-dependent
WIMP-nucleon scattering for points along the dark matter strips
for $\tan \beta = 5, 20, 40, 50, 55 \; and \; 56$ with $A_0 = 0$ and $\mu > 0$
compared with the LUX-ZEPLIN upper limit on WIMP-neutron scattering
(solid black line). The green and yellow bands are the 1$\sigma$ and 2$\sigma$ LZ sensitivities.}
\label{fig:sdxsec}
\end{figure}

We see in Fig.~\ref{fig:sdxsec} that all the dark matter strips for $\tan \beta \le 50$
are compatible with the spin-dependent scattering constraint from the LZ experiment.
However, there are regions of the funnel strips (purple) for $\tan \beta = 55$ and 56 
with $m_\chi \lesssim 100$~GeV that are excluded by this constraint, as seen in the bottom
panels of Fig.~\ref{fig:sdxsec}. We note in addition that there are some well-tempered/funnel
points (blue) with LSP masses $\sim 200$~GeV that are
marginally excluded by the LZ spin-dependent scattering constraint.

\subsection{Indirect Constraints}
\label{sec:indirect}

A variety of products of LSP annihilations
may offer detectable signals, including  $\gamma$ rays, neutrinos,
positrons, antiprotons and light antinuclei. We consider in this paper
$\gamma$ rays and neutrinos, whose production is relatively straightforward to model 
and may provide interesting observational constraints. Since annihilation rates are proportional 
to the square of the local DM density, we  study regions where the DM accumulates. These include 
dwarf spheroidal satellite galaxies (dSphs)
\cite{Fermi-LAT:2013sme,Fermi-LAT:2015ycq,Fermi-LAT:2015att,Boddy:2018qur,Hoof:2018hyn,Alvarez:2020cmw,MAGIC:2016xys,MAGIC:2017avy,HAWC:2017mfa,HAWC:2019jvm,HESS:2020zwn}
and the Galactic Center (GC)~\cite{HESS:2022ygk,Johnson:2019hsm,Abazajian:2020tww,HESS:2018cbt}, which are potential
sources of observable $\gamma$ fluxes, and the Sun, which is an interesting potential source
of energetic neutrinos~\cite{IceCube:2016yoy,IceCube:2016dgk,IceCube:2021xzo,Super-Kamiokande:2015xms,ANTARES:2016xuh,HAWC:2018szf}.

\subsubsection{Limits on $\gamma$ Fluxes from Dwarf Spheroidal Satellite Galaxies}
\label{sec:gammadSph}
 We first examine dSphs as
 possible sources of $\gamma$ rays from DM annihilations
 in the framework of the CMSSM, before revisiting the
 possible GC constraint on DM annihilations.

The annihilation of neutralinos can yield monochromatic photons via the one-loop processes $\chi \chi \rightarrow
\gamma \gamma$ and $\chi \chi \rightarrow \gamma Z$,
and also a continuous spectrum of photons via  the decays of $\pi ^0$s and other hadrons produced by the
fragmentation and hadronization of primary
annihilation products. The integrated  $\gamma$-ray
signal flux, $\phi_s$ (typically measured in photons/cm$^{2}$/s),  
expected from the annihilations of DM particles with a
density distribution $\rho_{DM}(r)$ is
\begin{equation}
\phi_s(\Delta \Omega)=\underbrace{\frac{1}{4\pi}\frac{\langle\sigma \upsilon\rangle}{2{m}_{DM}^2}\int_{E_{min}}^{E_{max}} \frac{dN_{\gamma}}{dE_{\gamma}}dE_{\gamma}}_{\substack{\Phi_{PP}}} \times \underbrace{ \int_{\Delta\Omega}\int_{l.o.s.} \rho^2_{DM}(r)dld\Omega'}_{\substack{\text{J factor}}}  \, ,
\label{eq:gamma_flux}
\end{equation}
whose components we analyze in the following. 

The first term of (\ref{eq:gamma_flux}), $\Phi_{PP}$,  depends on the particle physics properties. In particular, 
this term is dependent on  the thermal annihilation cross section $\langle\sigma\upsilon\rangle$, 
the mass of the dark matter particle, $m_{DM}$, and the differential $\gamma$ ray yield per annihilation, $ {dN_{\gamma}}/{dE_{\gamma}}$,
integrated over the experimental energy range. The differential yield $ {dN_{\gamma}}/{dE_{\gamma}}$ is a sum over specific final states
\begin{equation}
{dN_{\gamma}}/{dE_{\gamma}}=\sum_f {B_{f} } {dN_{\gamma}^f}/{dE_{\gamma}} \, ,
\end{equation}
where ${B_{f} }$ is the branching fraction into a given final state.  
Branching fractions for $\chi$ pair annihilation 
in the CMSSM were studied in \cite{Ellis:2011du}. As described in 
detail there,  we have used  {\tt PYTHIA}  tool~\cite{pythia} in order to simulate 
the gamma fluxes produced in  $\chi$ pair annihilations.  

The second term in (\ref{eq:gamma_flux}), the J factor, is the integral  along the  light of sight (l.o.s.) \cite{Bergstrom:1997fj}
through the DM density profile, $\rho_{DM}$, which is   integrated over a solid angle, $\Delta \Omega$. 
The DM density profile may be given by the general expression
\begin{equation}
\rho_{DM} (r)=\frac{\rho_0}{ (r/R)^{c} [1+ (r/R)^{a} ] ^{(b-c)/a}} \, ,
\end{equation}
where $R$ is the characteristic length scale and $a,b,c$ are parameters. 
In our analysis we assume that the DM distribution is a cuspy Navarro-Frank-White (NFW) \cite{Navarro:1996gj} profile with $a=c=1.0$ and $\beta=3.0$, i.e.
\begin{equation}
\rho_{DM} (r)=\frac{\rho r_s^3}{r (R+r)^2} \, .
\end{equation}
We note that both the NFW and Burkert profiles~\cite{Burkert:1995yz} were studied in~\cite{Fermi-LAT:2013sme}, 
whereas only the NFW profile was used in a subsequent {\it Fermi}-LAT study \cite{Fermi-LAT:2015att},
where it was found that the J factors for dSphs were relatively insensitive to the choice of DM distribution. 

In our analysis, we have used the information on 25 Milky Way dSphs given in \cite{Fermi-LAT:2015att},
including the {\tt python} code provided by {\it Fermi} Tools for calculating the binned likelihood with energy dispersion.  
In order to use this code, we used the corresponding spacecraft data files for the  dSphs and a background model,  
which are also provided by {\it Fermi} Tools. We used the energy range from 500~MeV to 500~GeV, divided into  30 logarithmic bins,
and verified that our results are insensitive to various choices of the energy range and number of bins within the limits provided by {\it Fermi}-LAT. 
The times of the events we used range from 01/01/2009 to 01/01/2022.  The window radius for searching these data is 15 degrees 
and the galactic coordinates for each  dSph are taken from \cite{Fermi-LAT:2015att}. 
The selection of data sets is in accordance with the suggestion by the {\it Fermi}-LAT Collaboration.

After completing the analysis using {\it Fermi} Tools, we have evaluated the signals in the framework of the CMSSM,  
following the approach used previously in a study of the prospective $\gamma$ signal from the GC~\cite{Ellis:2011du}. 
We calculate the $\chi^2$ likelihood function as follows: 
\begin{equation}
\chi^2=\sum^{n_{bins}}_{i=1}\frac{(d_i -(b_i+s_i))^2}{{\sigma}_i^2} \, ,
\end{equation}
where ${n_{bins}} = 30$ is the number of logarithmic bins in the energy range analyzed,
$d_i$ is the number of data counts  per bin, $b_i$ is the expected background provided by {\it Fermi} Tools~\cite{Fermi-LAT:2015att},
and $s_i$ is the signal due to DM annihilations in each bin.
 The quantity $\sigma_i$ is defined from
\begin{equation}
{\sigma}_i^2=d_i+{\sigma}_{ea}^2 \, ,
\end{equation}
where ${\sigma}_{ea}$ is the systematic uncertainty in the effective area~\cite{Ellis:2011du}. 
We evaluate the $\chi^2$ contribution for each dSph separately.

\begin{figure}[ht!]
\includegraphics[width=0.45\textwidth]{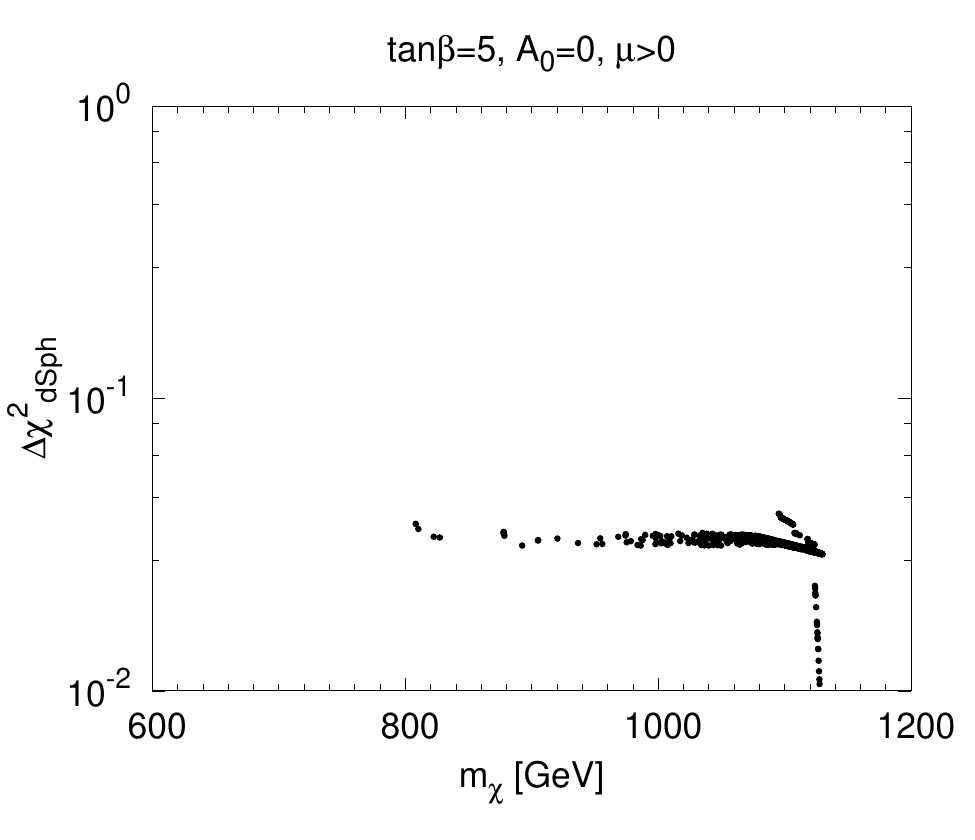} 
\includegraphics[width=0.45\textwidth]{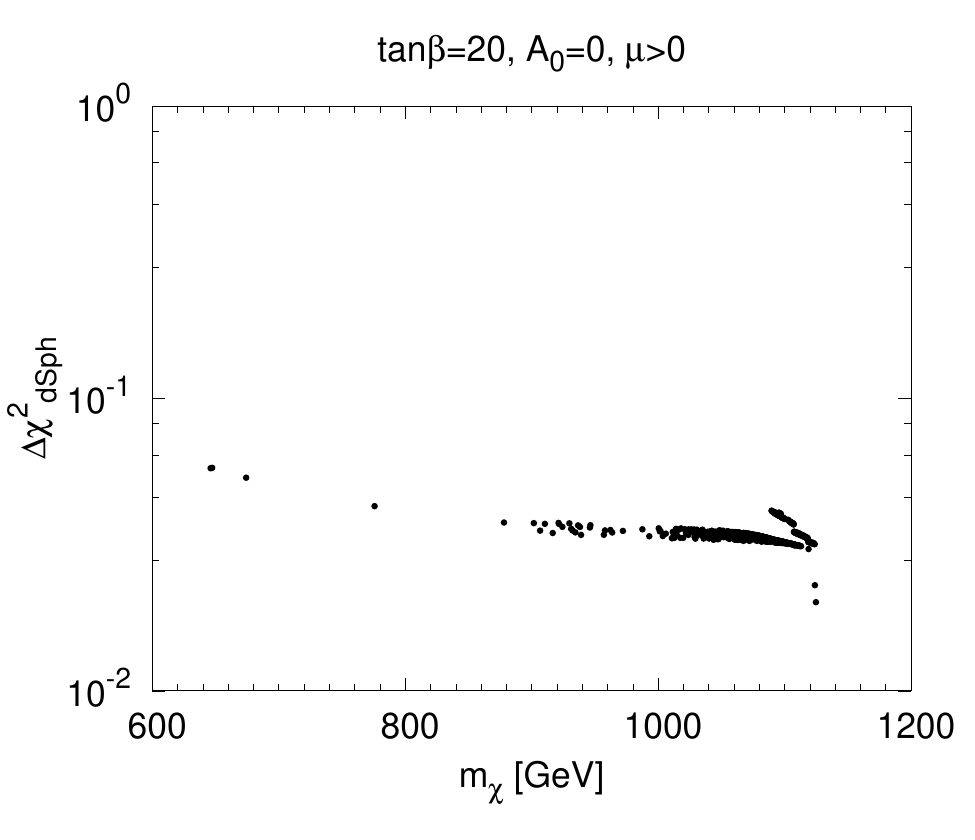} \\
\vspace{-2mm}
\includegraphics[width=0.45\textwidth]{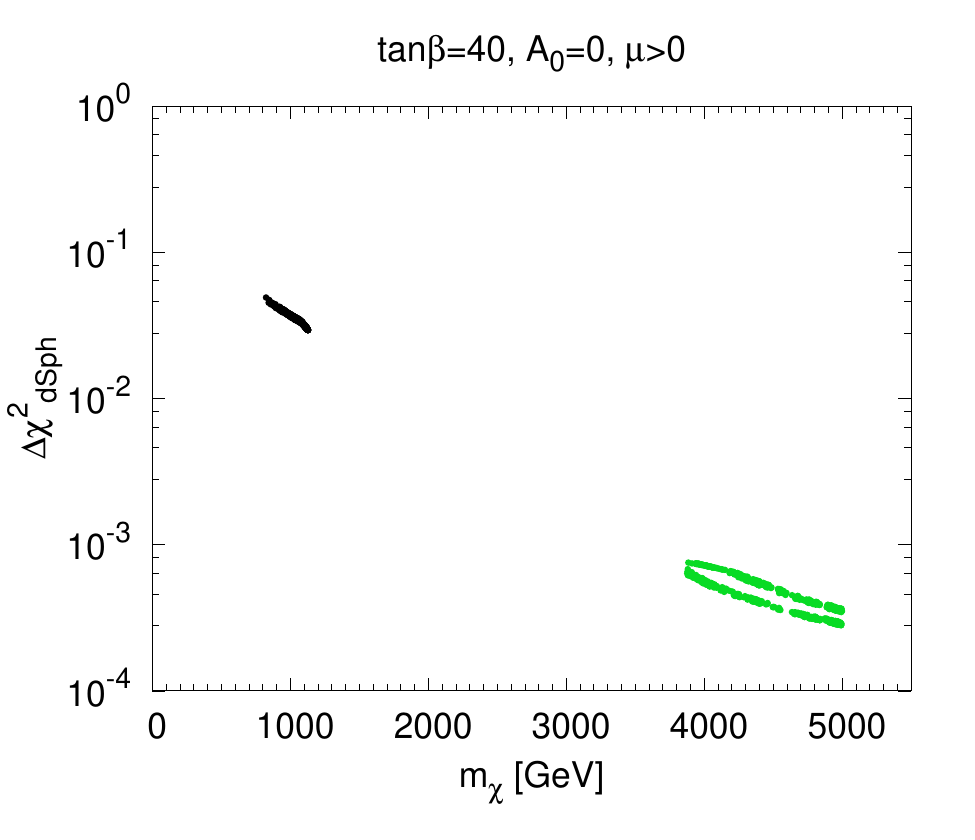}
\includegraphics[width=0.45\textwidth]{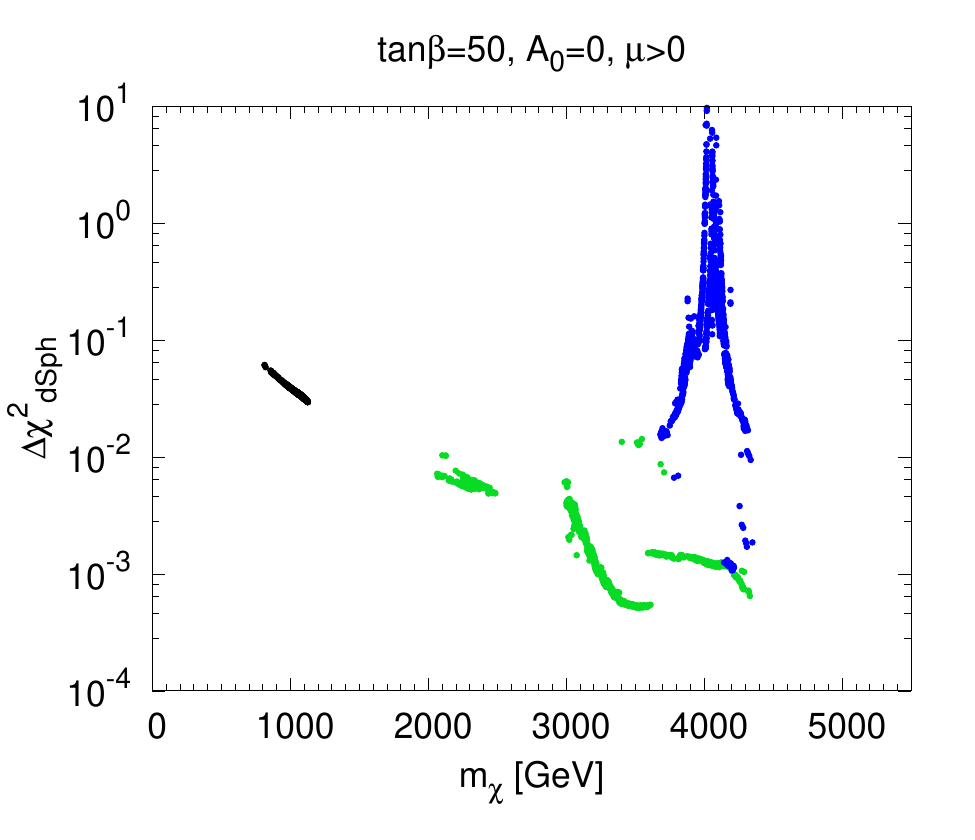} \\
\vspace{-2mm}
\includegraphics[width=0.45\textwidth]{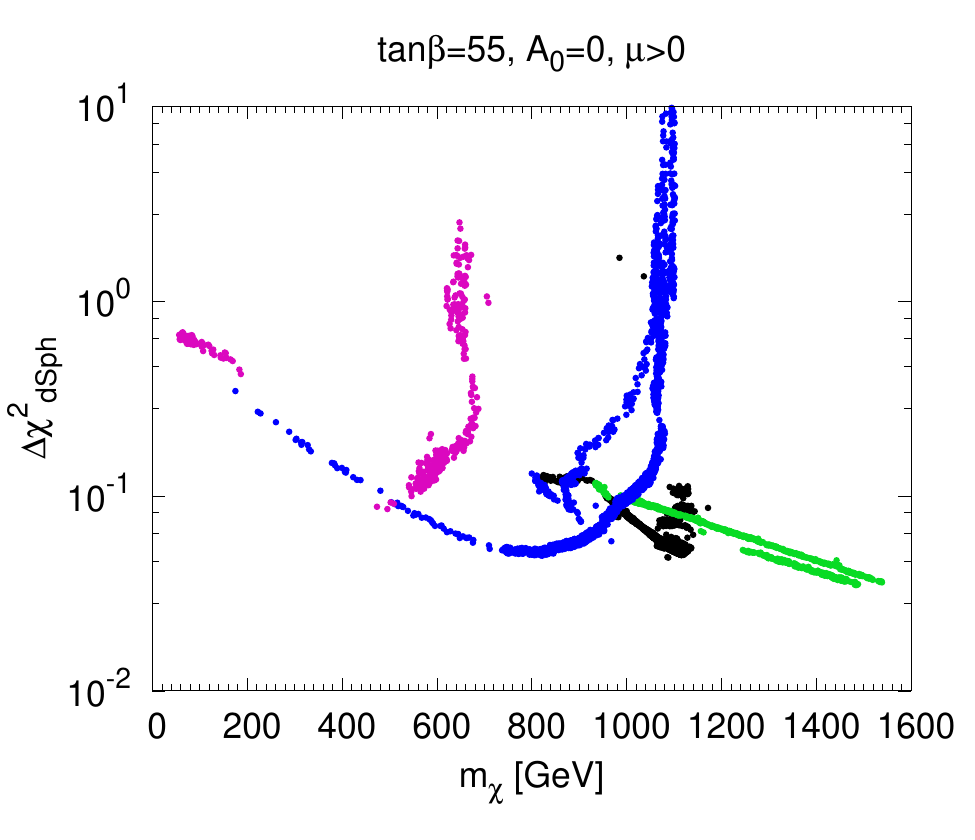} 
\includegraphics[width=0.45\textwidth]{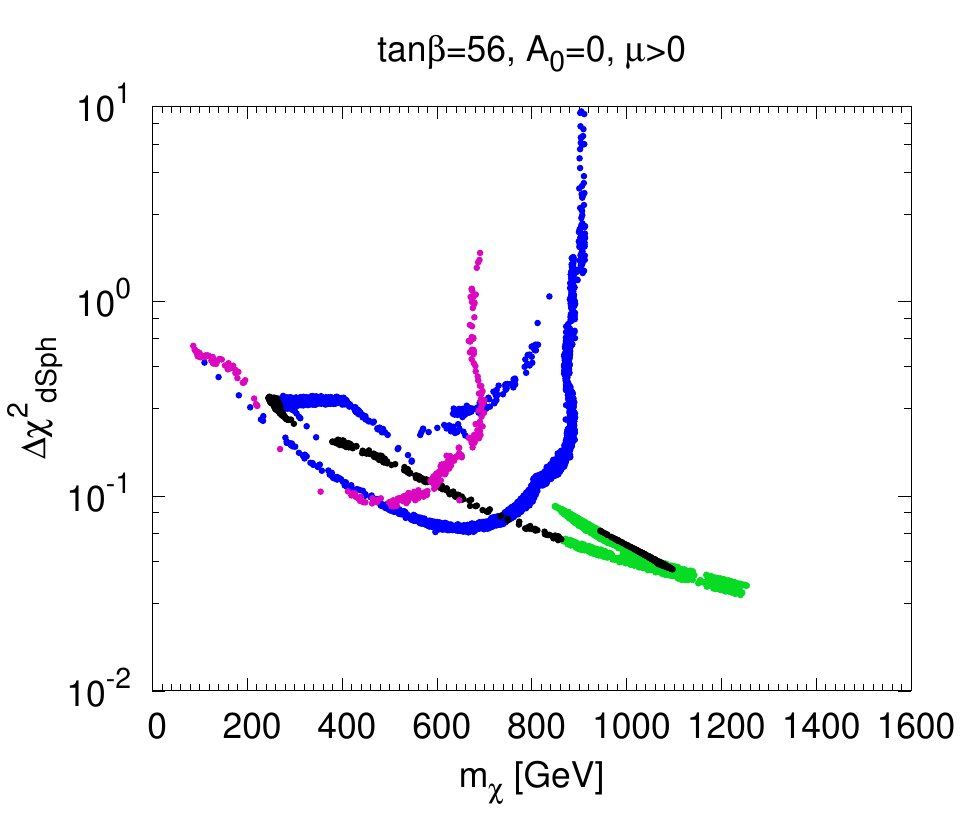} 
\vspace{-4mm}
\caption{\it Contributions to the global $\chi^2$ likelihood function
from a comparison of the {\it Fermi}-LAT upper limit on the flux of $\gamma$-rays from dwarf spheroidal satellite galaxies
with calculations for points along the dark matter strips
for $\tan \beta = 5, 20, 40, 50, 55 \; and \; 56$ with $A_0 = 0$ and $\mu > 0$.}
\label{fig:gamma}
\end{figure}

Fig.~\ref{fig:gamma} displays our evaluations of the contribution to the global likelihood function, $\Delta \chi^2$, 
from the aggregated {\it Fermi}-LAT data on $\gamma$-rays from dSphs for the same values of $\tan \beta$ as in previous
Figures. We see that $\Delta \chi^2 \lesssim 0.1$ for $\tan \beta \le 40$, and hence is negligible in these cases.
However, the situation is quite different for $\tan \beta \ge 50$, as seen in the middle right and bottom panels of Fig.~\ref{fig:gamma}.
Values of $\Delta \chi^2 \sim 10$ are reached in the
well-tempered/funnel regions (blue) where $m_\chi \simeq 4$~TeV for $\tan \beta = 50$, 
$\simeq 1.1$~TeV for $\tan \beta = 55$ and $m_\chi \simeq 0.9$~TeV for $\tan \beta = 56$, 
excluding points in these regions, where the LSP annihilates mainly into $\bar b b$ final states
through direct-channel $H/A$ resonances. However, these points are already excluded by
the calculation of $m_h$, as seen in Fig.~\ref{fig:mh}.
We note also that when $\tan \beta = 55$ and 56, $\Delta \chi^2$ reaches $\sim 2$ for funnel points with 
$m_\chi \simeq 650$~GeV. This is significant though not sufficient to exclude these points, which are in
any case excluded by the constraint on $m_h$.

\subsubsection{Limit on $\gamma$ Flux from the Galactic Centre}
\label{sec:gammadGC}

The production of  $\gamma$ rays near the GC was studied
as a possible signature of the CMSSM in~\cite{Ellis:2011du}, whose analysis we follow here. 
Modelling the DM density near the GC entails significant uncertainties.
The NFW model is frequently taken as a default,
but other possibilities have been considered for comparison.
For example, the H.E.S.S. Collaboration considered recently~\cite{HESS:2022ygk} an
Einasto profile that would strengthen the prospective flux limit on
annihilations of an LSP weighing 1~TeV by a factor $> 2$, and~\cite{Dessert:2022evk}
considered a FIRE simulation that would strengthen the limit by a factor $> 4$.
In a spirit of conservatism, here we use an NFW profile for setting constraints
from data on the $\gamma$ flux from near the GC.

Assuming the NFW profile, the recent H.E.S.S. analysis~\cite{HESS:2022ygk} finds a 95\% CL
upper limit on the cross section for LSP-LSP annihilation to $W^+ W^-$
that is stronger than that derived from the {\it Fermi}-LAT data on dSphs
for $m_{\rm LSP} \gtrsim 0.4$~TeV, which is the mass range of most interest 
for our analysis. However, in the CMSSM the branching fraction for $W^+ W^-$
final states is $< 100$\%, so other annihilation modes must be taken into
account, which we do when comparing the sensitivities of the {\it Fermi}-LAT dSph 
and H.E.S.S. GC data in our analysis.

\begin{figure}[ht!]
\includegraphics[width=0.45\textwidth]{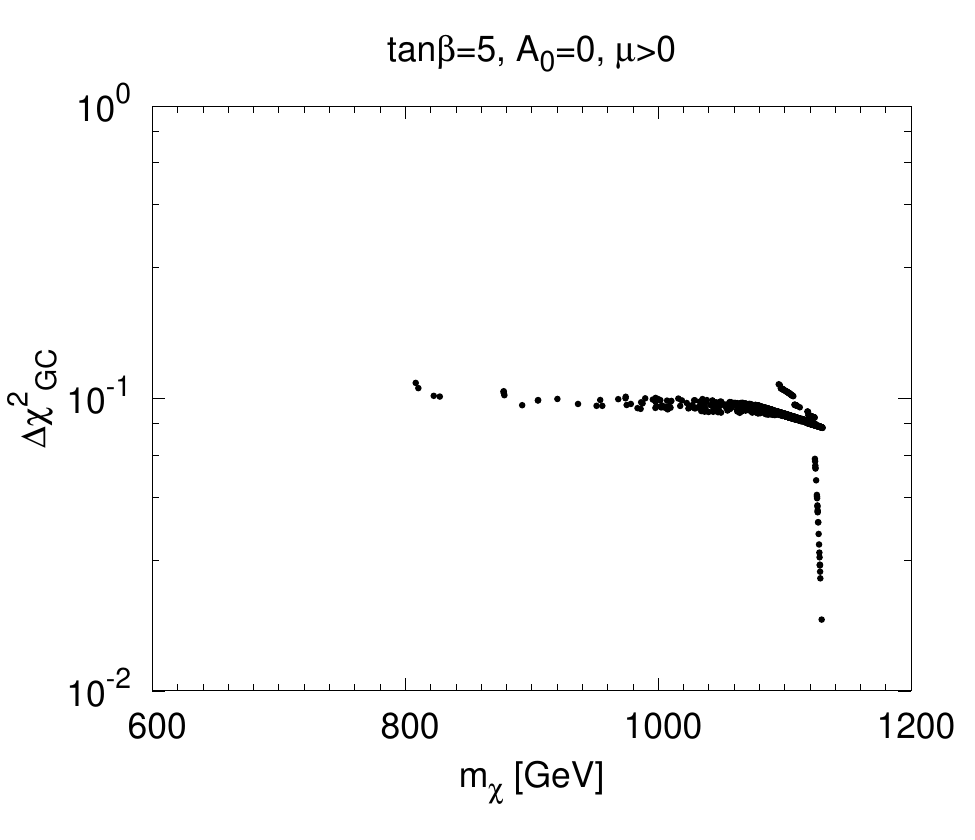} 
\includegraphics[width=0.45\textwidth]{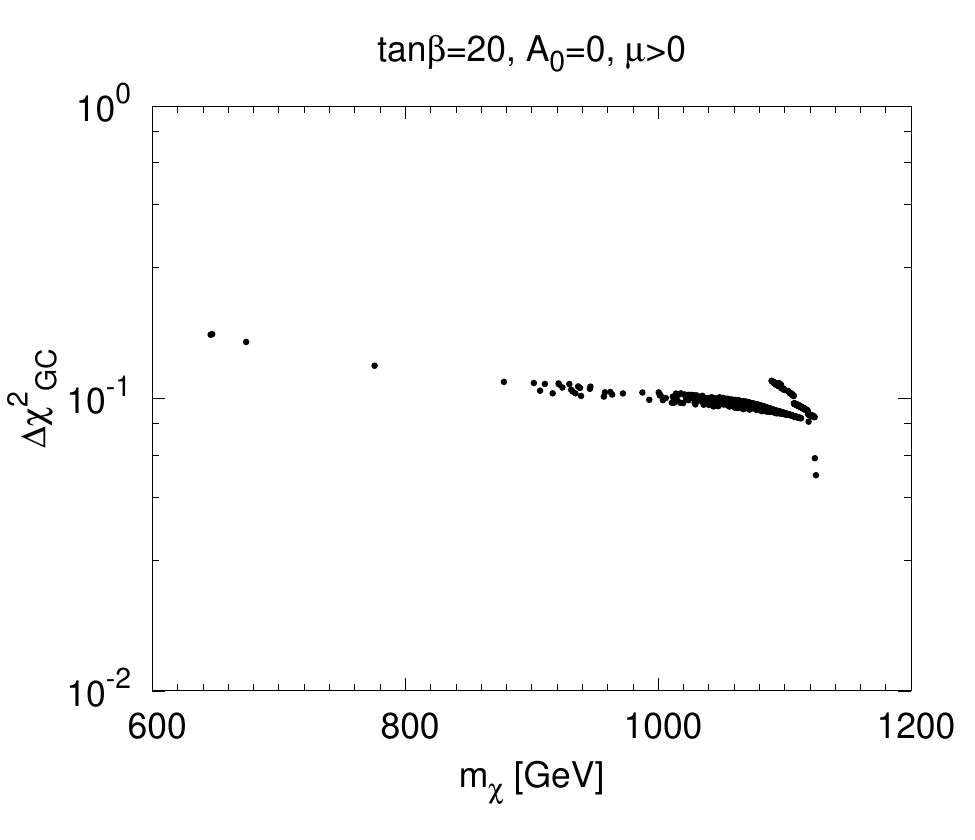} \\
\vspace{-2mm}
\includegraphics[width=0.45\textwidth]{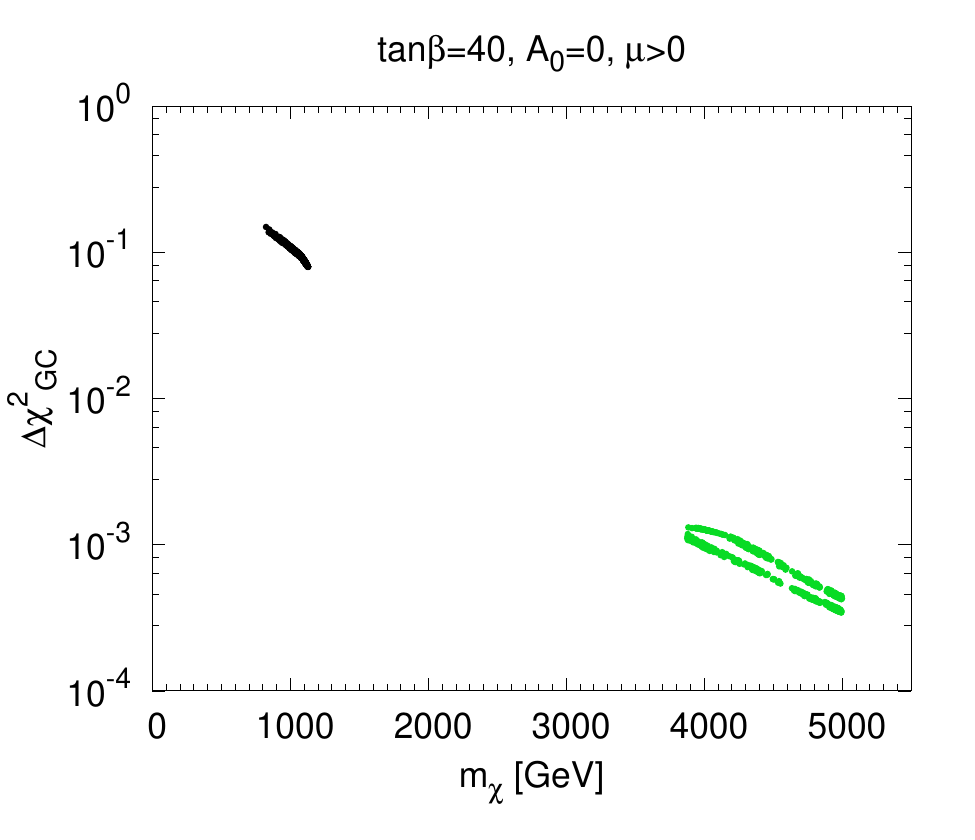}
\includegraphics[width=0.45\textwidth]{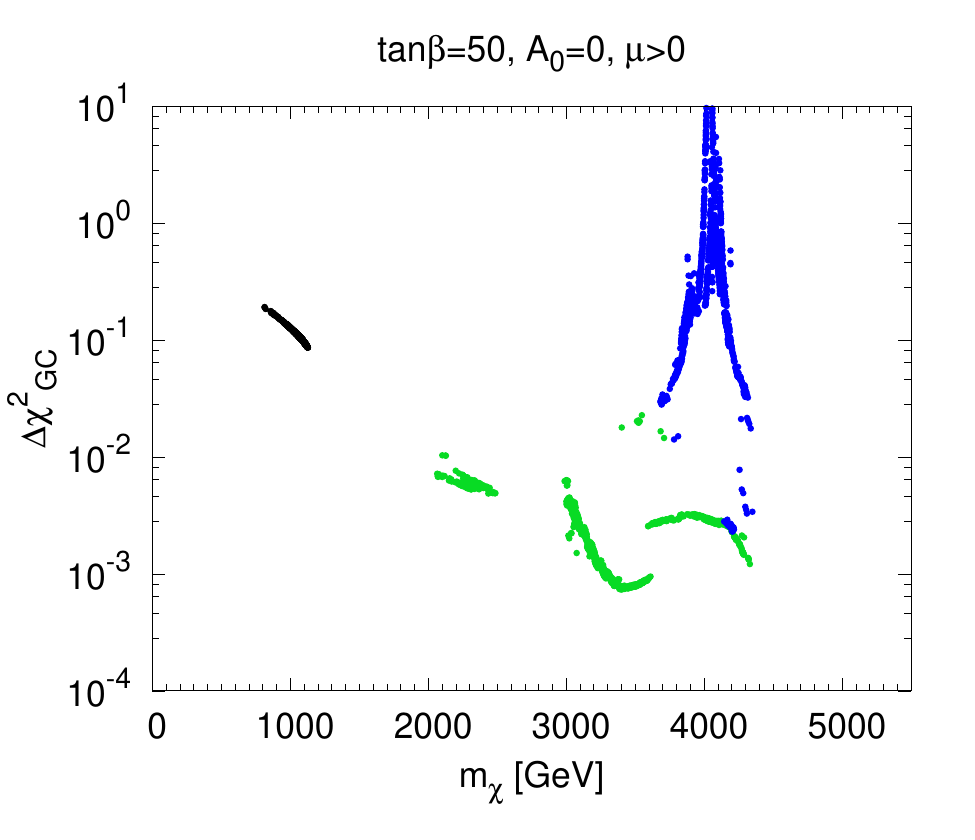} \\
\vspace{-2mm}
\includegraphics[width=0.45\textwidth]{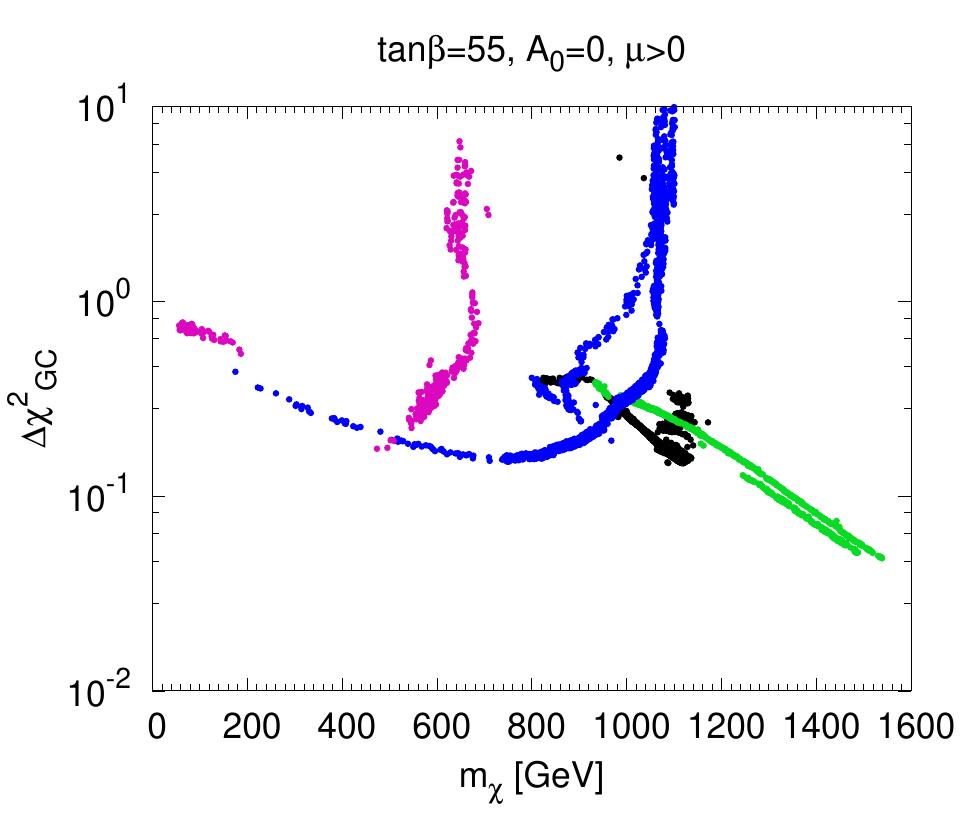} 
\includegraphics[width=0.45\textwidth]{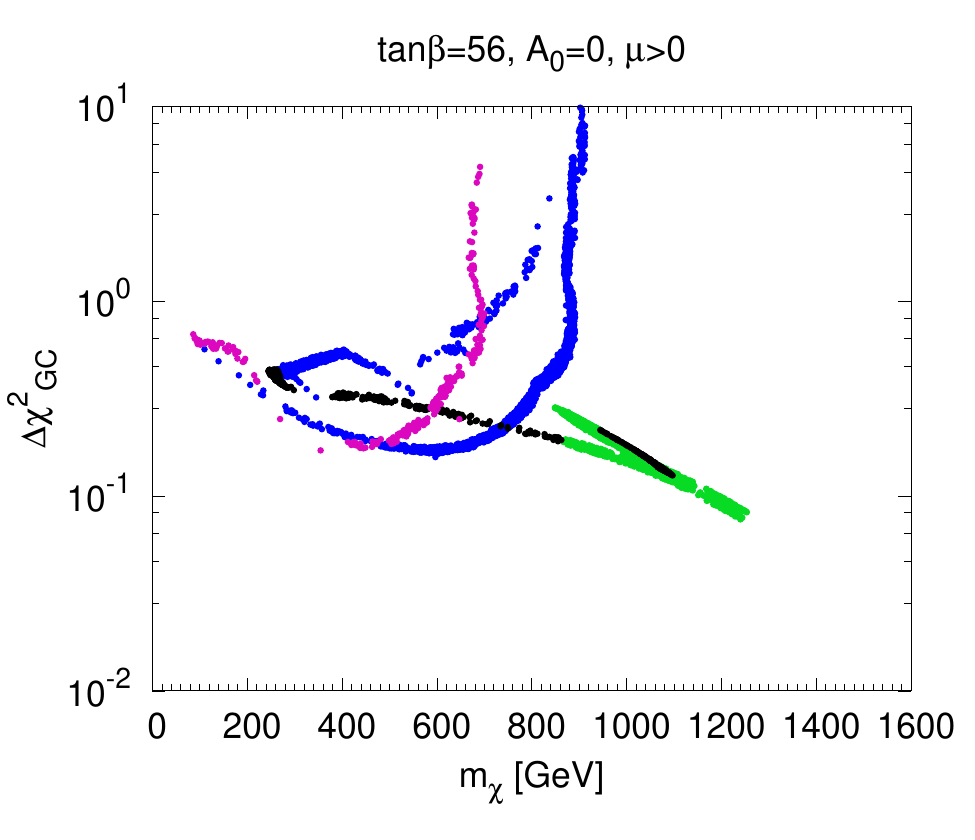} 
\vspace{-4mm}
\caption{\it Contributions to the global $\chi^2$ likelihood function
from a comparison of the H.E.S.S. upper limit on the flux of $\gamma$-rays from the Galactic Centre
with calculations for points along the dark matter strips
for $\tan \beta = 5, 20, 40, 50, 55 \; and \; 56$ with $A_0 = 0$ and $\mu > 0$.}
\label{fig:gammaGC}
\end{figure}

Fig.~\ref{fig:gammaGC} displays the contributions, $\Delta \chi^2$, to the global
likelihood function provided by H.E.S.S. searches for $\gamma$-rays from near the 
GC in the framework of the CMSSM for $A_0 = 0$, $\mu > 0$ and our
standard choices of $\tan \beta$. The general features are quite similar the those
of the corresponding plots for the {\it Fermi}-LAT dSph analysis shown in
Fig.~\ref{fig:gamma}. The values of $\Delta \chi^2$ provided by the H.E.S.S. data
are larger than for the {\it Fermi}-LAT dSph data for $\tan \beta \le 40$, though
still small. There are non-negligible peaks in $\Delta \chi^2$ for $\tan \beta \ge 50$,
which are again larger than those provided by the {\it Fermi}-LAT dSph data, but the two
sets of data yield similar constraints away from the peaks. Once again, the points
excluded by the H.E.S.S. GC $\gamma$-ray constraint also correspond to values of $m_h$
that are too small, as seen in Fig.~\ref{fig:mh}.~\footnote{{The 
projected increase in
sensitivity offered by CTA~\cite{CTA} would probe a larger range of funnel points.}}

\subsubsection{Limit on Energetic Neutrinos from the Sun}
\label{sec:neutr}

The principle of the search for energetic neutrinos from the Sun \cite{indirectdet:solar} is that an
LSP may scatter during passage through the Sun, losing energy and becoming
gravitational bound in an elliptical orbit with perihelion smaller than the
solar radius. The LSP will in general scatter again during subsequent passages
through the Sun, losing more energy each time and eventually settling into a
thermal distribution inside the Sun. This distribution eventually equilibrates
with a density that balances the LSP capture rate, $C$, with the loss of LSPs
via annihilation, $2 \Gamma_A$ where $\Gamma_A$ is the annihilation rate. The
annihilation rate in equilibrium is then determined by the capture rate:
\begin{equation}
\Gamma_A \; = \; \frac{1}{2} C \, .
\label{eq:solarannihilationrate}
\end{equation}
We note that this relation is not universal, as equilibrium may not be reached
if $C$ is too small. However, this is not an issue for the CMSSM scenarios we
consider here, for which Eq.~(\ref{eq:solarannihilationrate}) is a good approximation~\cite{Ellis:2009ka}.

Since nuclear matter in the Sun is largely composed of individual protons,
the capture rate receives an important contribution from spin-dependent
LSP-proton scattering, and limits on energetic solar neutrinos are occasionally
interpreted as upper limits on spin-dependent LSP-proton scattering. 
However, the Sun also contains a significant fraction by mass
of $^4$He, as well as trace amounts of heavier nuclei, so that spin-independent
LSP-nucleon scattering is also potentially important. Both spin-independent
and spin-dependent scattering are included in our calculation of the capture rate.

The calculation of the fluxes of the different neutrino species is straightforward,
and we follow the analysis in \cite{Ellis:2009ka}. The experimental observable is the flux of muons,
which may be produced by either muon or tau neutrinos reaching Earth.
We use the IceCube~\cite{IceCube:2016yoy,IceCube:2016dgk,IceCube:2021xzo}
upper limit on the flux of energetic muons from the Sun
to constrain CMSSM parameters, taking into account the branching fractions for
annihilations into different final states, e.g., $W^+ W^-, \tau^+ \tau^-, ...$
that are predicted at each point in the parameter space. 

For the calculation of the corresponding $\Delta \chi^2$ we are 
employing  the {\tt nulike} software~\cite{IceCube:2012fvn}, following the 
likelihood  estimation  described in \cite{IceCube:2012fvn,IceCube:2016yoy}.
In addition,  for the calculation of the likelihood 
 we are taking into  account   the 79-string IceCube data, that is known to 
provide improved limits on WIMP dark matter searches \cite{IceCube:2012ugg}.  
As the basis for  the $\Delta \chi^2$ we are plotting, 
we are using  the $p$-value output from the {\tt nulike} code.

\begin{figure}[ht!]
\includegraphics[width=0.45\textwidth]{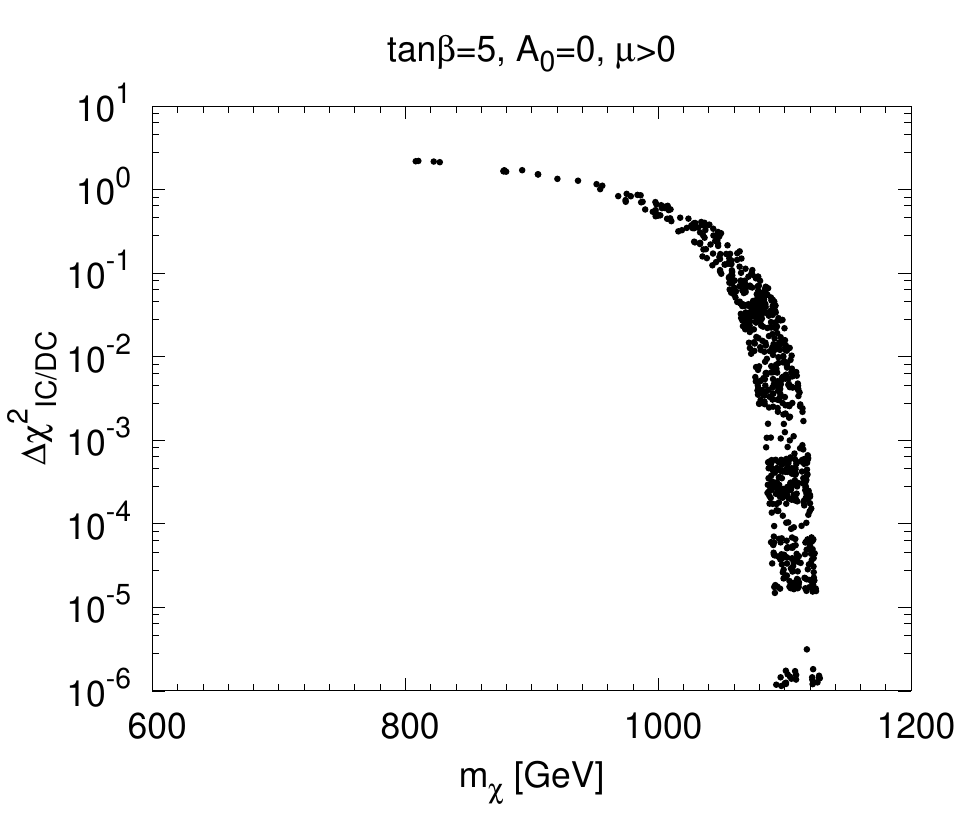} 
\includegraphics[width=0.45\textwidth]{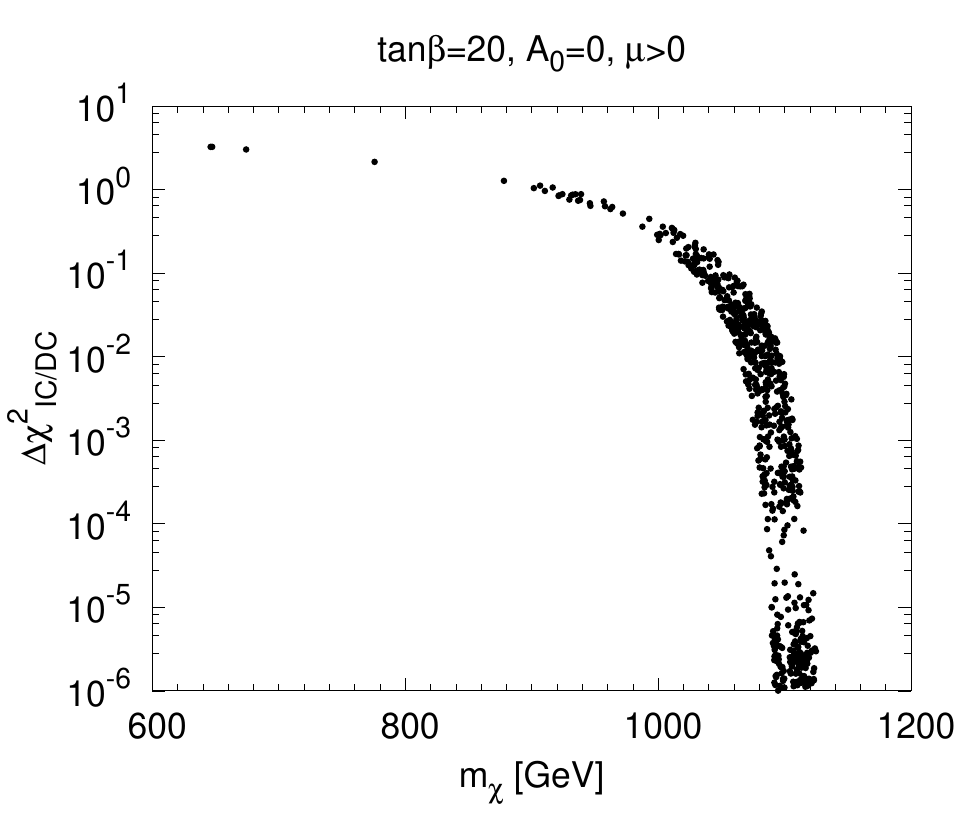} \\
\vspace{-2mm}
\includegraphics[width=0.45\textwidth]{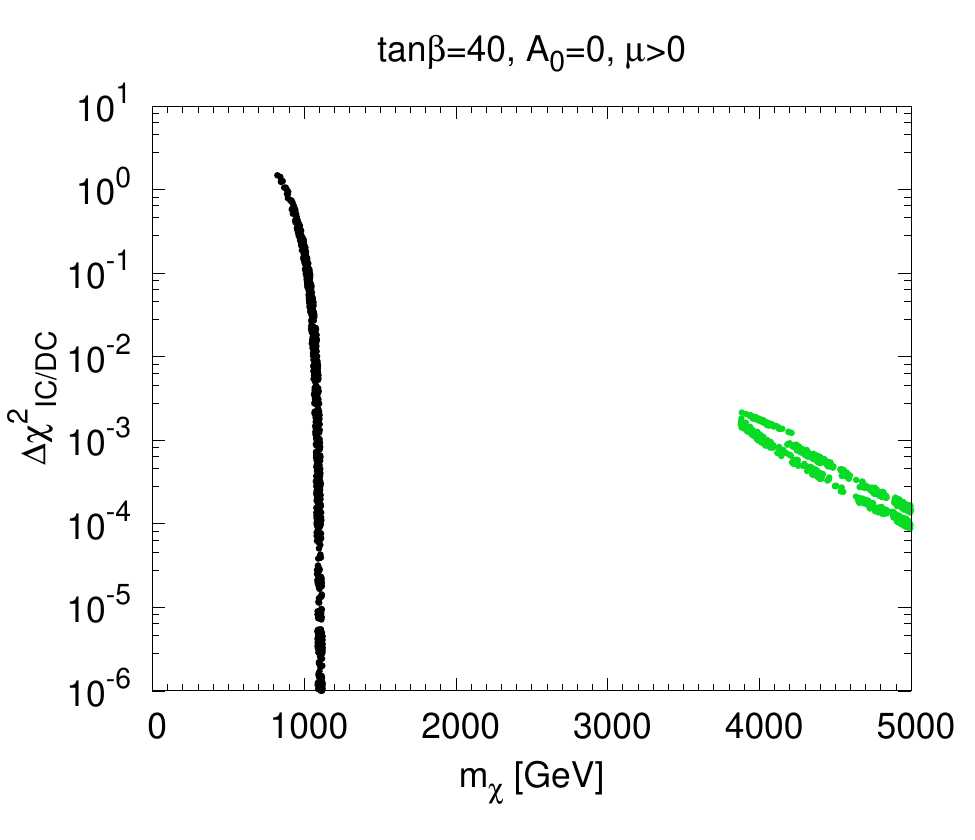}
\includegraphics[width=0.45\textwidth]{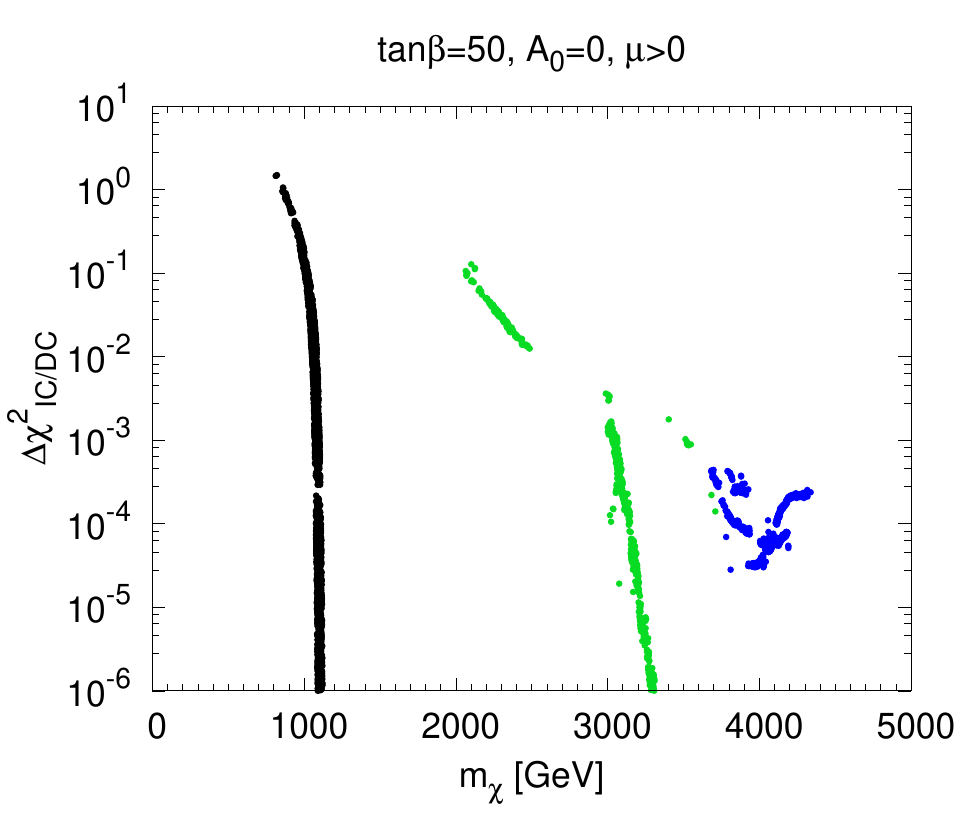} \\
\vspace{-2mm}
\includegraphics[width=0.45\textwidth]{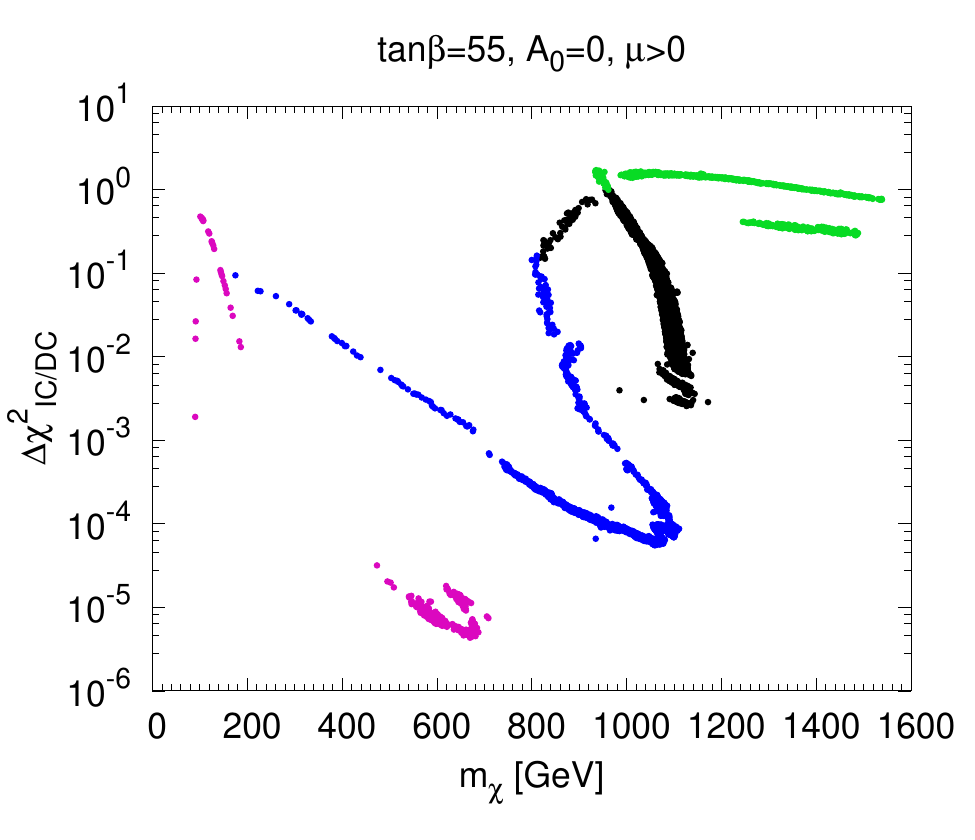} 
\includegraphics[width=0.45\textwidth]{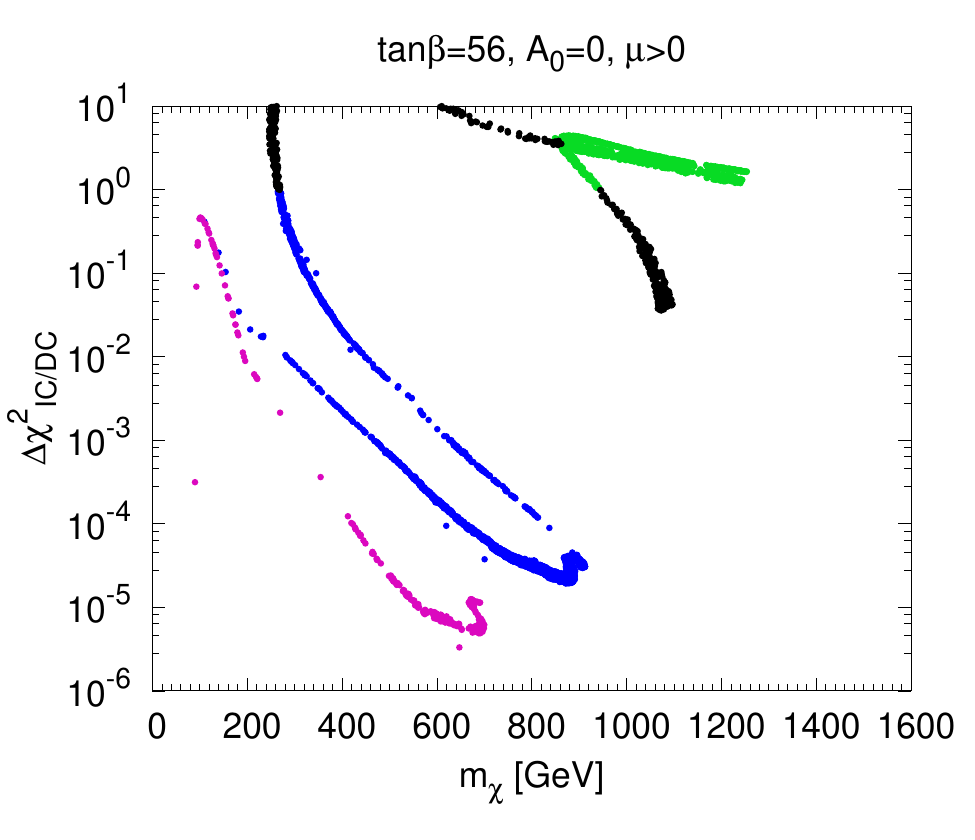} 
\vspace{-4mm}
\caption{\it Contributions to the global $\chi^2$ likelihood function
from a comparison of the IceCube upper limit on the muon flux generated by energetic neutrinos produced by WIMP annihilations in the core of the Sun~\cite{IceCube:2016yoy,IceCube:2016dgk,IceCube:2021xzo}
with calculations for points along the dark matter strips
for $\tan \beta = 5, 20, 40, 50, 55 \; and \; 56$ with $A_0 = 0$ and $\mu > 0$.}
\label{fig:muflux}
\end{figure}

The results of our analysis of the IceCube muon flux limits 
for $\tan \beta = 5, 20, 40, 50, 55$ and $56$ are shown in
Fig.~\ref{fig:muflux}. We see that for $m_\chi \lesssim 900$~GeV and $\tan \beta \le 50$
the IceCube constraints contribute $2 \gtrsim \Delta \chi^2 \gtrsim 1$, not negligible, but insufficient
to exclude any range of parameters. The $\tan \beta \le 50$ points with larger LSP masses
make negligible contributions to the $\chi^2$ function. For $\tan \beta = 55$ there are points with 
900~GeV $\lesssim m_\chi \lesssim 1400$~GeV that have $\Delta \chi^2 \gtrsim 1$, 
but there is again no exclusion. The only points excluded by the IceCube muon flux limits
are focus points for $\tan \beta = 56$ with 250~GeV $\lesssim m_\chi \lesssim 850$~GeV
and some well-tempered points with larger $m_\chi$.
However, these points were already excluded because they correspond to values of $m_h$
that are too low, as seen in Fig.~\ref{fig:mh}, and by spin-independent cross sections that
are too high as seen in Fig.~\ref{fig:xsec}.

\section{Combined Results}
\label{sec:results}

\begin{figure}[ht!]
\includegraphics[width=0.49\textwidth]{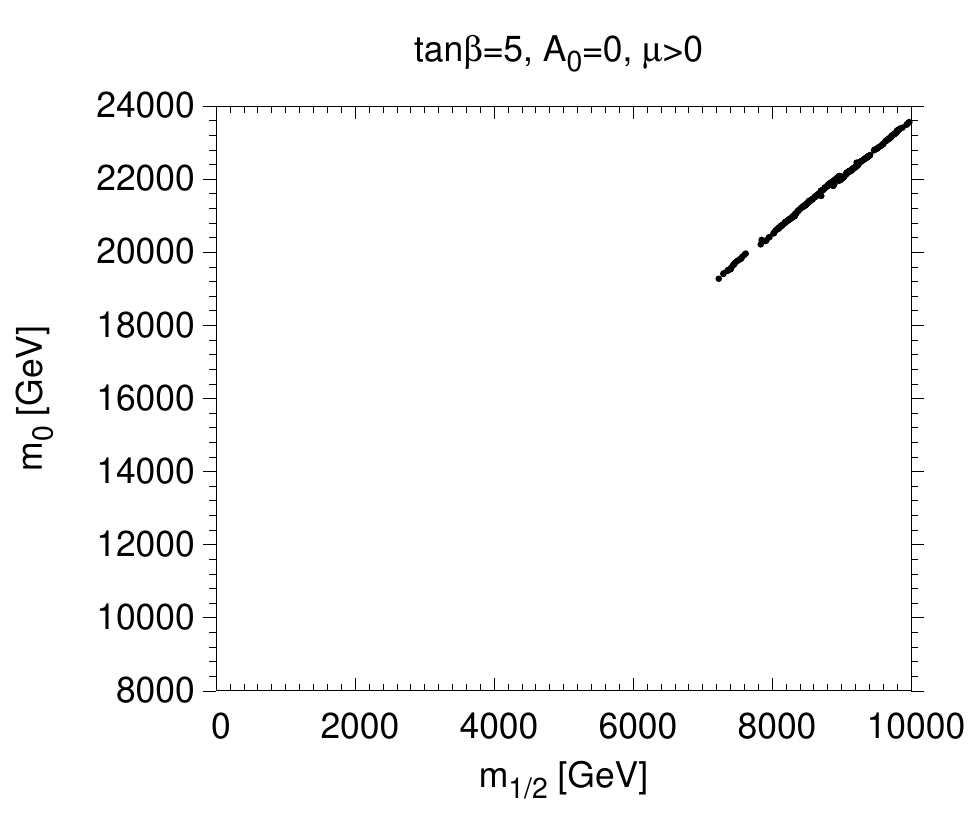} 
\includegraphics[width=0.49\textwidth]{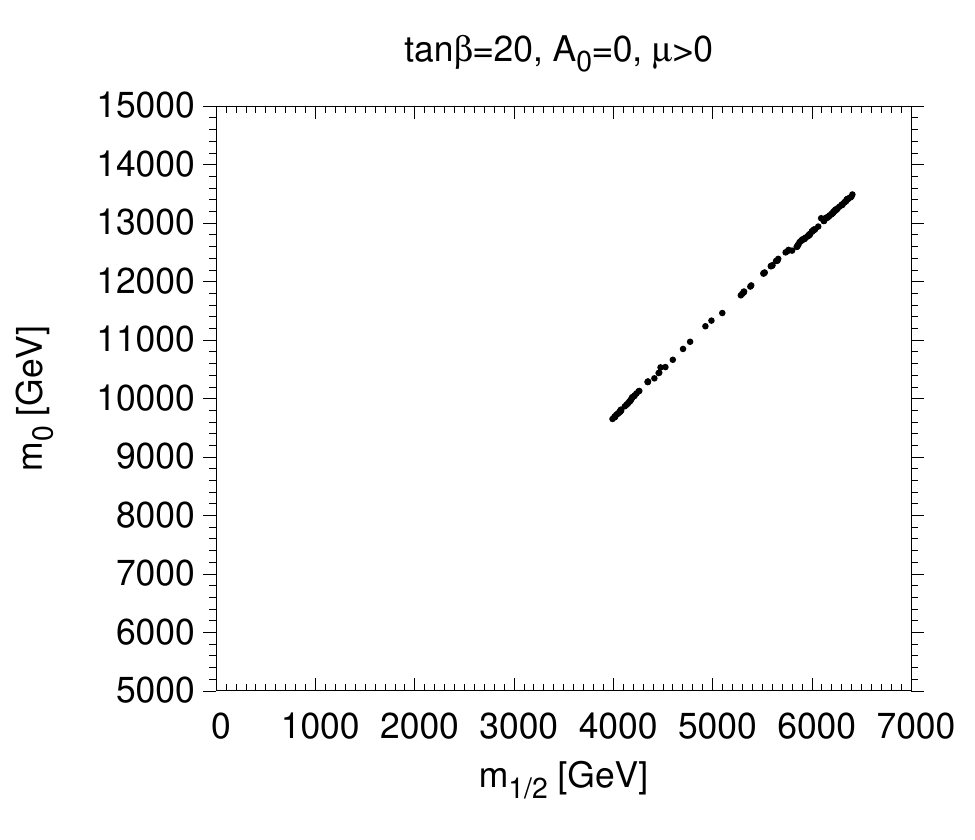} \\
\includegraphics[width=0.49\textwidth]{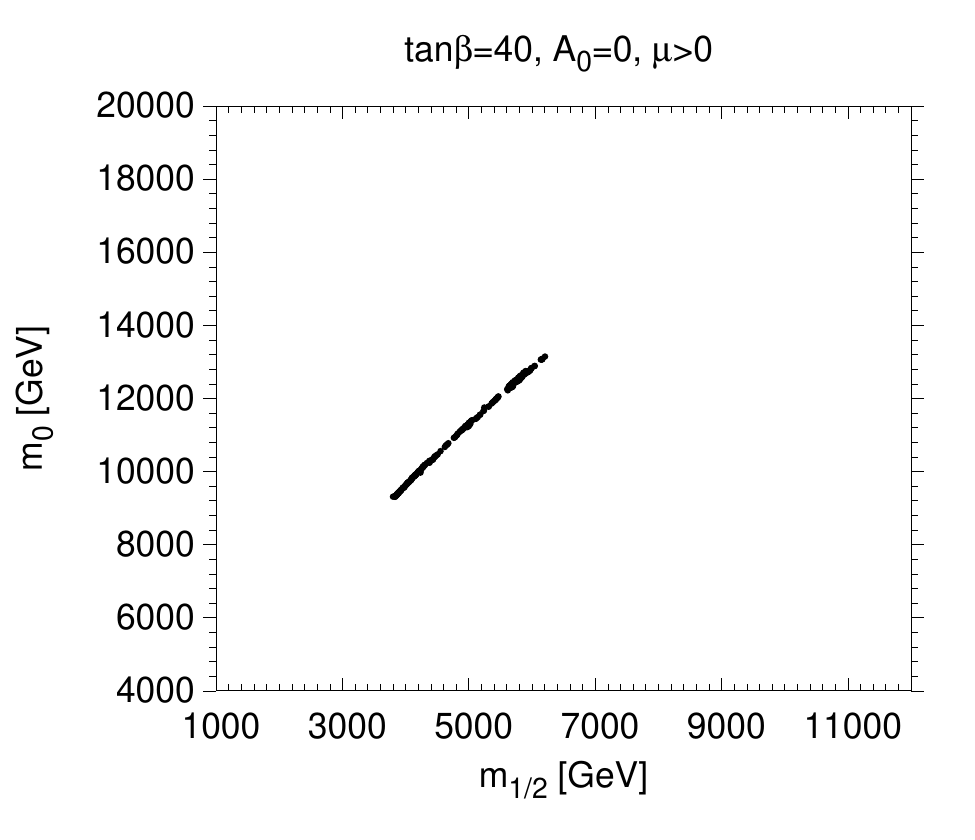}
\includegraphics[width=0.49\textwidth]{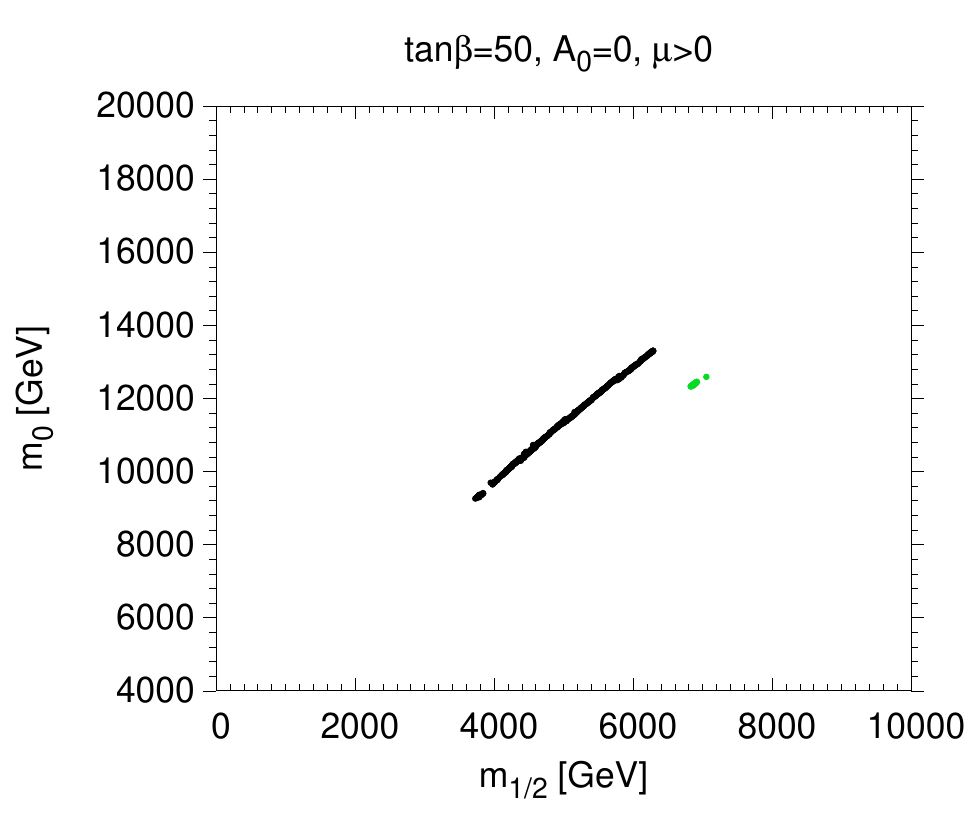} \\
\vspace{-4mm}
\caption{\it The portions of the dark matter strips
for $\tan \beta = 5, 20, 40$ and $50$,
calculated assuming $A_0 = 0$ and $\mu > 0$, that are
allowed by all the constraints, assuming an uncertainty of $1.5$~GeV in the calculation of $m_h$. There are no allowed regions
for $\tan \beta = 55$ or $56$.}
\label{fig:combined}
\end{figure}

\begin{figure}[ht!]
\includegraphics[width=0.49\textwidth]{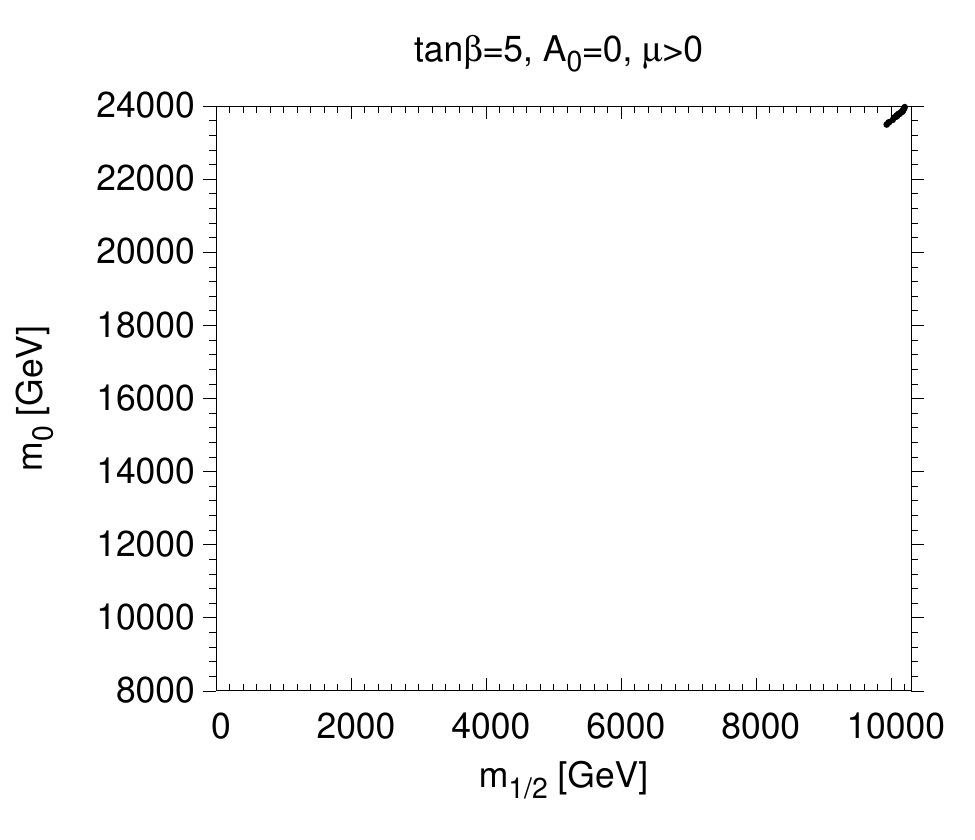} 
\includegraphics[width=0.49\textwidth]{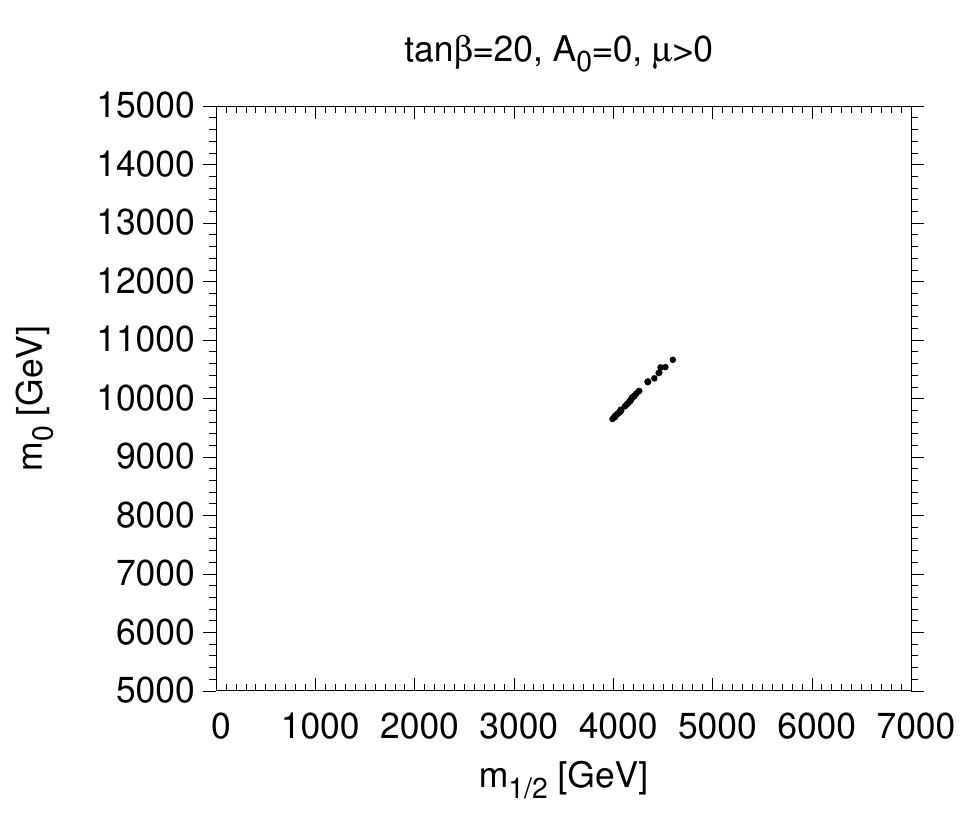} \\
\includegraphics[width=0.49\textwidth]{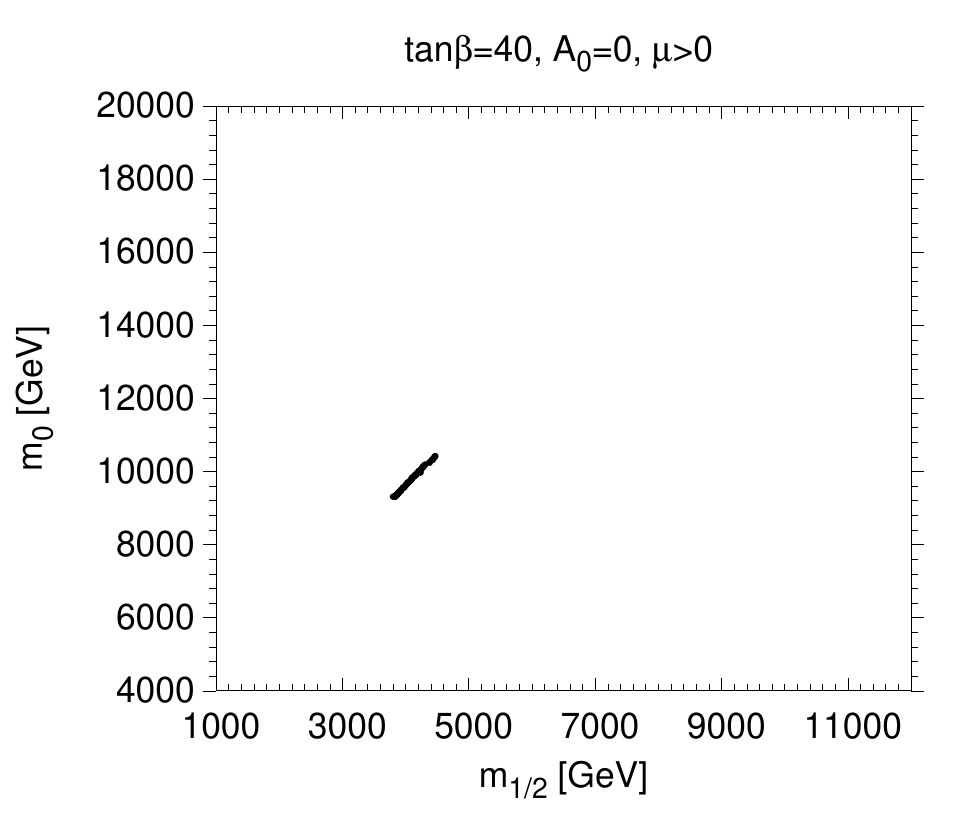}
\includegraphics[width=0.49\textwidth]{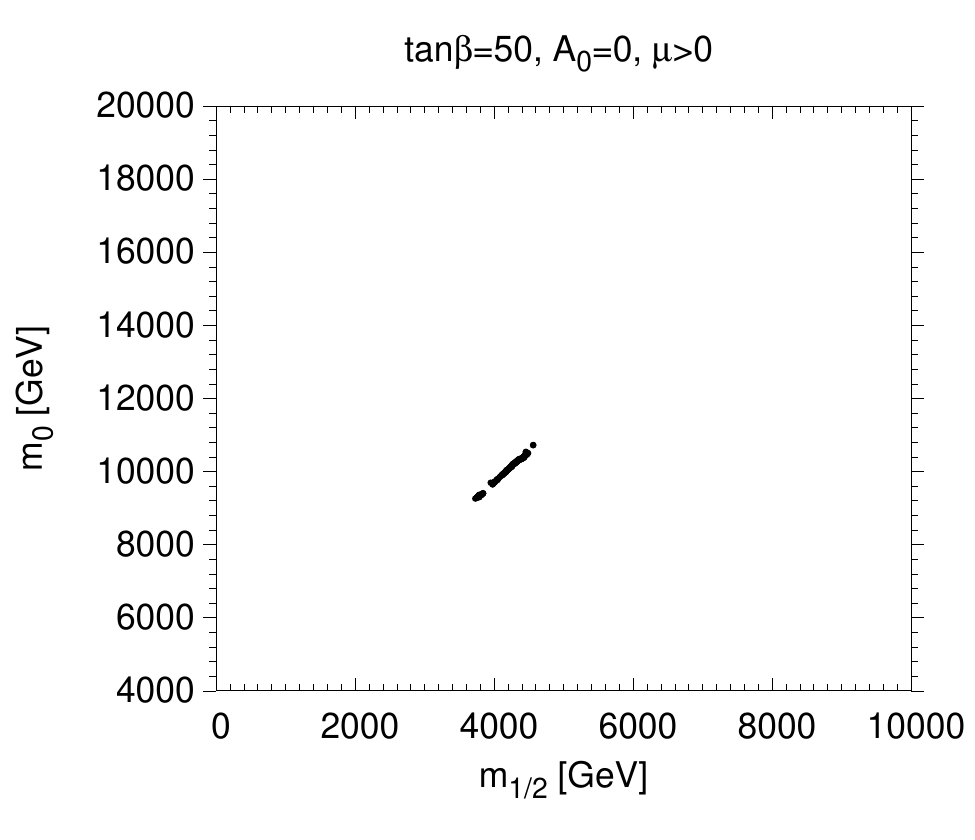} \\
\vspace{-4mm}
\caption{\it As in Fig.~\ref{fig:combined}, but assuming an uncertainty of $0.5$~GeV in the calculation of $m_h$.}
\label{fig:combined2}
\end{figure}

Figs.~\ref{fig:combined} and \ref{fig:combined2} display for $A_0 = 0$,
$\mu > 0$ and our standard set of
$\tan \beta$ values the results that we obtain by combining all the individual 
constraints discussed above, allowing uncertainties in the calculation
of the Higgs mass $\Delta m_h = 1.5$~GeV and $0.5$~GeV,
respectively. In the case of $\tan \beta = 5$ and $\Delta m_h = 1.5$~GeV
(top left panel of Fig.~\ref{fig:combined}) the $m_h$ and spin-independent scattering constraints
combine to exclude values of $m_{1/2} \lesssim 7$~TeV, increasing to
$\sim 9$~TeV for $\Delta m_h \sim 0.5$~GeV (top left panel of Fig.~\ref{fig:combined2}).~{\footnote{Note that the calculated uncertainty for $\tan \beta = 5$ is approximately 0.7~GeV, but $\sim 0.5$~GeV for all higher values of $\tan \beta$.}} 
For $\tan \beta = 5$, the Higgs mass constraint is dominant. There are
analogous exclusions for $\tan \beta = 20$ and 40, but extending
only to lower values of $m_{1/2} \lesssim 4$~TeV, where the lower limit is determined by the spin-independent scattering constraint.  In the case of
$\tan \beta = 50$, these constraints combine to exclude the funnel region and 
also the part of the focus-point strip with $m_{1/2} \lesssim 3.5$~TeV.
There is also a small region of the focus-point strip with
$m_{1/2} \gtrsim 6.5$~TeV that is excluded by the LHC measurement
of $m_h$. We also see that most of the well-tempered strip allowed by $m_h$ is excluded by $\sigma^{SI}$, leaving only a small set of points near $m_{1/2} \simeq 7$ TeV. Finally, we find that all of the strips for $\tan \beta = 55$
and 56 are excluded: the former by a combination of $m_h$ and
spin-independent scattering, and the latter by $m_h$ alone. The other
constraints considered also exclude independently parts of the regions
excluded by $m_h$ and spin-independent scattering, but no supplementary
regions.

\section{The Case $A_0 = 3 \, m_0$, $\mu > 0$}
\label{sec:3m0}

\begin{figure}[ht!]
\includegraphics[width=0.45\textwidth]{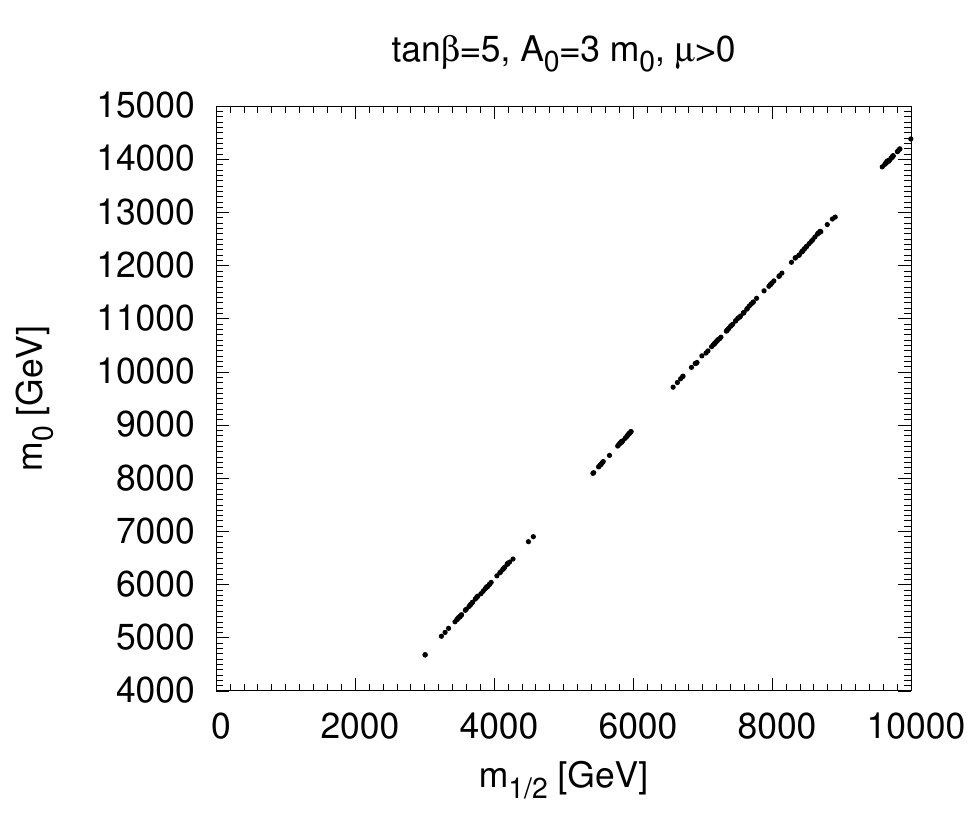} 
\includegraphics[width=0.45\textwidth]{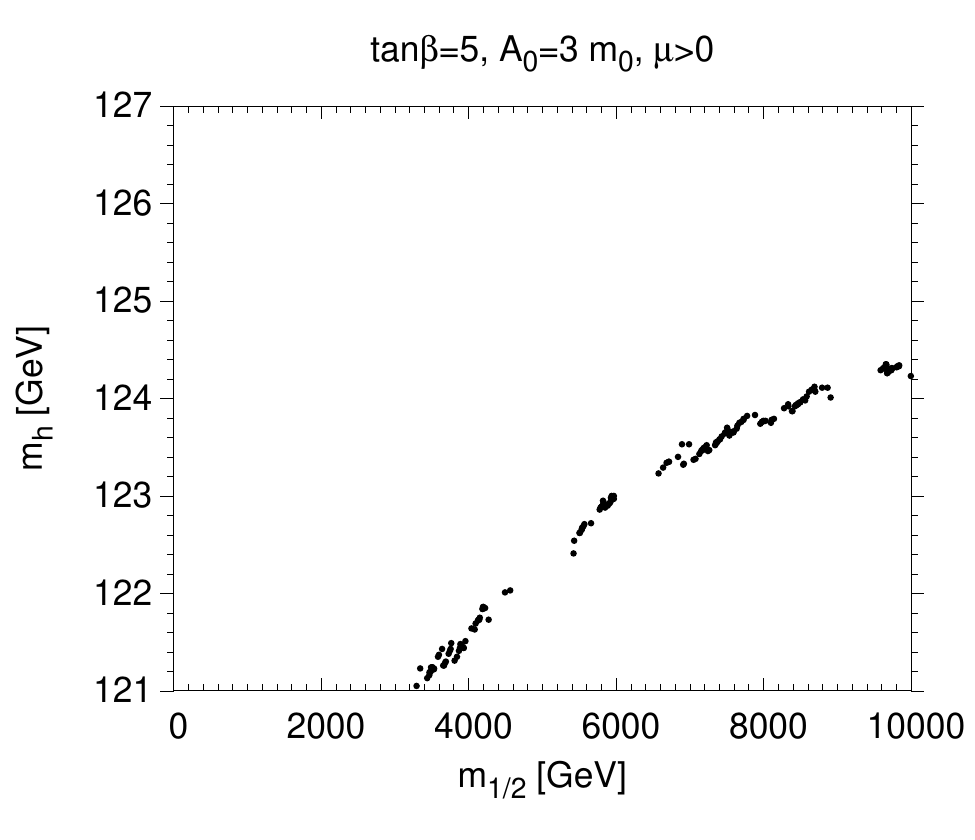} \\
\vspace{-2mm}
\includegraphics[width=0.45\textwidth]{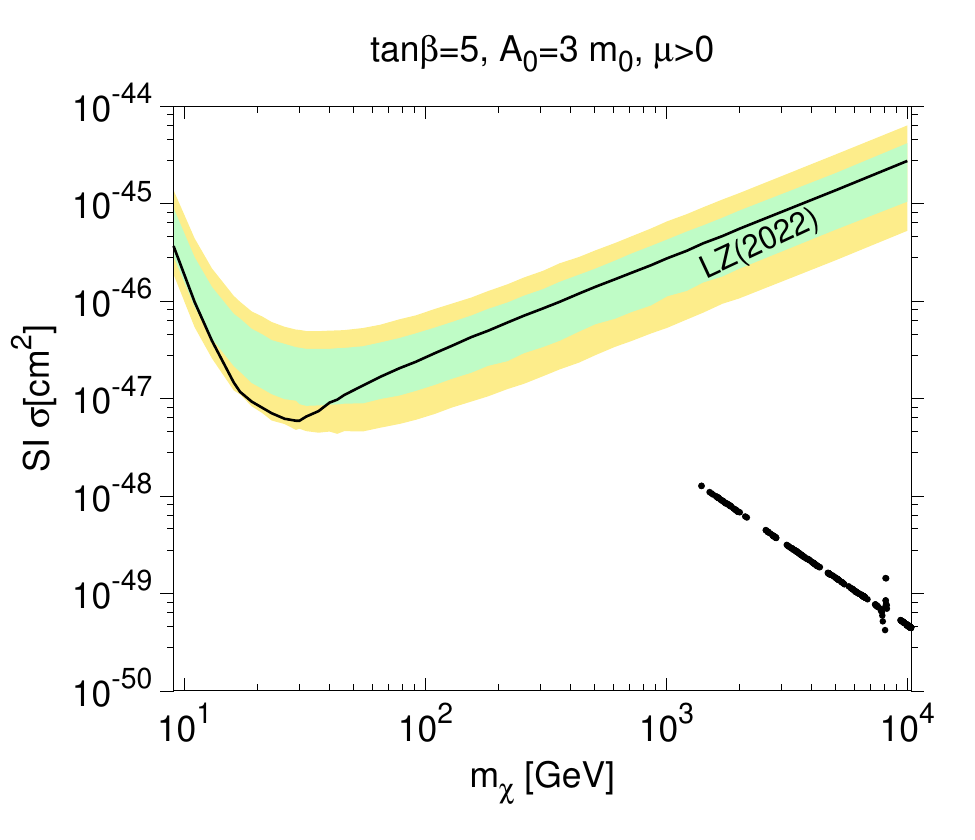} 
\includegraphics[width=0.45\textwidth]{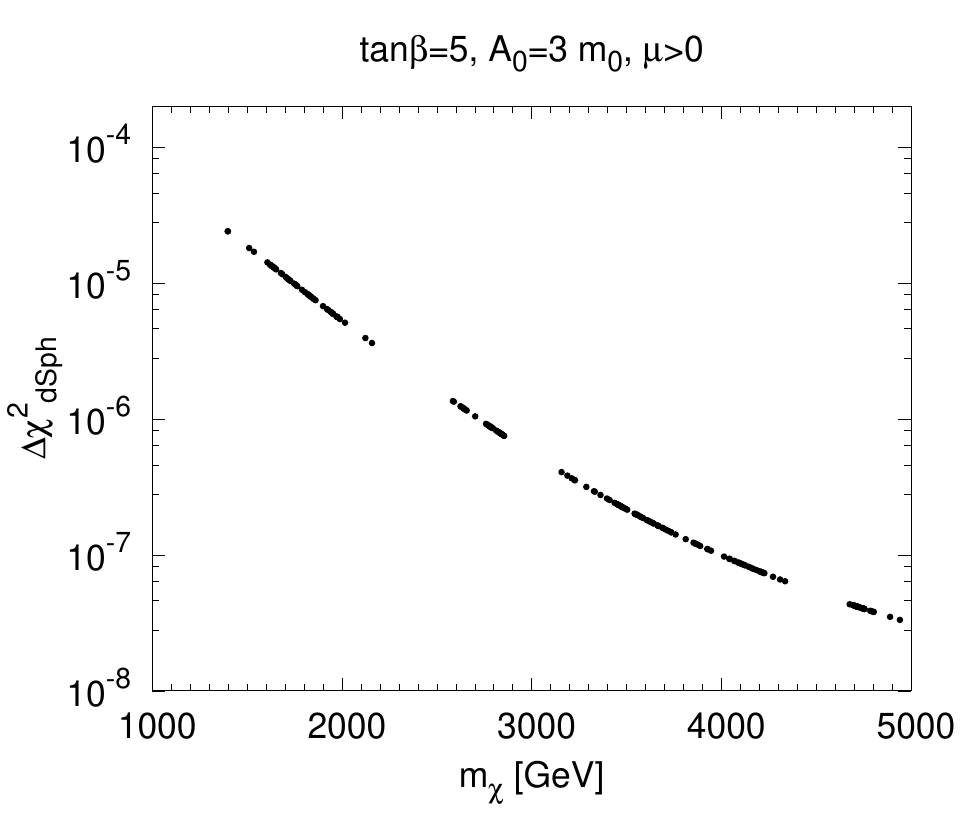}\\
\vspace{-2mm}
\includegraphics[width=0.45\textwidth]{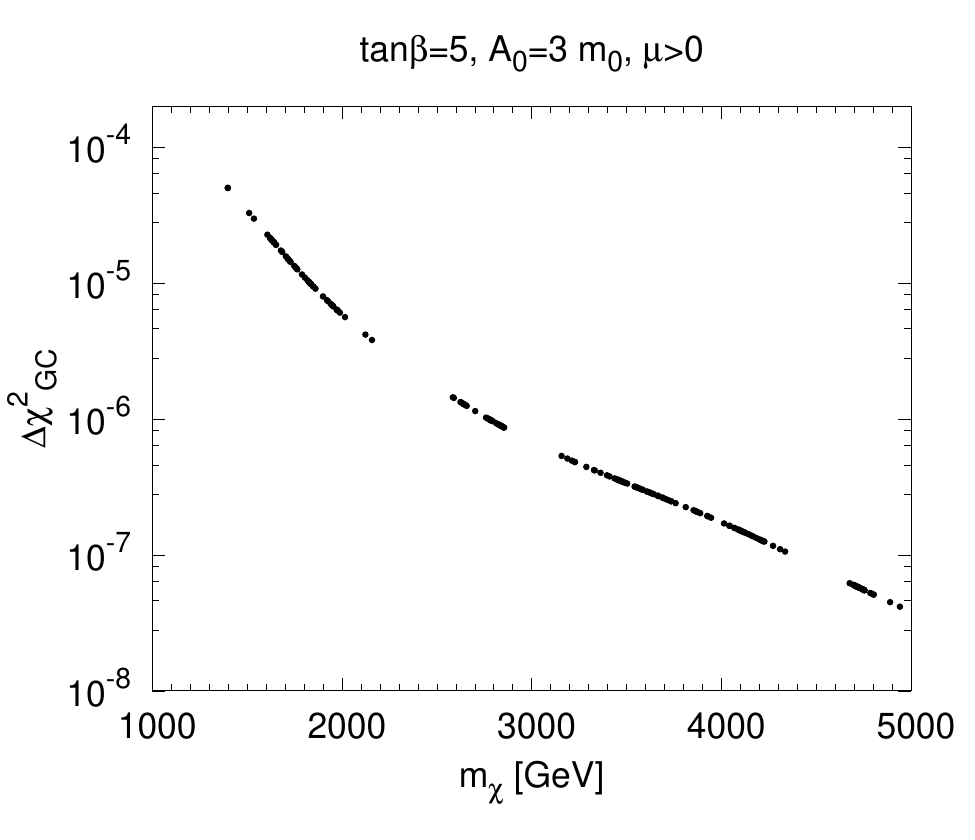} 
\includegraphics[width=0.45\textwidth]{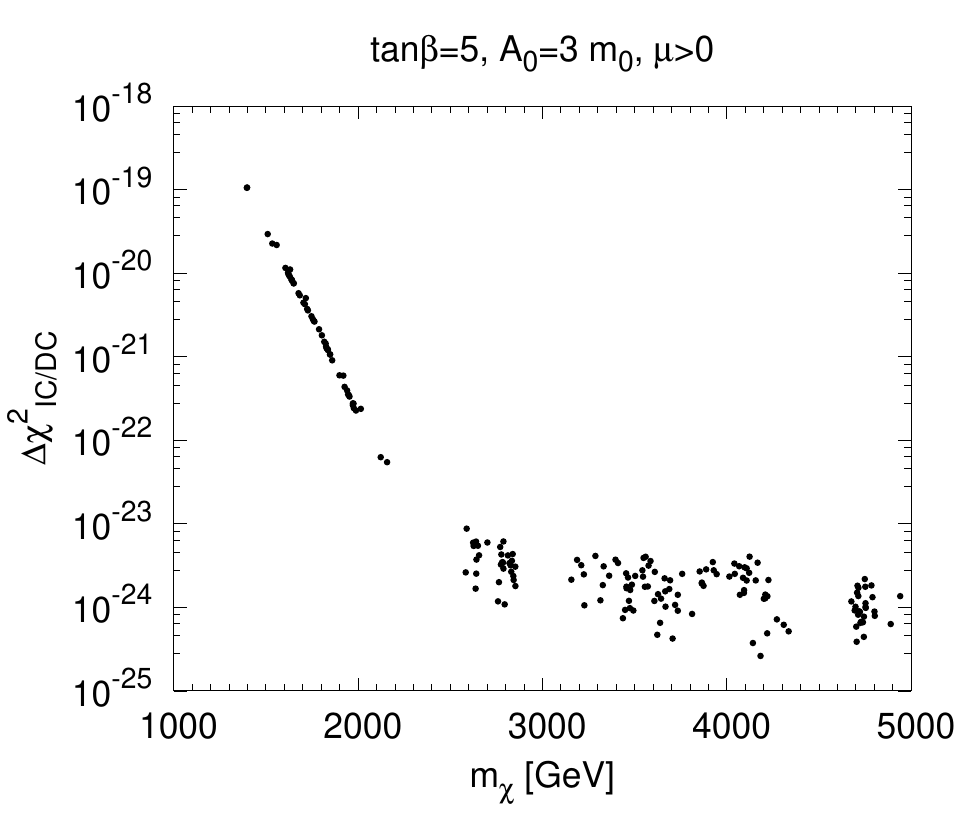} \\
\vspace{-4mm}
\caption{\it The dark matter strip, $m_h$ constraint, spin-independent direct scattering constraint, $\Delta \chi^2$ from
{\it Fermi}-LAT dSph and H.E.S.S. GC $\gamma$-ray and Icecube solar $\nu$ signals for the case
$\tan\beta=5$, $A_0=3 m_0$ and $\mu>0$.}
\label{fig:combined3}
\end{figure}

\begin{figure}[ht!]
\includegraphics[width=0.45\textwidth]{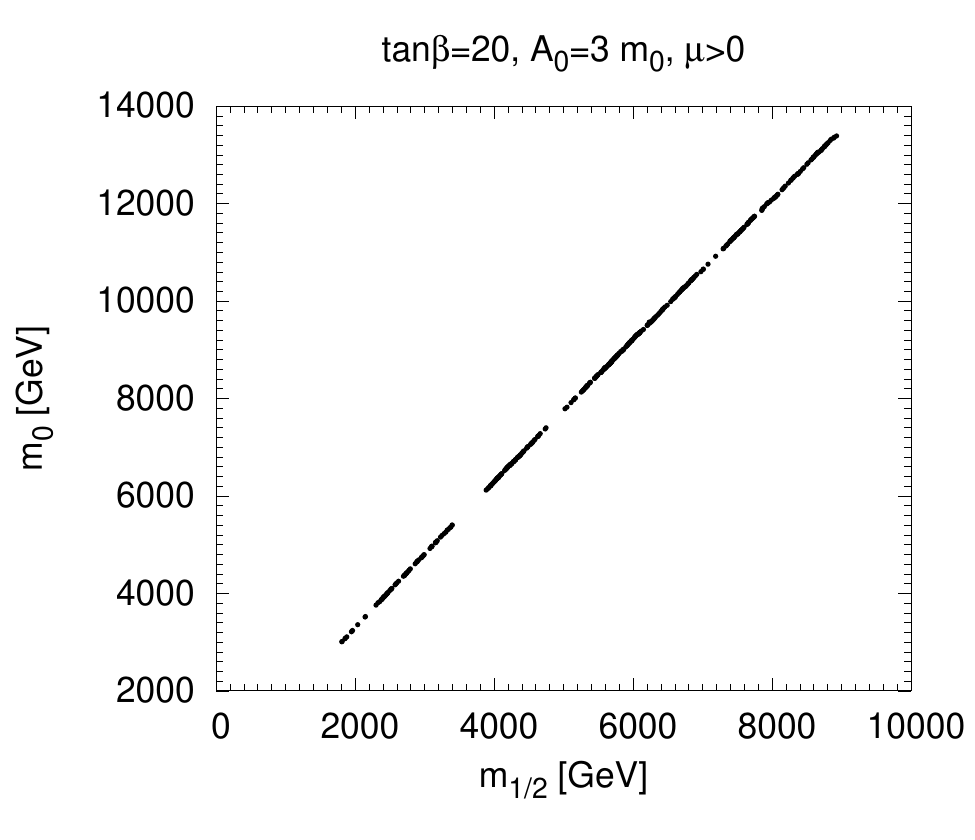} 
\includegraphics[width=0.45\textwidth]{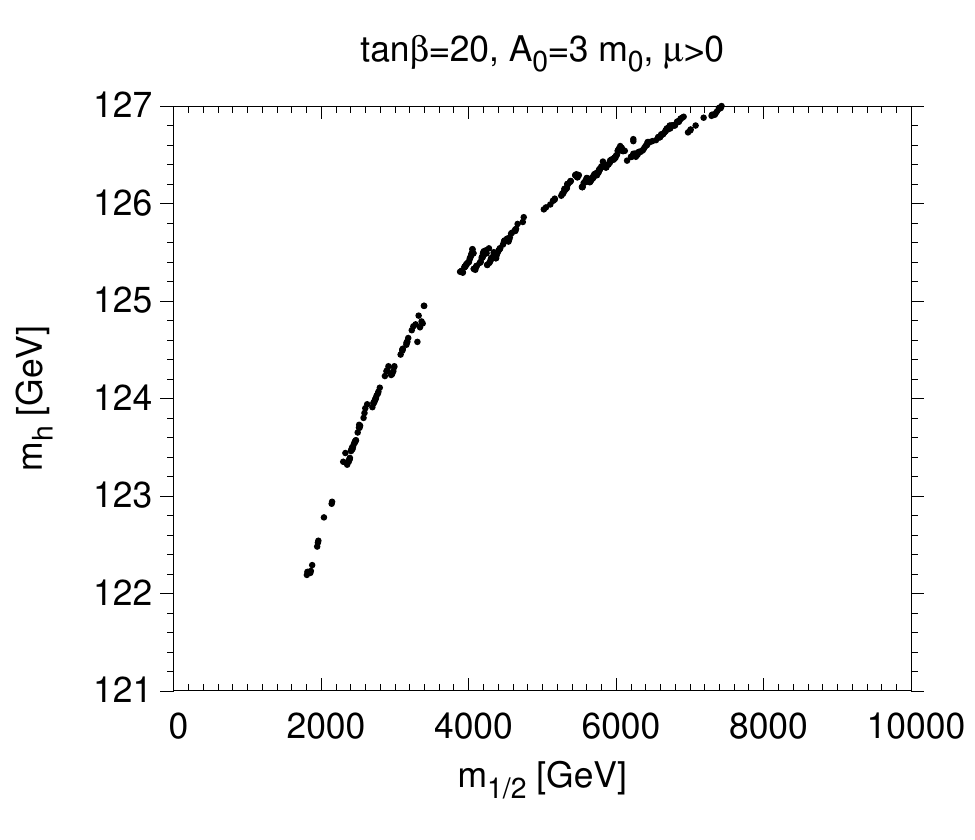} \\
\vspace{-2mm}
\includegraphics[width=0.45\textwidth]{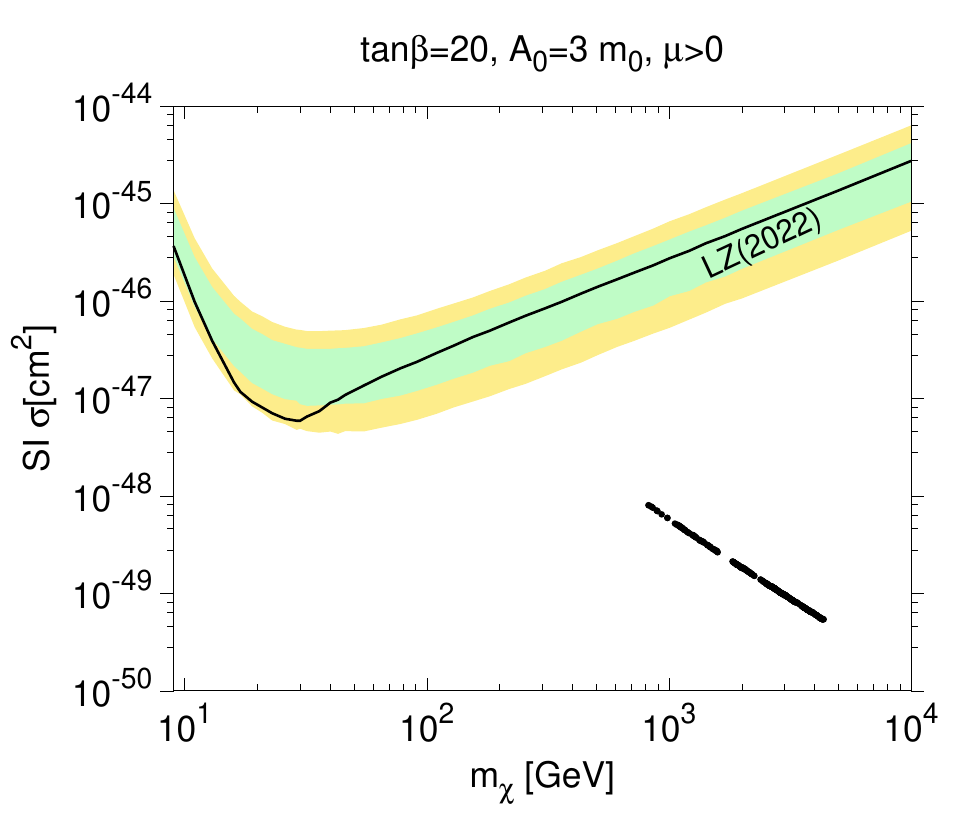} 
\includegraphics[width=0.45\textwidth]{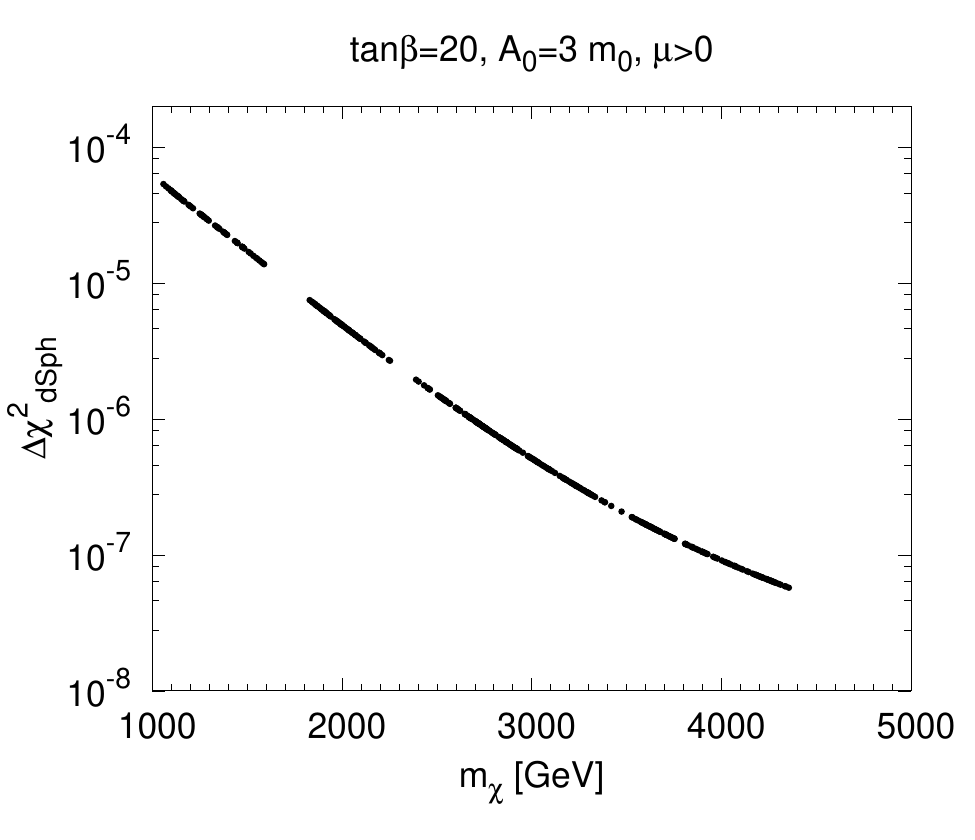}\\
\vspace{-2mm}
\includegraphics[width=0.45\textwidth]{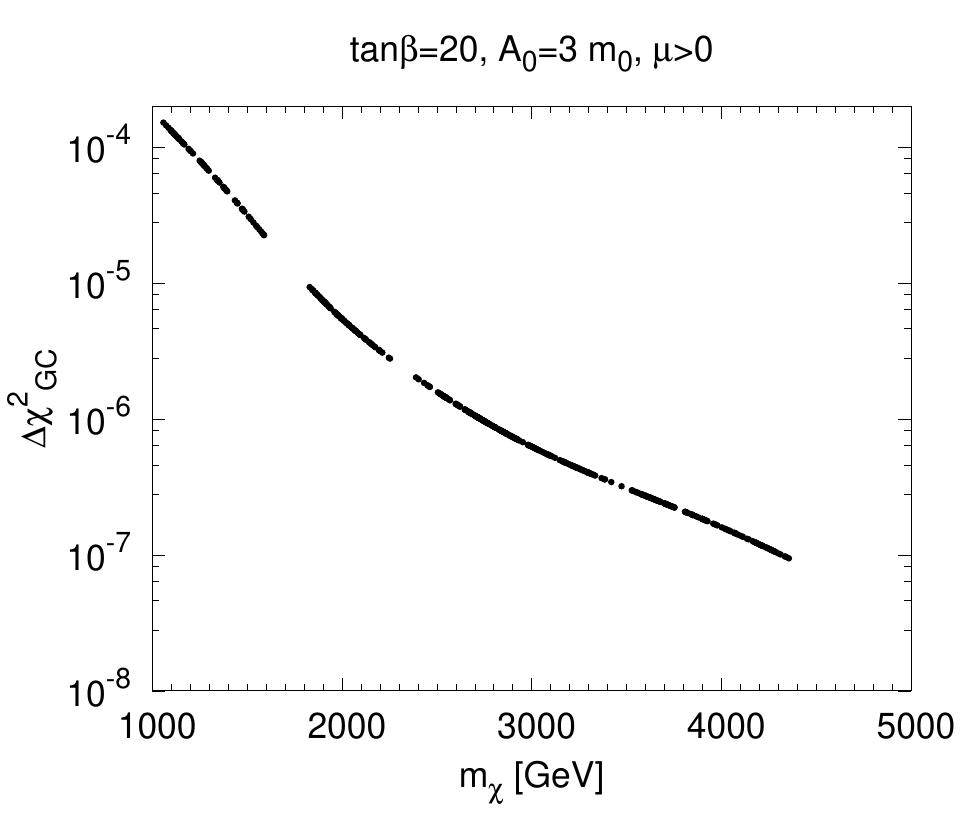} 
\includegraphics[width=0.45\textwidth]{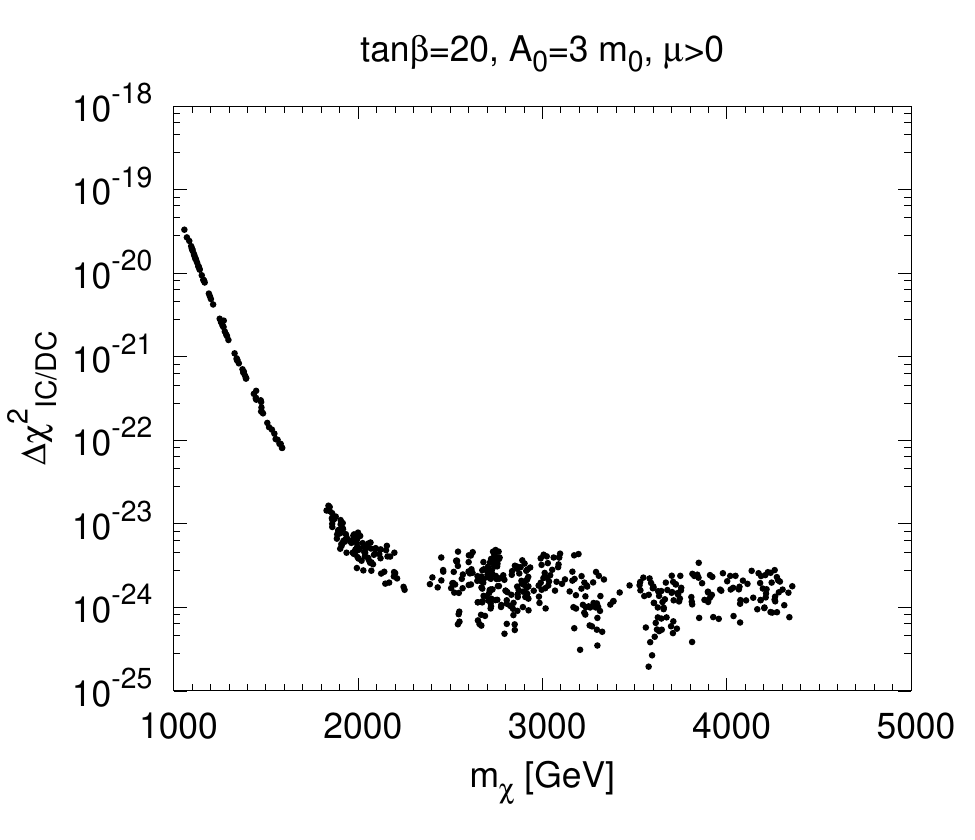} \\
\vspace{-4mm}
\caption{\it The dark matter strip, $m_h$ constraint, spin-independent direct scattering constraint, $\Delta \chi^2$ from
{\it Fermi}-LAT dSph and H.E.S.S. GC $\gamma$-ray and Icecube solar $\nu$ signals for the case
$\tan\beta=20$, $A_0=3 m_0$ and $\mu>0$.}
\label{fig:combined4}
\end{figure}

For $A_0 = 0$ and any fixed value of $m_{1/2}$, radiative electroweak symmetry breaking, i.e. a solution for the Higgs vevs by minimizing the Higgs potential, is no longer possible for sufficiently large $m_0$. This boundary is adjacent to the focus-point region where $\mu$ is driven to zero.
This boundary moves to higher values of $m_0$ as $A_0/m_0$ increases.  In addition, when $A_0/m_0$ is increased, there is increased splitting in the squark sector, and most notably, one of the stop masses becomes relatively light and comparable to the LSP mass allowing for the possibility that LSP-stop coannihilations determine the relic density \cite{stopco,esug,Ellis:2018jyl}. When this occurs, 
there is again a thin dark matter strip 
adjacent to the boundary of the region where the light stop becomes the LSP. At still higher $m_0$, the lighter stop becomes tachyonic. Examples of the stop coannihilation strips for the representative choice $A_0/m_0 = 3$ with $\tan\beta=5$ and $\tan \beta = 20$
are shown in the top left panels of Figs.~\ref{fig:combined3} and \ref{fig:combined4}. Along these strips, the complicated interplay of dark matter mechanisms such
as annihilation via $s$-channel resonances, well-tempered neutralino composition, etc., 
does not recur, and there are no other dark matter strips of interest. The uncertainty in the {\tt FeynHiggs~2.18.1}
calculation of $m_h$ is much larger for $A_0 = 3 \, m_0$ than for $A_0 = 0$, with $\Delta m_h \gtrsim 3$~GeV.
{This should be borne in mind when interpreting the top right panels of Figs.~\ref{fig:combined3} and \ref{fig:combined4}.}
In particular, it implies that none of the displayed portions of the stop-coannihilation strips can be excluded on the basis of $m_h$.
When $\tan \beta = 5$, the 1-$\sigma$ range of $m_h$ along the dark matter strip extends down to $m_{1/2} \sim 3$~TeV,
well above the current reach of the LHC. On the other hand, the 1-$\sigma$ range of $m_h$ along the dark matter strip 
for $\tan \beta = 20$ extends down to $m_{1/2} \sim 1.4$~TeV, where $m_\chi \simeq 630$~GeV and $m_{\tilde t_1} \simeq 700$~GeV,
coinciding with the current reach of LHC stop searches for the case of a compressed spectrum \cite{stopbound}, and we find similar results for larger $\tan\beta$.
However, the constraints from direct and indirect searches for astrophysical dark matter are not relevant
along the stop coannihilation strip, as seen in the bottom 4 panels of Figs.~\ref{fig:combined3} and \ref{fig:combined4}.

We note that 
the $\Delta \chi^2$ related to neutrino searches at IceCube is significantly 
suppressed in this case. 
This is due to the fact that in the stop  coannihilation region  both the scalar and 
the spin-dependent direct cross-sections are almost five orders of magnitude smaller than in  the cosmologically 
acceptable  higgsino region for the case $A_0=0$. This results in an annihilation rate
in the Sun which is about eleven orders of magnitude smaller and the same 
applies to  the muon fluxes. Thus,  the resulting  $\Delta \chi^2$ is suppressed by 
almost nineteen orders of magnitude.

\section{The Case $\mu < 0$}
\label{sec:muneg}

\begin{figure}[ht!]
\includegraphics[width=0.45\textwidth]{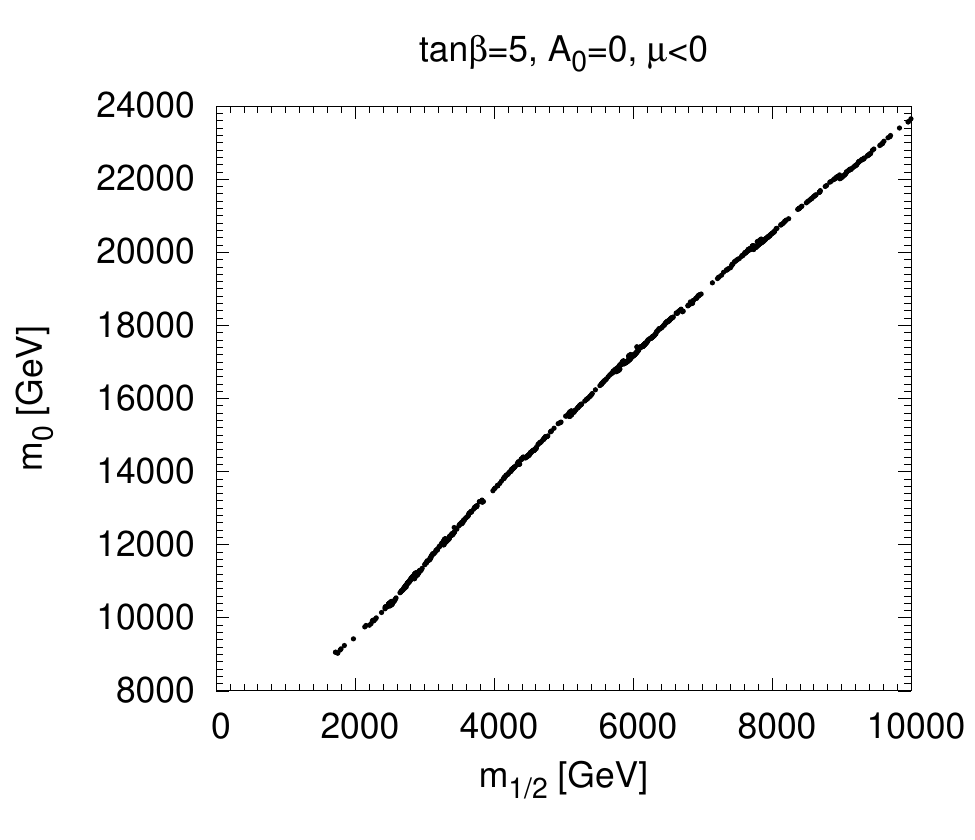}
\includegraphics[width=0.45\textwidth]{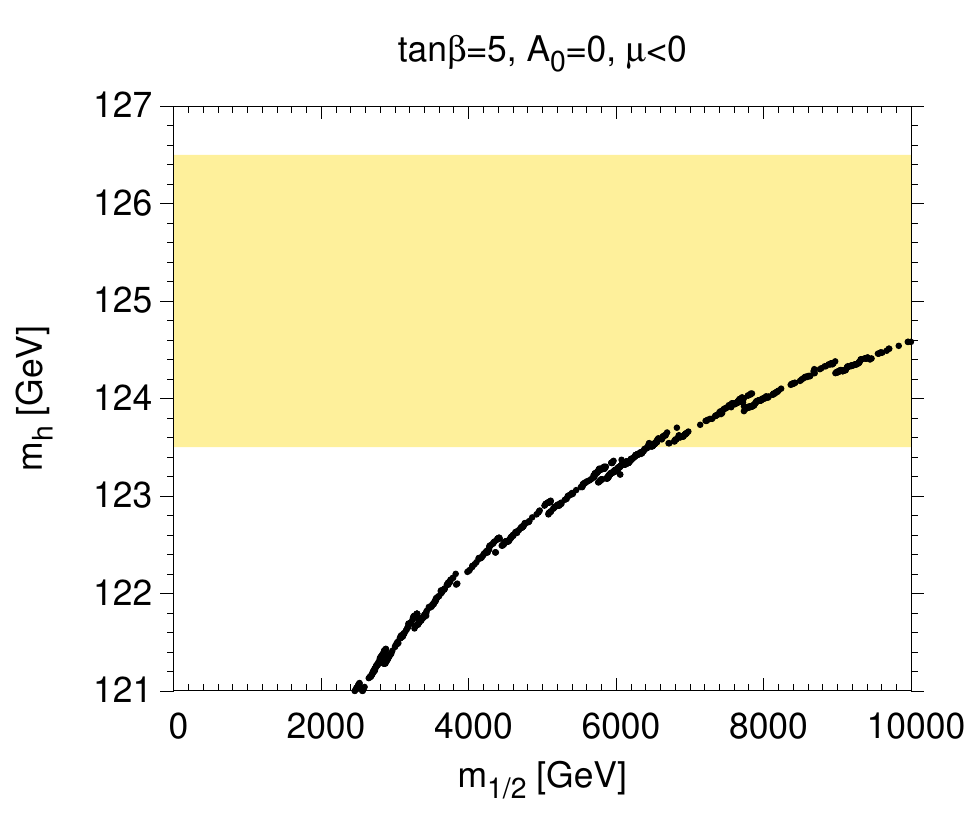} \\
\vspace{-2mm}
\includegraphics[width=0.45\textwidth]{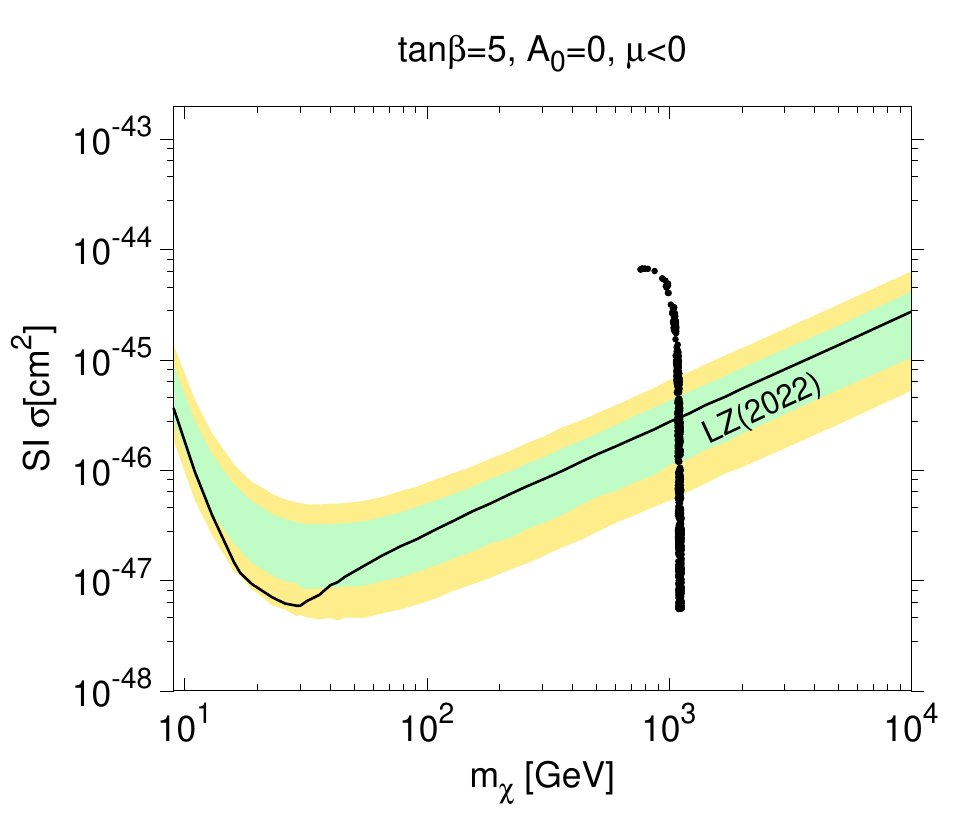}
\includegraphics[width=0.45\textwidth]{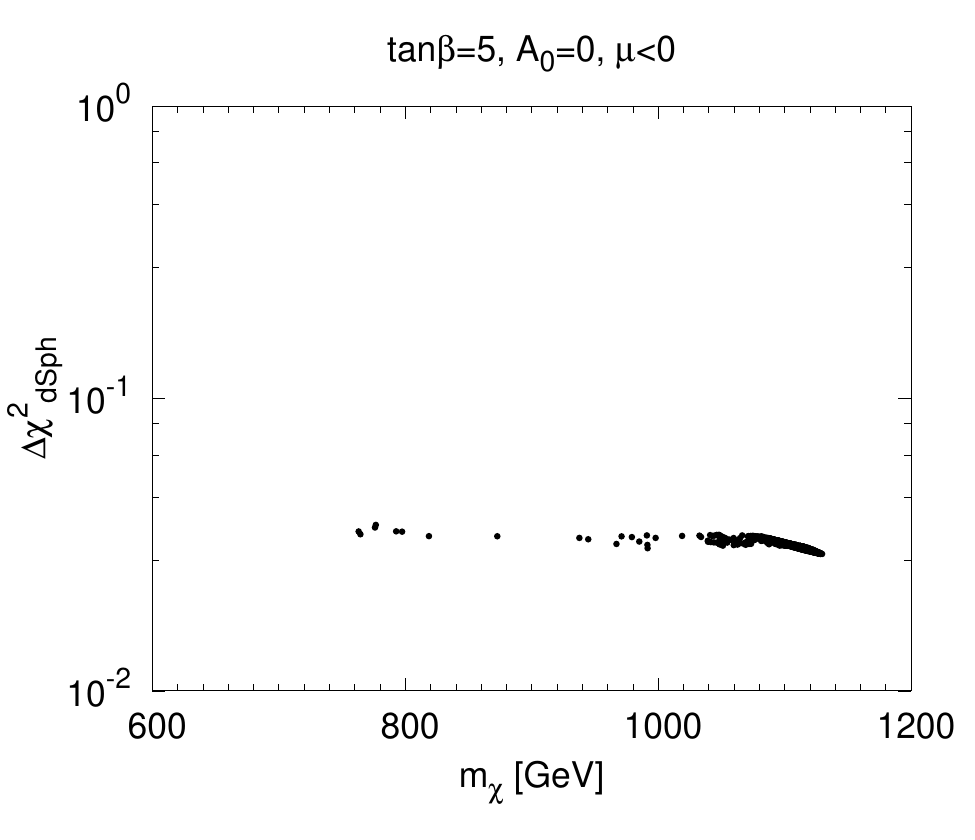} \\
\vspace{-2mm}
\includegraphics[width=0.45\textwidth]{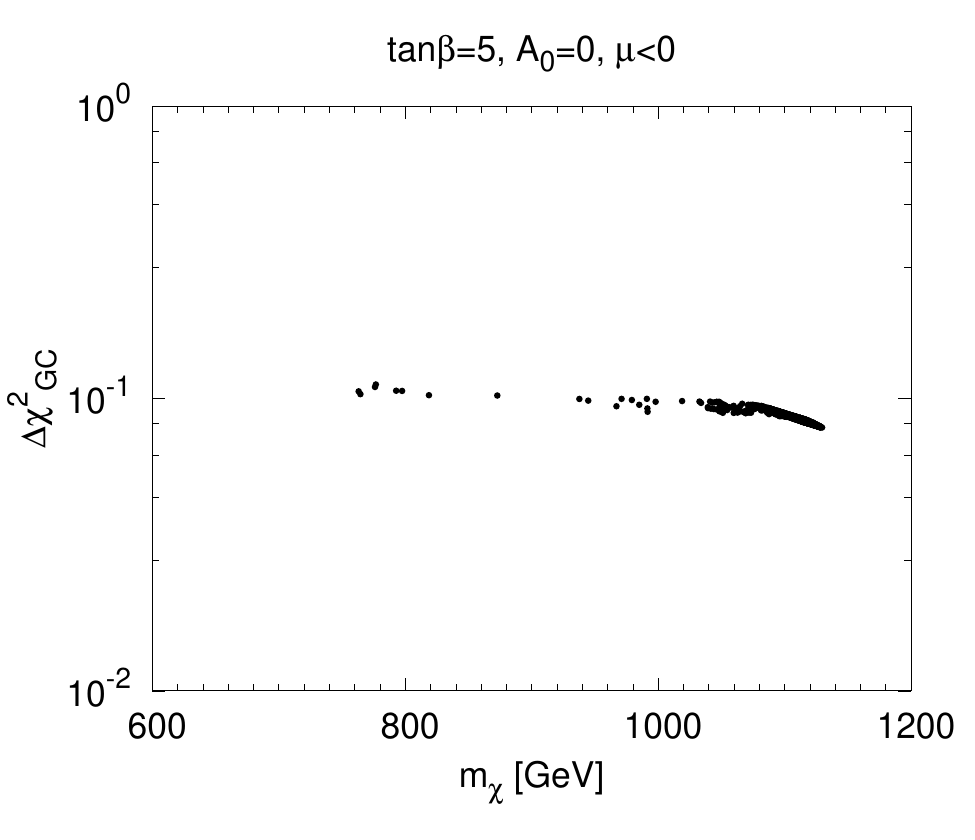} 
\includegraphics[width=0.45\textwidth]{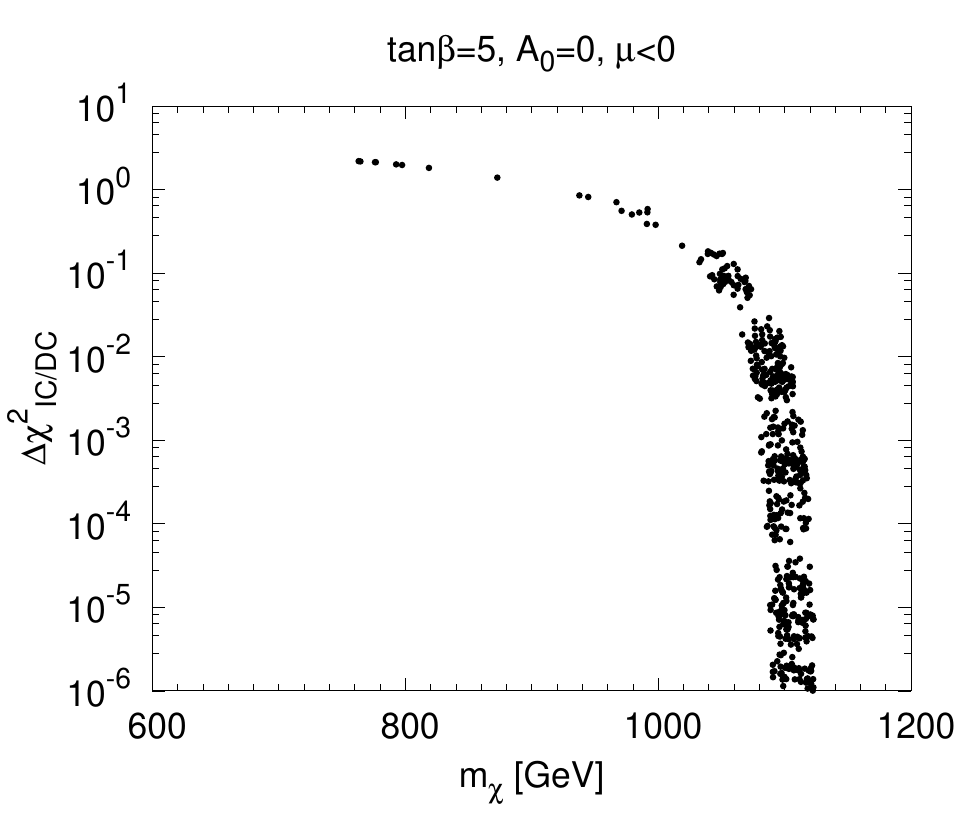} 
\vspace{-4mm}
 \caption{\it The $(m_{1/2}, m_0)$ plane, the $m_h$, spin-independent direct scattering
and indirect constraints for the case
$\tan\beta=5$, $A_0= 0$ and $\mu<0$.
}
\label{fig:5n}
\end{figure}

\begin{figure}[ht!]
\includegraphics[width=0.45\textwidth]{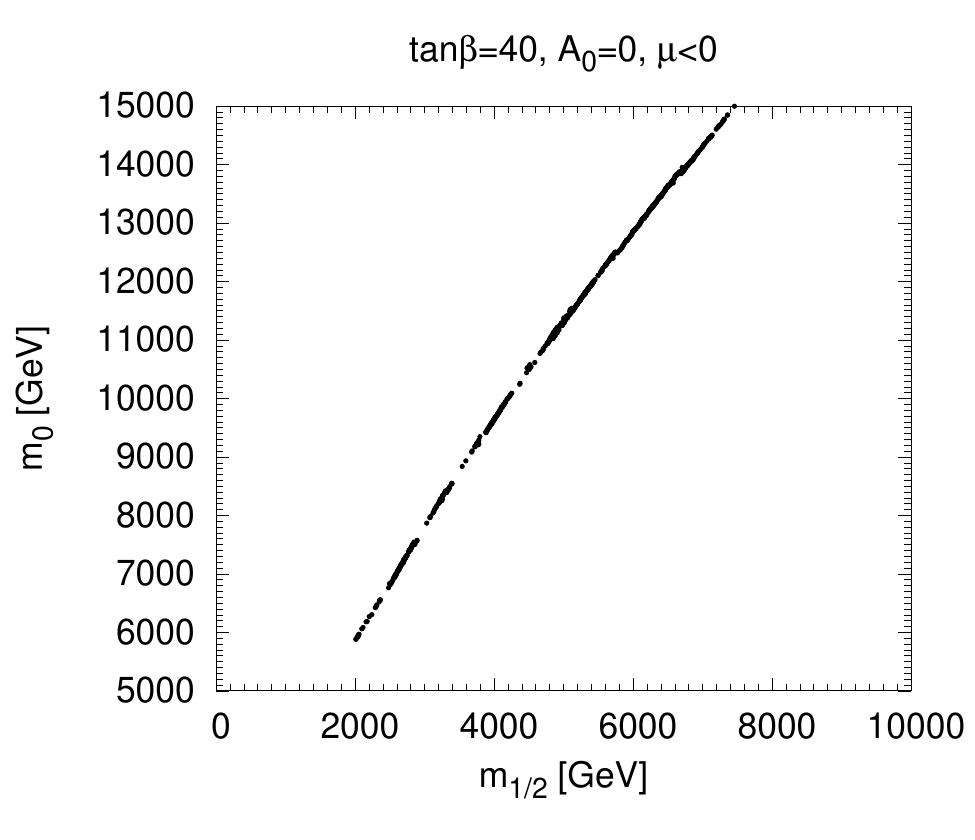}
\includegraphics[width=0.45\textwidth]{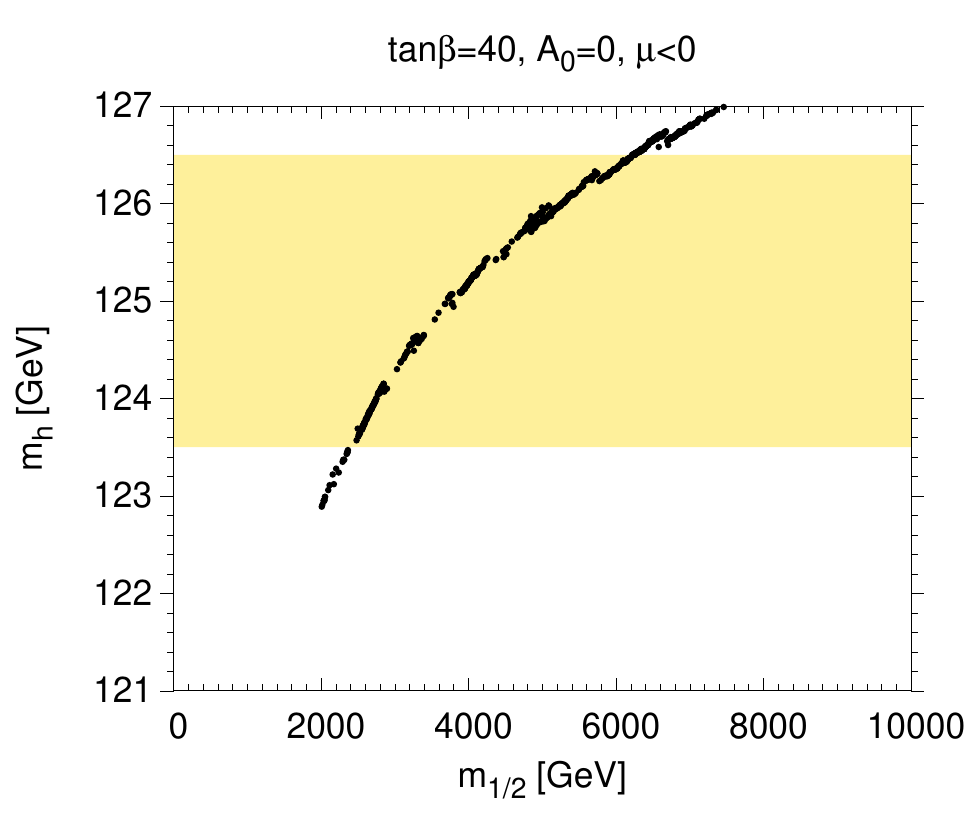} \\
\vspace{-2mm}
\includegraphics[width=0.45\textwidth]{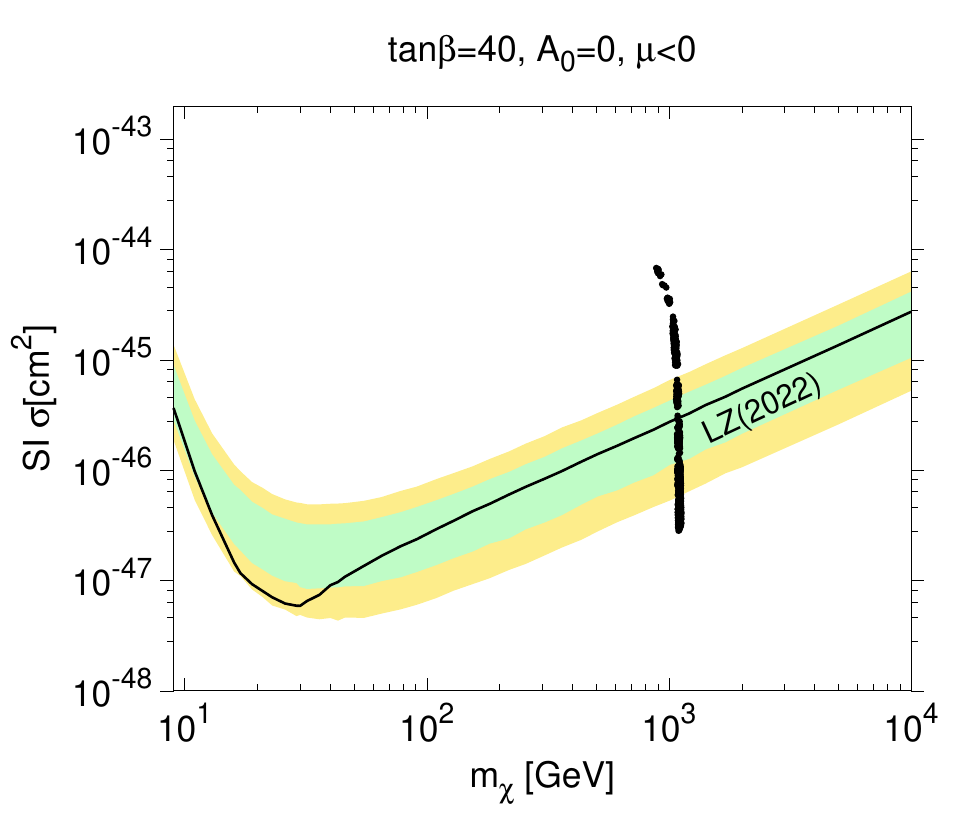}
\includegraphics[width=0.45\textwidth]{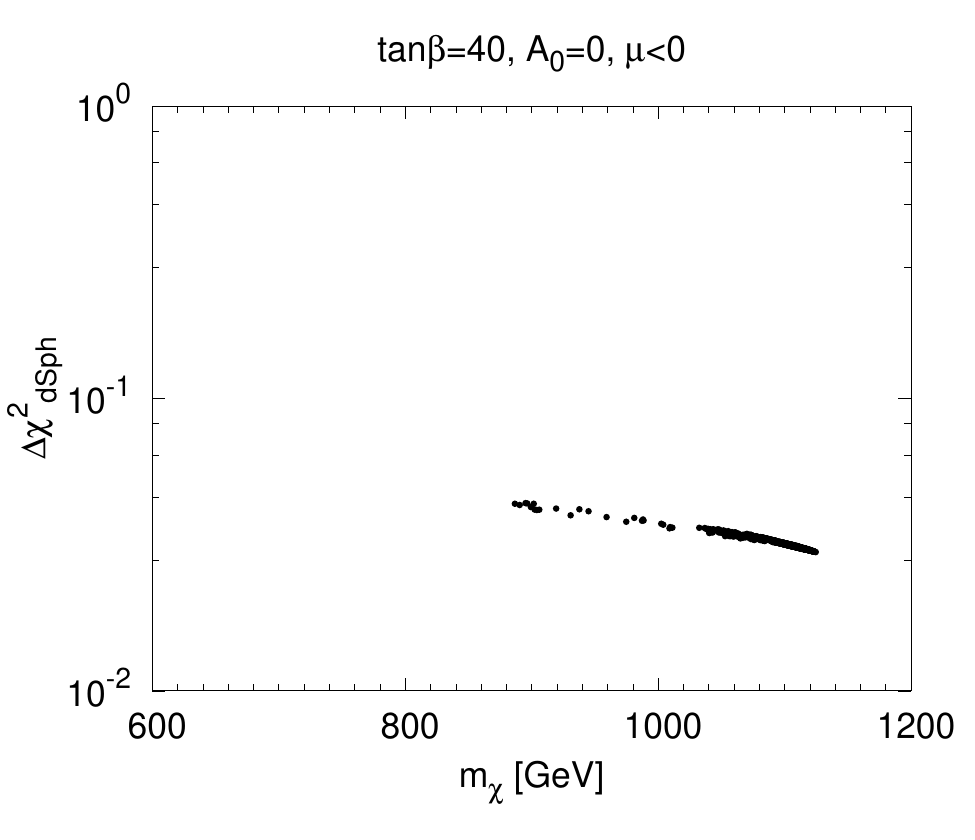} \\
\vspace{-2mm}
\includegraphics[width=0.45\textwidth]{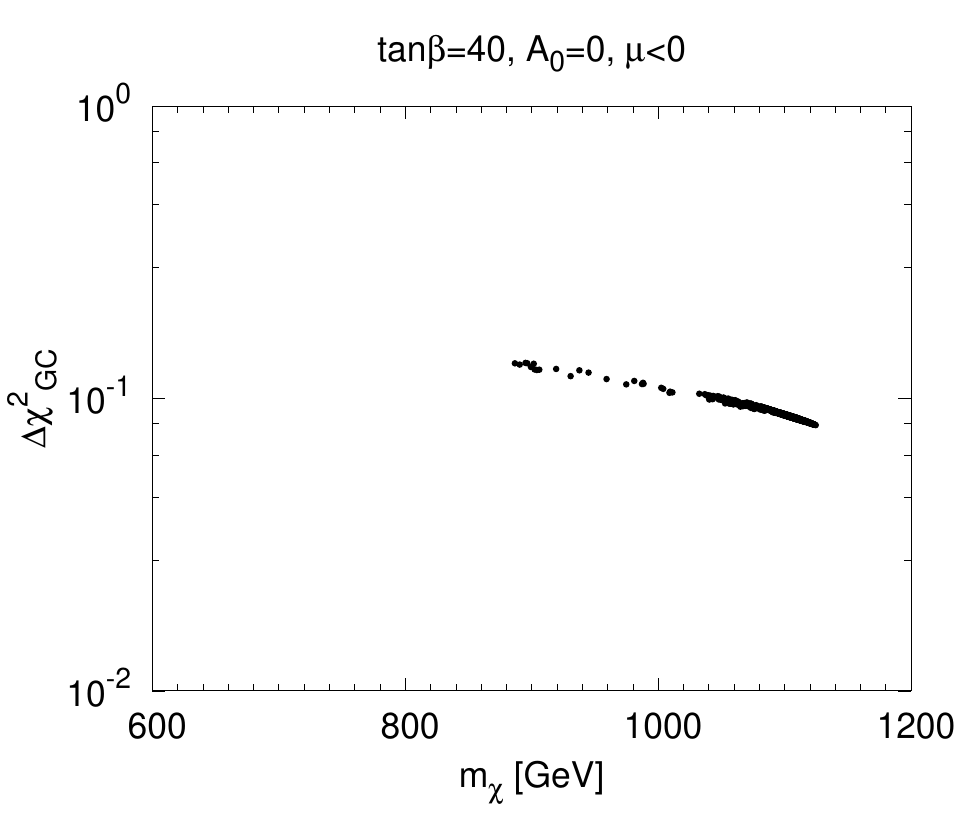} 
\includegraphics[width=0.45\textwidth]{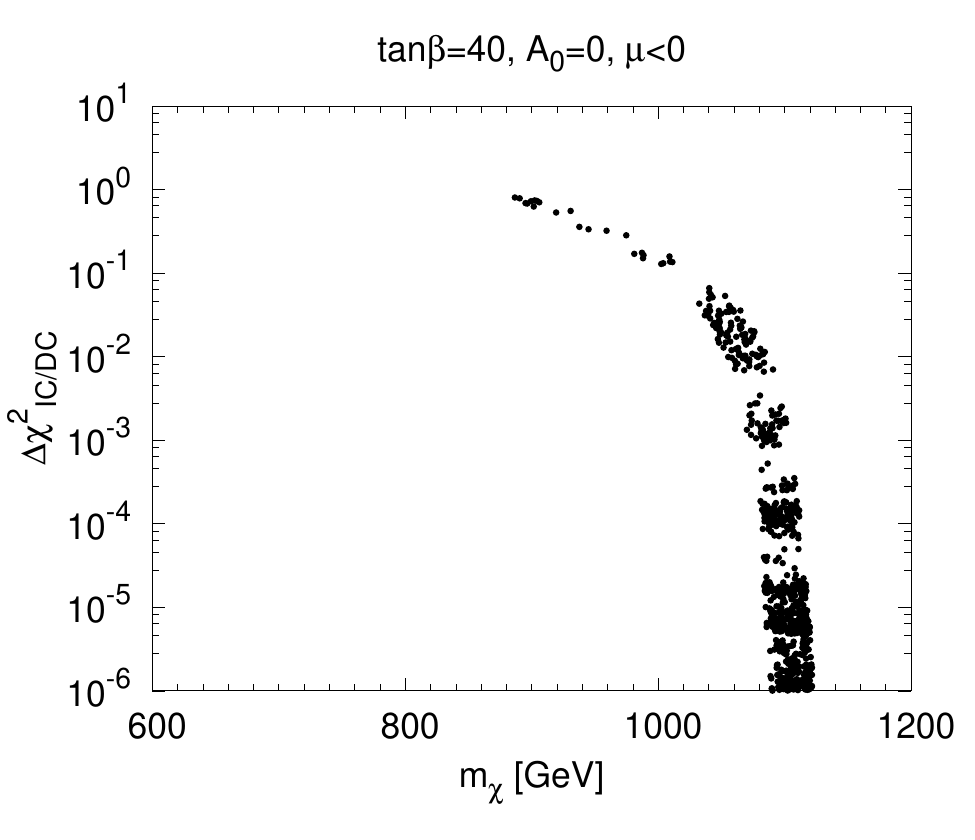} 
\vspace{-4mm}
 \caption{\it As in Fig.~\ref{fig:5n}, for the case
$\tan\beta=40$, $A_0= 0$ and $\mu<0$.}
\label{fig:40n}
\end{figure}

\begin{figure}[ht!]
\includegraphics[width=0.49\textwidth]{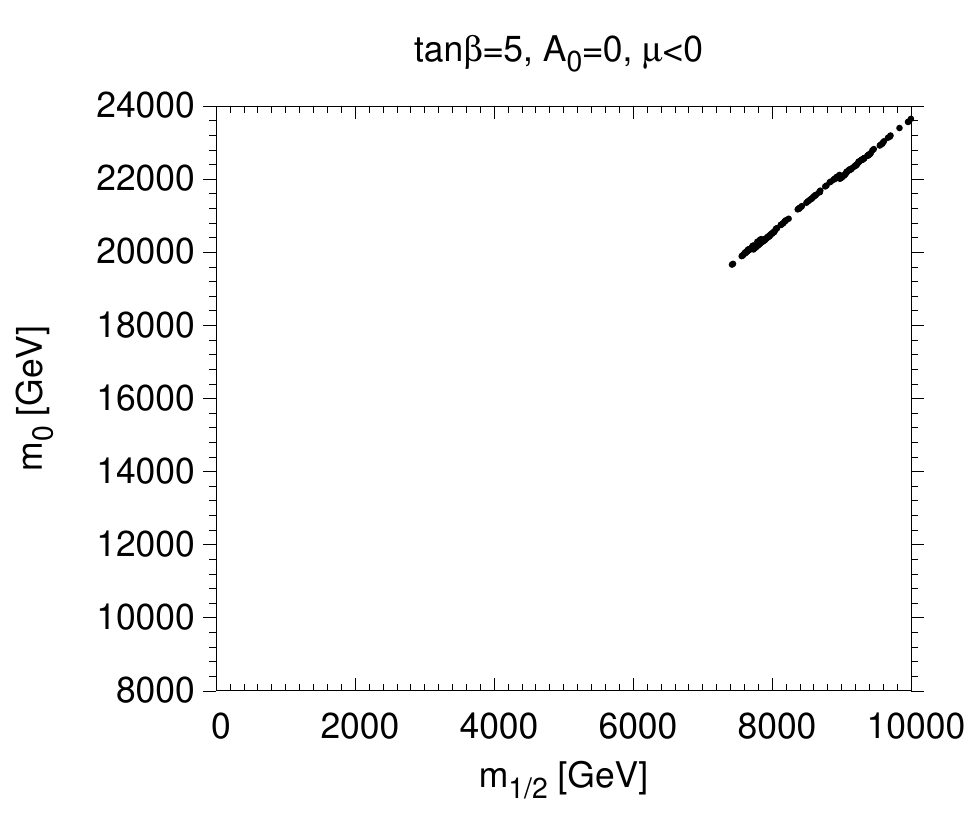} 
\includegraphics[width=0.49\textwidth]{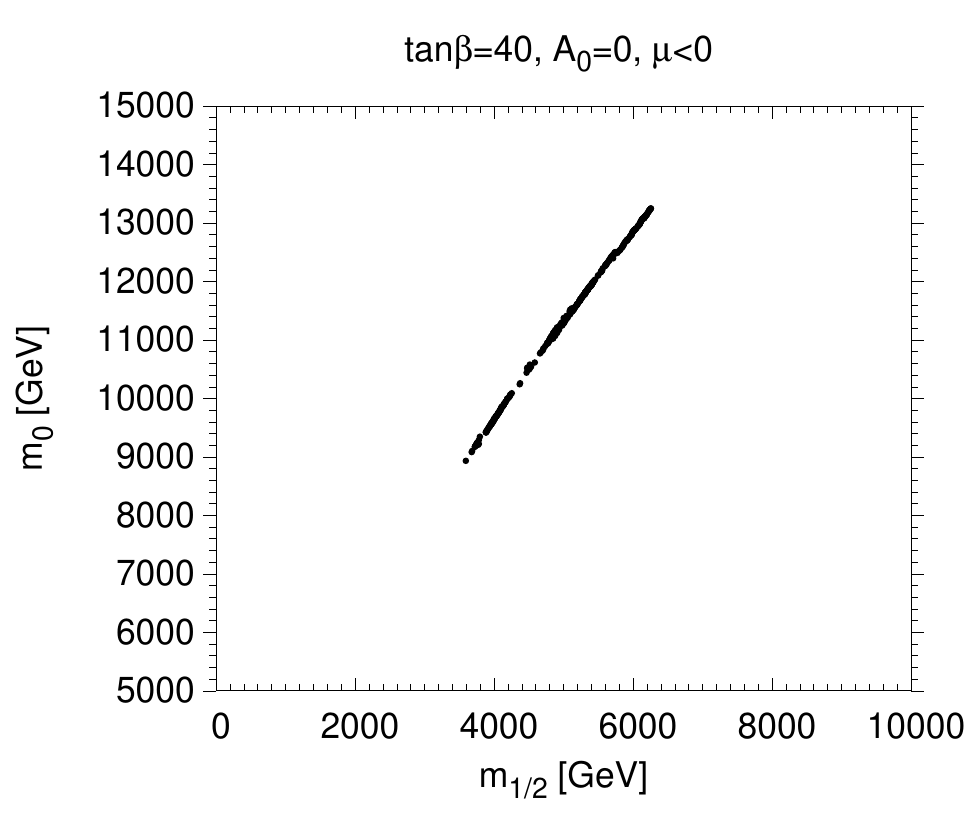} \\
\includegraphics[width=0.49\textwidth]{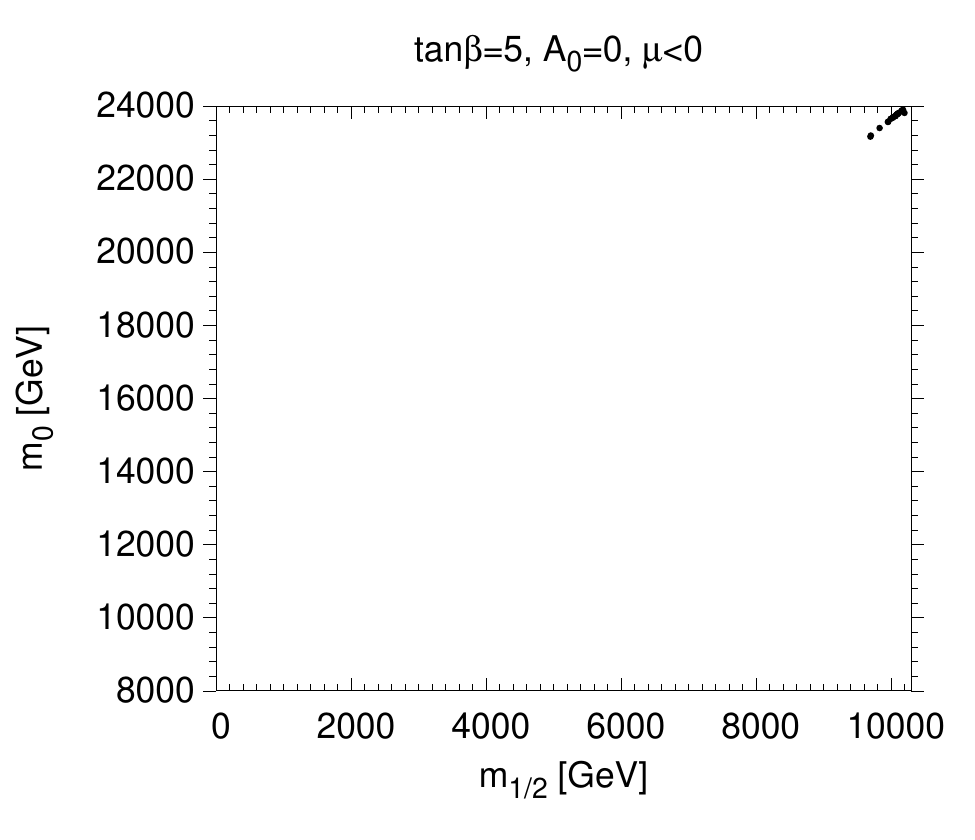} 
\includegraphics[width=0.49\textwidth]{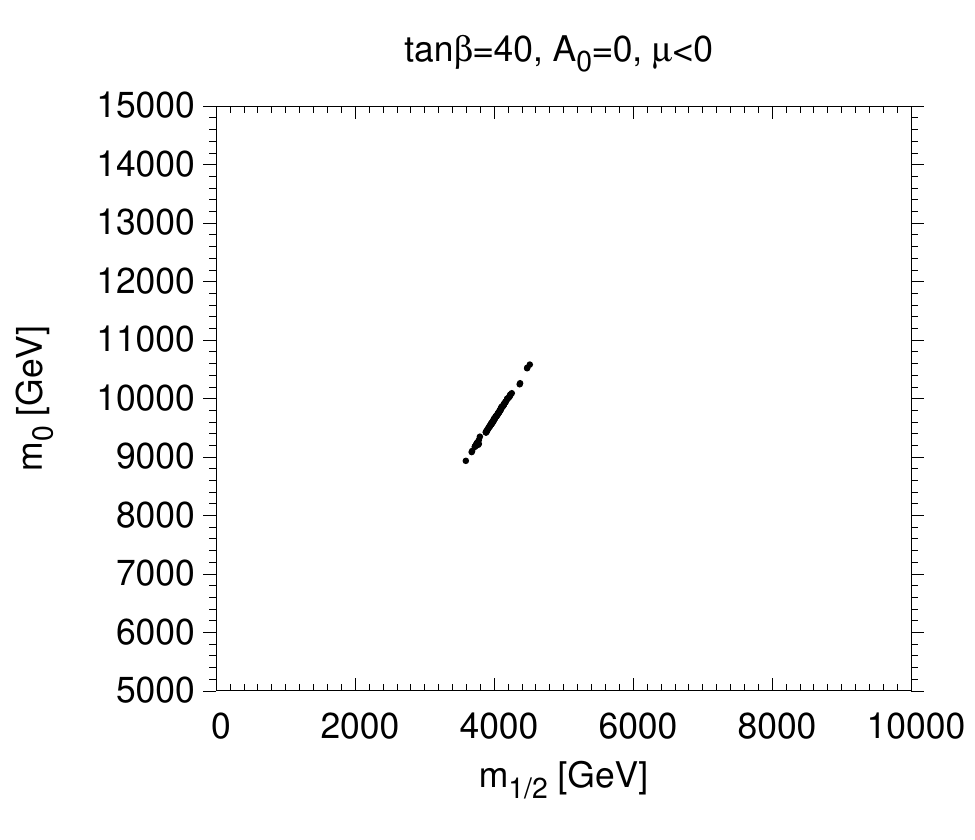} \\
\caption{\it The portions of the dark matter strips
for $\tan \beta = 5$ and $40$,
calculated assuming $A_0 = 0$ and $\mu < 0$, that are
allowed by all the constraints, assuming an uncertainty of $1.5 \ (0.5)$~GeV in the calculation of $m_h$
in the upper (lower) pair of panels.}
\label{fig:allowedmuneg}
\end{figure}

{The case $A_0 = 0$ and $\mu < 0$ is also simpler than $A_0 = 0$ and $\mu >0$.
As seen in the illustrative examples in Fig.~\ref{fig:5n} for $\tan \beta = 5$ and Fig.~\ref{fig:40n}
for $\tan \beta = 40$, there is only a single dark matter strip close to the boundary of
electroweak symmetry breaking. Overall, the results are very similar to those for $\mu > 0$
and the same values of $\tan \beta$, apart from the absence of the green
well-tempered/funnel strip that appears below the focus-point strip in the
$(m_{1/2}, m_0)$ plane when $\mu > 0$ and $\tan \beta = 40$. As for $\mu > 0$, 
$m_h$ again provides the strongest constraint for $A_0 = 0$ and $\mu < 0$, as seen
in the top right panels of Figs.~\ref{fig:5n} and Fig.~\ref{fig:40n}. The
direct search for spin-independent dark matter scattering also imposes an interesting constraint, 
as seen in the middle left panels of Figs.~\ref{fig:5n} and Fig.~\ref{fig:40n}, excluding a
range of LSP masses $\lesssim 1$~TeV that are, however, also excluded by $m_h$. The searches for
$\gamma$-rays from dSphs and the Galactic Centre are not constraining, as seen in the middle
right and bottom left panels of these figures. Finally, we note that a range of
neutralino masses $\sim 1$~TeV that are excluded by $m_h$ comes under some pressure 
from the IceCube search for energetic solar neutrinos, as seen in the bottom right panels 
of Figs.~\ref{fig:5n} and Fig.~\ref{fig:40n}.}
The portions of the dark matter strips
for $\tan \beta = 5$ and $40$
that are
allowed by all the constraints, assuming an uncertainty of $1.5 \ (0.5)$~GeV in the calculation of $m_h$, are shown
in the upper (lower) pair of panels of Fig.~\ref{fig:allowedmuneg}.

\section{Conclusions}
\label{sec:conclusion}

We have analyzed in this paper the viability of the CMSSM 
in light of the available phenomenological constraints, 
assuming that $R$-parity is conserved and the lightest 
supersymmetric particle is a neutralino that provides the
cosmological dark matter.  
The CMSSM framework assumes universality of all scalar masses, gaugino masses and $A$-terms separately at the GUT scale.
We also assume a standard thermal freeze-out scenario for the relic density. Relaxing these assumptions would correspondingly relax the constraints derived here, and enlarge the regions of parameters where supersymmetry can survive.
However, these assumptions restrict the
allowed CMSSM parameter space to narrow strips where the
relic dark matter density is compatible with the narrow
range permitted by {\it Planck} and other measurements. As we
have discussed in detail in Section~2, there are several
such strips when the ratio of Higgs vevs 
$\tan \beta \ge 40$, the trilinear supersymmetry-breaking
parameter $A_0 = 0$, and the Higgsino mixing parameter 
$\mu > 0$. On the other hand, when $\tan \beta < 40$
 or $\mu < 0$ there is a single CMSSM
dark matter strip running close to the boundary of the
region where electroweak symmetry breaking occurs. Similarly when $A_0 = 3 \, m_0$, there is a single CMSSM
dark matter strip running close to the boundary where the LSP and lighter stop are degenerate. 
As we have discussed in detail, when $A_0 = 0$ and
$\mu > 0$ several different mechanisms may come into play 
simultaneously to bring the dark matter density into the
allowed cosmological range, including focus-point and
well-tempered neutralino effects, $s$-channel annihilations 
via heavy Higgs bosons. In contrast, the dominant dark 
matter mechanism when $A_0 = 3 \, m_0$ is LSP coannihilation 
with stop squarks.

In Section~3 we have discussed the LHC constraints on 
parameter points along the CMSSM dark matter strips. We
have found that direct LHC searches for sparticles do not
constrain the CMSSM as strongly as the indirect constraint
provided by the LHC measurement of the Higgs mass and
calculations of $m_h$ using {\tt FeynHiggs~2.18.1}. These
favour relatively large values of the soft 
supersymmetry-breaking parameters.

In Section~4 we have discussed the constraints on the CMSSM
provided by searches for astrophysical dark matter. These
include the direct searches for dark matter scattering on
nuclei via spin-independent and -dependent interactions.
As we have shown, the non-detection of spin-independent
dark matter scattering imposes a stronger constraint.
It excludes significant parts of the dark matter strips, while
allowed parts are in general incompatible with the $m_h$ constraint.
We have also considered the constraints provided by searches
for $\gamma$-rays from dwarf spheroidal galaxies and the
Galactic Centre. These both exclude portions of the funnel
strips for large $\tan \beta$, $A_0 = 0$ and $\mu > 0$,
though these are also excluded by $m_h$ constraint. The
IceCube search for muons produced by energetic neutrinos
generated by the annihilations of dark matter particles
trapped inside the Sun excludes a range of CMSSM parameters
at large $\tan \beta$, $A_0 = 0$ and $\mu > 0$ that are,
however, also excluded by the $m_h$ constraint.

The overall result of our analysis is that, whilst the
searches for astrophysical dark matter do exclude portions
of the CMSSM dark matter strips, these are mostly also covered by
the $m_h$ constraint, which is the most powerful when $A_0 = 0$. 
In this case, as seen in Figs.~\ref{fig:combined}, \ref{fig:combined2}
and \ref{fig:allowedmuneg}, only very
restricted portions of the CMSSM dark matter strips are allowed
by $m_h$. The lengths of these allowed portions are sensitive
primarily to the assumed uncertainty in the calculation of
$m_h$, which is typically $\sim 0.5$~GeV when $A_0 = 0$. 
This is significantly larger than the measurement uncertainty, which is between 
0.1 and 0.2~GeV. {We emphasize that the uncertainty in the $m_h$ calculation is
much greater when $A_0 = 3 \ m_0$, $\gtrsim 3$~GeV, rendering it an
ineffective constraint for $A_0 = 3 \ m_0$, and note that the direct and indirect searches for
astrophysical dark matter are also ineffective for this value of $A_0$.}

For the moment the CMSSM survives {\it Planck}, 
the LHC, LUX-ZEPLIN, {\it Fermi}-LAT, H.E.S.S. and IceCube.
However, our  constraints push the supersymmetry breaking mass scale to $\mathcal{O}(10)$ TeV, larger than the scale originally associated with supersymmetry when proposed as a solution to the hierarchy problem.
Nevertheless, supersymmetry at this scale still alleviates the hierarchy problem to some extent.
In the future, the most immediate prospect for putting pressure on the
CMSSM may be provided by future direct searches for spin-independent
dark matter scattering, which could probe essentially all the
strips for $A_0 = 0$ and either sign of $\mu$ if their sensitivity reaches
down to the neutrino `floor'~\cite{floor}.
In the longer run, searches at FCC-hh or SppC offer good prospects for discovering supersymmetry within the CMSSM framework also when $A_0 \ne 0$~\cite{FCC}. On the other hand, the astrophysical
searches for dark matter are relatively unpromising for
$A_0 = 3 m_0$, and also have lacunae when $\mu < 0$.
The CMSSM may survive a while yet.

\newpage 
\section*{Acknowledgements}

 The work of J.E. was supported partly by the United Kingdom STFC Grant ST/T000759/1 and partly by the Estonian Research Council via a Mobilitas Pluss grant.
  The work of K.A.O. was supported in part by DOE grant DE-SC0011842 at the University of Minnesota.
The work of V.C.S.  was supported by the Hellenic Foundation for Research 
   and Innovation (H.F.R.I.) under the ``First Call for H.F.R.I. Research Projects to support 
   Faculty members and Researchers and the procurement of high-cost research equipment grant'' (Project Number: 824).


\begin{thebibliography}{99}

\bibitem{DN}
M.~Drees and M.~M.~Nojiri,
Phys.\ Rev.\ D {\bf 47} (1993) 376 [arXiv:hep-ph/9207234].

\bibitem{cmssm}
 G.~L.~Kane, C.~F.~Kolda, L.~Roszkowski and J.~D.~Wells,
  Phys.\ Rev.\  D {\bf 49} (1994) 6173
  [arXiv:hep-ph/9312272];
J.~R.~Ellis, K.~A.~Olive, Y.~Santoso and V.~C.~Spanos,
Phys.\ Lett.\ B {\bf 565} (2003) 176
[arXiv:hep-ph/0303043];
H.~Baer and C.~Balazs,
  JCAP {\bf 0305}, 006 (2003)
  [arXiv:hep-ph/0303114];
  A.~B.~Lahanas and D.~V.~Nanopoulos,
  Phys.\ Lett.\  B {\bf 568}, 55 (2003)
  [arXiv:hep-ph/0303130];
U.~Chattopadhyay, A.~Corsetti and P.~Nath,
  Phys.\ Rev.\  D {\bf 68}, 035005 (2003)
  [arXiv:hep-ph/0303201];
     J.~Ellis and K.~A.~Olive,
  arXiv:1001.3651 [astro-ph.CO], published in {\it Particle dark matter}, ed. G.~Bertone, pp. 142-163;
  J.~Ellis and K.~A.~Olive,
  Eur.\ Phys.\ J.\ C {\bf 72}, 2005 (2012)
  [arXiv:1202.3262 [hep-ph]];
  J.~Cao, Z.~Heng, D.~Li and J.~M.~Yang,
Phys. Lett. B \textbf{710}, 665-670 (2012)
[arXiv:1112.4391 [hep-ph]];
O.~Buchmueller {\it et al.},
  Eur.\ Phys.\ J.\ C {\bf 74} (2014) 3,  2809
  [arXiv:1312.5233 [hep-ph]];
  E.~Bagnaschi, H.~Bahl, J.~Ellis, J.~Evans, T.~Hahn, S.~Heinemeyer, W.~Hollik, K.~Olive, S.~Passehr, H.~Rzehak, I.~Sobolev, G.~Weiglein and J.~Zheng,
Eur. Phys. J. C \textbf{79}, no.2, 149 (2019)
[arXiv:1810.10905 [hep-ph]].


\bibitem{interplay}
J.~Ellis, F.~Luo, K.~A.~Olive and P.~Sandick,
Eur. Phys. J. C \textbf{73}, no.4, 2403 (2013)
[arXiv:1212.4476 [hep-ph]];
O.~Buchmueller, M.~Citron, J.~Ellis, S.~Guha, J.~Marrouche, K.~A.~Olive, K.~de Vries and J.~Zheng,
Eur. Phys. J. C \textbf{75}, no.10, 469 (2015)
[erratum: Eur. Phys. J. C \textbf{76}, no.4, 190 (2016)]
[arXiv:1505.04702 [hep-ph]].


\bibitem{Ellis:2015rya}
J.~Ellis, J.~L.~Evans, F.~Luo, N.~Nagata, K.~A.~Olive and P.~Sandick,
Eur. Phys. J. C \textbf{76}, no.1, 8 (2016)
[arXiv:1509.08838 [hep-ph]].

\bibitem{Ellis:2018jyl}
J.~Ellis, J.~L.~Evans, F.~Luo, K.~A.~Olive and J.~Zheng,
Eur. Phys. J. C \textbf{78}, no.5, 425 (2018)
[arXiv:1801.09855 [hep-ph]].

\bibitem{Ellis:2019fwf}
J.~Ellis, J.~L.~Evans, N.~Nagata, K.~A.~Olive and L.~Velasco-Sevilla,
Eur. Phys. J. C \textbf{80}, no.4, 332 (2020)
[arXiv:1912.04888 [hep-ph]].

 \bibitem{nosusy}
  M.~Aaboud {\it et al.} [ATLAS Collaboration],
  JHEP {\bf 1806}, 107 (2018)
  [arXiv:1711.01901 [hep-ex]];
  M.~Aaboud {\it et al.} [ATLAS Collaboration],
  Phys.\ Rev.\ D {\bf 97}, no. 11, 112001 (2018)
  [arXiv:1712.02332 [hep-ex]];
  A.~M.~Sirunyan {\it et al.} [CMS Collaboration],
  Eur.\ Phys.\ J.\ C {\bf 77}, no. 10, 710 (2017)
  [arXiv:1705.04650 [hep-ex]];
  A.~M.~Sirunyan {\it et al.} [CMS Collaboration],
  JHEP {\bf 1805}, 025 (2018)
  [arXiv:1802.02110 [hep-ex]].


\bibitem{XENON}
 E.~Aprile {\it et al.} [XENON Collaboration],
  Phys.\ Rev.\ Lett.\  {\bf 121} (2018) no.11,  111302
  [arXiv:1805.12562 [astro-ph.CO]].
  

  
  \bibitem{LUX}
  D.~S.~Akerib {\it et al.} [LUX Collaboration],
  Phys.\ Rev.\ Lett.\  {\bf 118} (2017) no.2,  021303
  [arXiv:1608.07648 [astro-ph.CO]].
  
   \bibitem{PANDAX}
Q.~Wang \textit{et al.} [PandaX-II Collaboration],
Chin. Phys. C \textbf{44} (2020) no.12, 125001
[arXiv:2007.15469 [astro-ph.CO]];
Y.~Meng \textit{et al.} [PandaX-4T Collaboration],
Phys. Rev. Lett. \textbf{127}, no.26, 261802 (2021)
[arXiv:2107.13438 [hep-ex]].



\bibitem{LZ}
J.~Aalbers \textit{et al.} [LUX-ZEPLIN Collaboration],
arXiv:2207.03764 [hep-ex].



  \bibitem{Planck}
  N.~Aghanim \textit{et al.} [Planck Collaboration],
Astron. Astrophys. \textbf{641}, A6 (2020)
[arXiv:1807.06209 [astro-ph.CO]].

\bibitem{Gelmini:2006pw}
G.~B.~Gelmini and P.~Gondolo,
Phys. Rev. D \textbf{74}, 023510 (2006)
[arXiv:hep-ph/0602230 [hep-ph]];
G.~Kane, K.~Sinha and S.~Watson,
Int. J. Mod. Phys. D \textbf{24}, no.08, 1530022 (2015)
[arXiv:1502.07746 [hep-th]].

\bibitem{Aad:2012tfa}
  G.~Aad {\it et al.} [ATLAS Collaboration],
  Phys.\ Lett.\ B {\bf 716} (2012) 1
  [arXiv:1207.7214 [hep-ex]];
  S.~Chatrchyan {\it et al.} [CMS Collaboration],
  Phys.\ Lett.\ B {\bf 716} (2012) 30
  [arXiv:1207.7235 [hep-ex]];
   G.~Aad {\it et al.} [ATLAS and CMS Collaborations],
  Phys.\ Rev.\ Lett.\  {\bf 114} (2015) 191803
  [arXiv:1503.07589 [hep-ex]].

\bibitem{FH}
H.~Bahl, T.~Hahn, S.~Heinemeyer, W.~Hollik, S.~Passehr, H.~Rzehak and G.~Weiglein,
Comput. Phys. Commun. \textbf{249}, 107099 (2020)
[arXiv:1811.09073 [hep-ph]].

\bibitem{nuhm}
J.~Ellis, K.~Olive and Y.~Santoso,
Phys.\ Lett.\  B~{\bf 539}, 107 (2002)
[arXiv:hep-ph/0204192];
J.~R.~Ellis, T.~Falk, K.~A.~Olive and Y.~Santoso,
Nucl.\ Phys.\ B {\bf 652}, 259 (2003)
[arXiv:hep-ph/0210205];
H.~Baer, A.~Mustafayev, S.~Profumo, A.~Belyaev and X.~Tata,
Phys.\ Rev.\  D {\bf 71}, 095008 (2005)[arXiv:hep-ph/0412059];
 H.~Baer, A.~Mustafayev, S.~Profumo, A.~Belyaev and X.~Tata,
 {\em JHEP} {\bf 0507} (2005) 065,
   hep-ph/0504001;
J.~R.~Ellis, K.~A.~Olive and P.~Sandick,
Phys.\ Rev.\  D {\bf 78}, 075012 (2008)
[arXiv:0805.2343 [hep-ph]].

\bibitem{Craig:2013cxa}
N.~Craig,
[arXiv:1309.0528 [hep-ph]].

\bibitem{mc12}
E.~A.~Bagnaschi, O.~Buchmueller, R.~Cavanaugh, M.~Citron, A.~De Roeck, M.~J.~Dolan, J.~R.~Ellis, H.~Fl\"acher, S.~Heinemeyer and G.~Isidori, \textit{et al.}
Eur. Phys. J. C \textbf{75}, 500 (2015)
[arXiv:1508.01173 [hep-ph]].

\bibitem{Maiani:1979cx}
L.~Maiani,
in Proceedings, Gif-sur-Yvette Summer School On Particle Physics,
  1979, 1-52;
Gerard 't~Hooft and others (eds.),
{\it Recent Developments in Gauge Theories, Proceedings of the Nato Advanced
  Study Institute, Cargese, France, August 26 - September 8, 1979},
Plenum press, New York, USA, 1980, Nato Advanced Study Institutes
  Series: Series B, Physics, 59.;
Edward Witten,
{\em Phys. Lett.} B105, 267, 1981.


\bibitem{Strigari:2012acq}
L.~E.~Strigari,
Phys. Rept. \textbf{531}, 1-88 (2013)
[arXiv:1211.7090 [astro-ph.CO]];
T.~R.~Slatyer,
SciPost Phys. Lect. Notes \textbf{53}, 1 (2022)
[arXiv:2109.02696 [hep-ph]].

\bibitem{Fermi-LAT:2013sme}
M.~Ackermann \textit{et al.} [Fermi-LAT Collaboration],
Phys. Rev. D \textbf{89} (2014), 042001
[arXiv:1310.0828 [astro-ph.HE]].



\bibitem{Fermi-LAT:2015ycq}
A.~Drlica-Wagner \textit{et al.} [Fermi-LAT and DES Collaborations],
Astrophys. J. Lett. \textbf{809} (2015) no.1, L4
[arXiv:1503.02632 [astro-ph.HE]].

\bibitem{Fermi-LAT:2015att}
M.~Ackermann \textit{et al.} [Fermi-LAT Collaboration],
Phys. Rev. Lett. \textbf{115} (2015) no.23, 231301
[arXiv:1503.02641 [astro-ph.HE]].

\bibitem{Boddy:2018qur}
K.~Boddy, J.~Kumar, D.~Marfatia and P.~Sandick,
Phys. Rev. D \textbf{97}, no.9, 095031 (2018)
[arXiv:1802.03826 [hep-ph]].

\bibitem{Hoof:2018hyn}
S.~Hoof, A.~Geringer-Sameth and R.~Trotta,
JCAP \textbf{02}, 012 (2020)
[arXiv:1812.06986 [astro-ph.CO]].

\bibitem{Alvarez:2020cmw}
A.~Alvarez, F.~Calore, A.~Genina, J.~Read, P.~D.~Serpico and B.~Zaldivar,
JCAP \textbf{09}, 004 (2020)
[arXiv:2002.01229 [astro-ph.HE]].




\bibitem{Ellis:2011du}
J.~Ellis, K.~A.~Olive and V.~C.~Spanos,
JCAP \textbf{10} (2011), 024
[arXiv:1106.0768 [hep-ph]].



\bibitem{HESS:2022ygk}
H.~Abdalla \textit{et al.} [H.E.S.S. Collaboration],
Phys. Rev. Lett. \textbf{129} (2022) no.11, 111101
[arXiv:2207.10471 [astro-ph.HE]];
H.~Abdallah \textit{et al.} [HESS],
Phys. Rev. Lett. \textbf{120}, no.20, 201101 (2018)
[arXiv:1805.05741 [astro-ph.HE]].

\bibitem{Johnson:2019hsm}
C.~Johnson, R.~Caputo, C.~Karwin, S.~Murgia, S.~Ritz and J.~Shelton,
Phys. Rev. D \textbf{99}, no.10, 103007 (2019)
[arXiv:1904.06261 [astro-ph.HE]].

\bibitem{Abazajian:2020tww}
K.~N.~Abazajian, S.~Horiuchi, M.~Kaplinghat, R.~E.~Keeley and O.~Macias,
Phys. Rev. D \textbf{102}, no.4, 043012 (2020)
[arXiv:2003.10416 [hep-ph]].

\bibitem{indirectdet:solar}
  J.~Silk, K.~A.~Olive and M.~Srednicki,
  Phys.\ Rev.\ Lett.\  {\bf 55}, 257 (1985);
  M.~Srednicki, K.~A.~Olive and J.~Silk,
  Nucl.\ Phys.\  B {\bf 279}, 804 (1987);
  J.~S.~Hagelin, K.~W.~Ng and K.~A.~Olive,
  Phys.\ Lett.\  B {\bf 180}, 375 (1986);
  K.~W.~Ng, K.~A.~Olive and M.~Srednicki,
  Phys.\ Lett.\  B {\bf 188}, 138 (1987);
  T.~K.~Gaisser, G.~Steigman and S.~Tilav,
  Phys.\ Rev.\  D {\bf 34}, 2206 (1986);
  F.~Halzen, T.~Stelzer and M.~Kamionkowski,
  Phys.\ Rev.\  D {\bf 45}, 4439 (1992);
  L.~Bergstrom, J.~Edsjo and P.~Gondolo,
  Phys.\ Rev.\  D {\bf 55}, 1765 (1997)
  [arXiv:hep-ph/9607237];
N.~Fornengo, A.~Masiero, F.~S.~Queiroz and C.~E.~Yaguna,
JCAP \textbf{12}, 012 (2017)
[arXiv:1710.02155 [hep-ph]].

\bibitem{IceCube:2016yoy}
M.~G.~Aartsen \textit{et al.} [IceCube Collaboration],
JCAP \textbf{04}, 022 (2016)
[arXiv:1601.00653 [hep-ph]].

\bibitem{IceCube:2016dgk}
M.~G.~Aartsen \textit{et al.} [IceCube Collaboration],
Eur. Phys. J. C \textbf{77}, no.3, 146 (2017)
[erratum: Eur. Phys. J. C \textbf{79}, no.3, 214 (2019)]
[arXiv:1612.05949 [astro-ph.HE]].

\bibitem{IceCube:2021xzo}
R.~Abbasi \textit{et al.} [IceCube Collaboration],
Phys. Rev. D \textbf{105}, no.6, 062004 (2022)
[arXiv:2111.09970 [astro-ph.HE]].

\bibitem{fp}
  J.~L.~Feng, K.~T.~Matchev and T.~Moroi,
  Phys.\ Rev.\ Lett.\  {\bf 84}, 2322 (2000)
  [arXiv:hep-ph/9908309];
  H.~Baer, T.~Krupovnickas, S.~Profumo and P.~Ullio,
  JHEP {\bf 0510} (2005) 020
  [hep-ph/0507282];
J.~L.~Feng, K.~T.~Matchev and D.~Sanford,
  Phys.\ Rev.\ D {\bf 85}, 075007 (2012)
  [arXiv:1112.3021 [hep-ph]];
  P.~Draper, J.~Feng, P.~Kant, S.~Profumo and D.~Sanford,
  Phys.\ Rev.\ D {\bf 88}, 015025 (2013)
  [arXiv:1304.1159 [hep-ph]].


\bibitem{floor}
   J.~Billard, L.~Strigari and E.~Figueroa-Feliciano,
  Phys.\ Rev.\ D {\bf 89}, no. 2, 023524 (2014)
  [arXiv:1307.5458 [hep-ph]];
P.~Cushman, C.~Galbiati, D.~N.~McKinsey, H.~Robertson, T.~M.~P.~Tait, D.~Bauer, A.~Borgland and B.~Cabrera {\it et al.},
{\it Snowmass Working Group Report: WIMP Dark Matter Direct Detection},
  arXiv:1310.8327 [hep-ex];
L.~E.~Strigari,
Phys. Rev. D \textbf{93}, no.10, 103534 (2016)
[arXiv:1604.00729 [astro-ph.CO]].
  

\bibitem{Theory}
T.~Aoyama, \textit{et al.}
Phys. Rept. \textbf{887} (2020), 1-166
[arXiv:2006.04822 [hep-ph]];
M.~Davier, A.~Hoecker, B.~Malaescu and Z.~Zhang,
Eur. Phys. J. C \textbf{71} (2011), 1515
[erratum: Eur. Phys. J. C \textbf{72} (2012), 1874]
[arXiv:1010.4180 [hep-ph]];
A.~Kurz, T.~Liu, P.~Marquard and M.~Steinhauser,
Phys. Lett. B \textbf{734}, 144-147 (2014)
[arXiv:1403.6400 [hep-ph]];
G.~Colangelo, M.~Hoferichter and P.~Stoffer,
JHEP \textbf{02}, 006 (2019)
[arXiv:1810.00007 [hep-ph]];
M.~Hoferichter, B.~L.~Hoid and B.~Kubis,
JHEP \textbf{08}, 137 (2019)
[arXiv:1907.01556 [hep-ph]];
M.~Davier, A.~Hoecker, B.~Malaescu and Z.~Zhang,
Eur. Phys. J. C \textbf{80} (2020) no.3, 241
[erratum: Eur. Phys. J. C \textbf{80} (2020) no.5, 410]
[arXiv:1908.00921 [hep-ph]];
A.~Keshavarzi, D.~Nomura and T.~Teubner,
Phys. Rev. D \textbf{101}, no.1, 014029 (2020)
[arXiv:1911.00367 [hep-ph]].


\bibitem{BNL1}
H.~N.~Brown \textit{et al.} [Muon g-2 Collaboration],
Phys. Rev. Lett. \textbf{86} (2001), 2227-2231
[arXiv:hep-ex/0102017 [hep-ex]].

\bibitem{BNL2}
G.~W.~Bennett \textit{et al.} [Muon g-2 Collaboration],
Phys. Rev. D \textbf{73} (2006), 072003
[arXiv:hep-ex/0602035 [hep-ex]].

\bibitem{FNAL}
B. Abi \textit{et al.} [Muon g-2 Collaboration]
Phys.~Rev.~Lett. \textbf{126} (2021), 141801
[arXiv:2104.03281 [hep-ex]].


\bibitem{gm2}
L.~L.~Everett, G.~L.~Kane, S.~Rigolin and L.~Wang,
Phys.\ Rev.\ Lett.\  {\bf 86} (2001) 3484 
[arXiv:hep-ph/0102145];
J.~L.~Feng and K.~T.~Matchev,
Phys.\ Rev.\ Lett.\  {\bf 86} (2001) 3480 
[arXiv:hep-ph/0102146];
E.~A.~Baltz and P.~Gondolo,   
Phys.\ Rev.\ Lett.\  {\bf 86} (2001) 5004 
[arXiv:hep-ph/0102147];
U.~Chattopadhyay and P.~Nath,
Phys.\ Rev.\ Lett.\  {\bf 86} (2001) 5854 
[arXiv:hep-ph/0102157];
S.~Komine, T.~Moroi and M.~Yamaguchi,
Phys.\ Lett.\ B {\bf 506} (2001) 93 
[arXiv:hep-ph/0102204];
J.~Hisano and K.~Tobe,
Phys. Lett. B \textbf{510}, 197-204 (2001)
[arXiv:hep-ph/0102315 [hep-ph]];
J.~R.~Ellis, D.~V.~Nanopoulos and K.~A.~Olive,
Phys. Lett. B \textbf{508} (2001), 65-73
[arXiv:hep-ph/0102331 [hep-ph]];
R.~Arnowitt, B.~Dutta, B.~Hu and Y.~Santoso,
Phys.\ Lett.\ B {\bf 505} (2001) 177  
[arXiv:hep-ph/0102344]
S.~P.~Martin and J.~D.~Wells,
Phys.\ Rev.\ D {\bf 64} (2001) 035003 
[arXiv:hep-ph/0103067];
H.~Baer, C.~Balazs, J.~Ferrandis and X.~Tata,
Phys.\ Rev.\ D {\bf 64} (2001) 035004 
[arXiv:hep-ph/0103280].

\bibitem{otherCMSSM}
M.~Chakraborti, S.~Heinemeyer and I.~Saha,
Eur. Phys. J. C \textbf{81}, no.12, 1114 (2021)
[arXiv:2104.03287 [hep-ph]];
P.~Cox, C.~Han and T.~T.~Yanagida,
Phys. Rev. D \textbf{104}, no.7, 075035 (2021)
[arXiv:2104.03290 [hep-ph]];
P.~Athron, C.~Bal\'azs, D.~H.~Jacob, W.~Kotlarski, D.~St\"ockinger and H.~St\"ockinger-Kim,
[arXiv:2104.03691 [hep-ph]];
F.~Wang, L.~Wu, Y.~Xiao, J.~M.~Yang and Y.~Zhang,
arXiv:2104.03262 [hep-ph];
M.~Chakraborti, L.~Roszkowski and S.~Trojanowski,
JHEP \textbf{05} (2021), 252
[arXiv:2104.04458 [hep-ph]];
J.~Ellis, J.~L.~Evans, N.~Nagata, D.~V.~Nanopoulos and K.~A.~Olive,
Eur. Phys. J. C \textbf{81}, no.12, 1079 (2021)
[arXiv:2107.03025 [hep-ph]];
J.~Ellis, J.~L.~Evans, N.~Nagata, D.~V.~Nanopoulos and K.~A.~Olive,
Eur. Phys. J. C \textbf{81}, no.12, 1109 (2021)
[arXiv:2110.06833 [hep-ph]].

\bibitem{bmw}
 S.~Borsanyi, 
 \textit{et al.}
Nature \textbf{593}, no.7857, 51-55 (2021)
[arXiv:2002.12347 [hep-lat]].

 \bibitem{gs}
    K.~Griest and D.~Seckel,
  Phys.\ Rev.\ D {\bf 43}, 3191 (1991).


\bibitem{stauco}
J. Ellis, T. Falk, and K.A. Olive, Phys. Lett.  {\bf B444} (1998) 367
[arXiv:hep-ph/9810360];
J. Ellis, T. Falk, K.A. Olive, and M. Srednicki, {\it Astr. Part. Phys.}
{\bf 13} (2000) 181
[Erratum-ibid.\  {\bf 15} (2001) 413]
[arXiv:hep-ph/9905481];
R.~Arnowitt, B.~Dutta and Y.~Santoso,
Nucl.\ Phys.\ B {\bf 606} (2001) 59
[arXiv:hep-ph/0102181];
M.~E.~G\'omez, G.~Lazarides and C.~Pallis,
Phys. Rev. D {\bf D61} (2000) 123512
[arXiv:hep-ph/9907261];
  Phys.\ Lett. {\bf B487} (2000) 313
[arXiv:hep-ph/0004028];
  Nucl. Phys. B {\bf B638} (2002) 165
[arXiv:hep-ph/0203131];
T.~Nihei, L.~Roszkowski and R.~Ruiz de Austri,
  JHEP {\bf 0207} (2002) 024
[arXiv:hep-ph/0206266];
 M.~Citron, J.~Ellis, F.~Luo, J.~Marrouche, K.~A.~Olive and K.~J.~de Vries,
  Phys.\ Rev.\ D {\bf 87} (2013) no.3,  036012
  [arXiv:1212.2886 [hep-ph]];
  N.~Desai, J.~Ellis, F.~Luo and J.~Marrouche,
  Phys.\ Rev.\ D {\bf 90}, no. 5, 055031 (2014)
  [arXiv:1404.5061 [hep-ph]].
 



\bibitem{stopco}
 C.~Boehm, A.~Djouadi and M.~Drees,
  Phys.\ Rev.\  D {\bf 62}, 035012 (2000)
  [arXiv:hep-ph/9911496];
  J.~R.~Ellis, K.~A.~Olive and Y.~Santoso,
  Astropart.\ Phys.\  {\bf 18}, 395 (2003)
  [arXiv:hep-ph/0112113];
       J.~L.~Diaz-Cruz, J.~R.~Ellis, K.~A.~Olive and Y.~Santoso,
  JHEP {\bf 0705}, 003 (2007)
  [arXiv:hep-ph/0701229];
   M.~A.~Ajaib, T.~Li and Q.~Shafi,
  Phys.\ Rev.\ D {\bf 85}, 055021 (2012)
  [arXiv:1111.4467 [hep-ph]];
  J.~Harz, B.~Herrmann, M.~Klasen, K.~Kovarik and Q.~L.~Boulc'h,
  Phys.\ Rev.\ D {\bf 87} (2013) 5,  054031
  [arXiv:1212.5241];
   J.~Ellis, K.~A.~Olive and J.~Zheng,
  Eur.\ Phys.\ J.\ C {\bf 74} (2014) 2947
  [arXiv:1404.5571 [hep-ph]];
  S.~Raza, Q.~Shafi and C.~S.~\"Un,
  Phys.\ Rev.\ D {\bf 92}, no. 5, 055010 (2015)
  [arXiv:1412.7672 [hep-ph]];
A.~Ibarra, A.~Pierce, N.~R.~Shah and S.~Vogl,
  Phys.\ Rev.\ D {\bf 91}, no. 9, 095018 (2015)
  [arXiv:1501.03164 [hep-ph]].

\bibitem{chaco}
  S.~Mizuta and M.~Yamaguchi,
Phys.\ Lett.\ B {\bf 298} (1993) 120
[arXiv:hep-ph/9208251];
  J.~Edsjo and P.~Gondolo,
  Phys.\ Rev.\ D {\bf 56}, 1879 (1997)
  [hep-ph/9704361];
   H.~Baer, C.~Balazs and A.~Belyaev,
  JHEP {\bf 0203}, 042 (2002)
  [hep-ph/0202076];
    A.~Birkedal-Hansen and E.~h.~Jeong,
  JHEP {\bf 0302}, 047 (2003)
  [hep-ph/0210041].

  \bibitem{esug}
 J.~Edsjo, M.~Schelke, P.~Ullio and P.~Gondolo,
  JCAP {\bf 0304}, 001 (2003)
  [hep-ph/0301106].


 
\bibitem{funnel}
H.~Baer and M.~Brhlik,
Phys.\ Rev.\ D {\bf 53} (1996) 597 [arXiv:hep-ph/9508321];
  Phys.\ Rev.\  D {\bf 57} (1998) 567
  [arXiv:hep-ph/9706509];
  H.~Baer, M.~Brhlik, M.~A.~Diaz, J.~Ferrandis, P.~Mercadante, P.~Quintana and X.~Tata,
    Phys.\ Rev.\  D {\bf 63} (2001) 015007
  [arXiv:hep-ph/0005027];
  J.~R.~Ellis, T.~Falk, G.~Ganis, K.~A.~Olive and M.~Srednicki,
Phys.\ Lett.\ B {\bf 510} (2001) 236
[arXiv:hep-ph/0102098].

 \bibitem{anom}
   M.~Dine and D.~MacIntire,
  Phys.\ Rev.\ D {\bf 46}, 2594 (1992)
  [hep-ph/9205227];
   L.~Randall and R.~Sundrum,
  Nucl.\ Phys.\  B {\bf 557}, 79 (1999)
  [arXiv:hep-th/9810155];
   G.~F.~Giudice, M.~A.~Luty, H.~Murayama and R.~Rattazzi,
  JHEP {\bf 9812}, 027 (1998)
  [arXiv:hep-ph/9810442];
    A.~Pomarol and R.~Rattazzi,
  JHEP {\bf 9905}, 013 (1999)
  [hep-ph/9903448];
    J.~A.~Bagger, T.~Moroi and E.~Poppitz,
  JHEP {\bf 0004}, 009 (2000)
  [arXiv:hep-th/9911029];
  P.~Binetruy, M.~K.~Gaillard and B.~D.~Nelson,
  Nucl.\ Phys.\  B {\bf 604}, 32 (2001)
  [arXiv:hep-ph/0011081].


\bibitem{mcmc}
N.~Metropolis, A.~W.~Rosenbluth, M.~N.~Rosenbluth, A.~H.~Teller and E.~Teller,
J. Chem. Phys. \textbf{21} (1953), 1087-1092
doi:10.1063/1.1699114;
W.~K.~Hastings,
Biometrika \textbf{57} (1970), 97-109
doi:10.1093/biomet/57.1.97.



\bibitem{wt}
N.~Arkani-Hamed, A.~Delgado and G.~F.~Giudice,
Nucl. Phys. B \textbf{741}, 108-130 (2006)
[arXiv:hep-ph/0601041 [hep-ph]];
H.~Baer, A.~Mustafayev, E.~K.~Park and X.~Tata,
JCAP \textbf{01}, 017 (2007)
[arXiv:hep-ph/0611387 [hep-ph]].

\bibitem{PDG}
R.~L.~Workman [Particle Data Group],
PTEP \textbf{2022}, 083C01 (2022)

\bibitem{DEAP}
R.~Ajaj \textit{et al.} [DEAP Collaboration],
Phys. Rev. D \textbf{100}, no.2, 022004 (2019)
[arXiv:1902.04048 [astro-ph.CO]].

\bibitem{sospin}
J.~Ellis, N.~Nagata and K.~A.~Olive,
Eur. Phys. J. C \textbf{78}, no.7, 569 (2018)
[arXiv:1805.09795 [hep-ph]];
J.~R.~Ellis, K.~A.~Olive and C.~Savage,
Phys. Rev. D \textbf{77}, 065026 (2008)
[arXiv:0801.3656 [hep-ph]].

\bibitem{PICO}
C.~Amole \textit{et al.} [PICO Collaboration],
Phys. Rev. D \textbf{100} (2019) no.2, 022001
[arXiv:1902.04031 [astro-ph.CO]].

\bibitem{MAGIC:2016xys}
M.~L.~Ahnen \textit{et al.} [MAGIC and {\it Fermi}-LAT Collaborations],
JCAP \textbf{02}, 039 (2016)
[arXiv:1601.06590 [astro-ph.HE]].

\bibitem{MAGIC:2017avy}
M.~L.~Ahnen \textit{et al.} [MAGIC Collaboration],
JCAP \textbf{03}, 009 (2018)
[arXiv:1712.03095 [astro-ph.HE]].

\bibitem{HAWC:2017mfa}
A.~Albert \textit{et al.} [HAWC Collaboration],
Astrophys. J. \textbf{853}, no.2, 154 (2018)
[arXiv:1706.01277 [astro-ph.HE]].

\bibitem{HAWC:2019jvm}
A.~Albert \textit{et al.} [HAWC Collaboration],
Phys. Rev. D \textbf{101}, no.10, 103001 (2020)
[arXiv:1912.05632 [astro-ph.HE]].

\bibitem{HESS:2020zwn}
H.~Abdallah \textit{et al.} [H.E.S.S. Collaboration],
Phys. Rev. D \textbf{102}, no.6, 062001 (2020)
[arXiv:2008.00688 [astro-ph.HE]].



\bibitem{HESS:2018cbt}
H.~Abdallah \textit{et al.} [H.E.S.S. Collaboration],
Phys. Rev. Lett. \textbf{120}, no.20, 201101 (2018)
[arXiv:1805.05741 [astro-ph.HE]].

\bibitem{Super-Kamiokande:2015xms}
K.~Choi \textit{et al.} [Super-Kamiokande Collaboration],
Phys. Rev. Lett. \textbf{114}, no.14, 141301 (2015)
[arXiv:1503.04858 [hep-ex]].

\bibitem{ANTARES:2016xuh}
S.~Adrian-Martinez \textit{et al.} [ANTARES Collaboration],
Phys. Lett. B \textbf{759}, 69-74 (2016)
[arXiv:1603.02228 [astro-ph.HE]].

\bibitem{HAWC:2018szf}
A.~Albert \textit{et al.} [HAWC Collaboration],
Phys. Rev. D \textbf{98}, 123012 (2018)
[arXiv:1808.05624 [hep-ph]].

\bibitem{pythia}
T.~Sjostrand, S.~Mrenna and P.~Z.~Skands,
JHEP \textbf{05} (2006), 026
doi:10.1088/1126-6708/2006/05/026
[arXiv:hep-ph/0603175 [hep-ph]];
T.~Sj\"ostrand, S.~Ask, J.~R.~Christiansen, R.~Corke, N.~Desai, P.~Ilten, S.~Mrenna, S.~Prestel, C.~O.~Rasmussen and P.~Z.~Skands,
Comput. Phys. Commun. \textbf{191} (2015), 159-177
doi:10.1016/j.cpc.2015.01.024
[arXiv:1410.3012 [hep-ph]].


\bibitem{Bergstrom:1997fj}
L.~Bergstrom, P.~Ullio and J.~H.~Buckley,
Astropart. Phys. \textbf{9}, 137-162 (1998)
[arXiv:astro-ph/9712318 [astro-ph]];
P.~Ullio, L.~Bergstrom, J.~Edsjo and C.~G.~Lacey,
Phys. Rev. D \textbf{66}, 123502 (2002)
[arXiv:astro-ph/0207125 [astro-ph]];
Y.~Mambrini and C.~Munoz,
Astropart. Phys. \textbf{24}, 208-230 (2005)
[arXiv:hep-ph/0407158 [hep-ph]].






\bibitem{Navarro:1996gj}
J.~F.~Navarro, C.~S.~Frenk and S.~D.~M.~White,
Astrophys. J. \textbf{490} (1997), 493-508
doi:10.1086/304888
[arXiv:astro-ph/9611107 [astro-ph]].



\bibitem{Burkert:1995yz}
A.~Burkert,
Astrophys. J. Lett. \textbf{447} (1995), L25
doi:10.1086/309560
[arXiv:astro-ph/9504041 [astro-ph]].

\bibitem{Dessert:2022evk}
C.~Dessert, J.~W.~Foster, Y.~Park, B.~R.~Safdi and W.~L.~Xu,
arXiv:2207.10090 [hep-ph].

\bibitem{CTA}
A.~Acharyya \textit{et al.} [CTA Collaboration],
JCAP \textbf{01} (2021), 057
doi:10.1088/1475-7516/2021/01/057
[arXiv:2007.16129 [astro-ph.HE]].


\bibitem{Ellis:2009ka}
J.~Ellis, K.~A.~Olive, C.~Savage and V.~C.~Spanos,
Phys. Rev. D \textbf{81} (2010), 085004
[arXiv:0912.3137 [hep-ph]];
J.~Ellis, K.~A.~Olive, C.~Savage and V.~C.~Spanos,
Phys. Rev. D \textbf{83}, 085023 (2011)
[arXiv:1102.1988 [hep-ph]].




\bibitem{IceCube:2012fvn}
P.~Scott \textit{et al.} [IceCube Collaboration],
JCAP \textbf{11} (2012), 057
doi:10.1088/1475-7516/2012/11/057
[arXiv:1207.0810 [hep-ph]].




\bibitem{IceCube:2012ugg}
M.~G.~Aartsen \textit{et al.} [IceCube Collaboration],
Phys. Rev. Lett. \textbf{110} (2013) no.13, 131302
doi:10.1103/PhysRevLett.110.131302
[arXiv:1212.4097 [astro-ph.HE]].


\bibitem{stopbound}
ATLAS Collaboration, {\tt https://twiki.cern.ch/twiki/bin/view/AtlasPublic/} {\tt SupersymmetryPublicResults}.

\bibitem{FCC}
A.~Abada \textit{et al.} [FCC Collaboration],
Eur. Phys. J. C \textbf{79} (2019) no.6, 474
doi:10.1140/epjc/s10052-019-6904-3.




\end{thebibliography}
\end{document}